\documentclass[10pt]{article}
\usepackage{amsmath,amsthm,latexsym,amssymb,amsfonts,epsfig, psfrag,color}
\usepackage{graphicx}
\usepackage{caption}
\usepackage{url}
\usepackage{subcaption}
\usepackage{sidecap}
\usepackage{wrapfig}
\usepackage{multirow}
\usepackage[small,nohug,heads=vee]{diagrams}
\diagramstyle[labelstyle=\scriptstyle]

\makeatletter \@addtoreset{equation}{section} \makeatother

\newtheorem{Theorem}{Theorem}[section]

\newtheorem{lem}{Lemma}[section]

\newtheorem{Example}{Example}[section]

\newtheorem*{R}{Remark}

\def\be{\begin{equation}}
\def\ee{\end{equation}}
\def\ba{\begin{eqnarray}}
\def\ea{\end{eqnarray}}

\newcommand\nn{\nonumber}
\newcommand\q{\quad}
\def\Nl{{\mathchoice
{\setbox0=\hbox{$\displaystyle\rm N$}\hbox{\hbox to0pt
{\kern0.4\wd0\vrule height0.9\ht0\hss}\box0}}
{\setbox0=\hbox{$\textstyle\rm N$}\hbox{\hbox to0pt
{\kern0.4\wd0\vrule height0.9\ht0\hss}\box0}}
{\setbox0=\hbox{$\scriptstyle\rm N$}\hbox{\hbox to0pt
{\kern0.4\wd0\vrule height0.9\ht0\hss}\box0}}
{\setbox0=\hbox{$\scriptscriptstyle\rm N$}\hbox{\hbox to0pt
{\kern0.4\wd0\vrule height0.9\ht0\hss}\box0}}}}
\def\Zl{{\mathchoice
{\setbox0=\hbox{$\displaystyle\rm Z$}\hbox{\hbox to0pt
{\kern0.4\wd0\vrule height0.9\ht0\hss}\box0}}
{\setbox0=\hbox{$\textstyle\rm Z$}\hbox{\hbox to0pt
{\kern0.4\wd0\vrule height0.9\ht0\hss}\box0}}
{\setbox0=\hbox{$\scriptstyle\rm Z$}\hbox{\hbox to0pt
{\kern0.4\wd0\vrule height0.9\ht0\hss}\box0}}
{\setbox0=\hbox{$\scriptscriptstyle\rm Z$}\hbox{\hbox to0pt
{\kern0.4\wd0\vrule height0.9\ht0\hss}\box0}}}}
\def\Ql{{\mathchoice
{\setbox0=\hbox{$\displaystyle\rm Q$}\hbox{\hbox to0pt
{\kern0.4\wd0\vrule height0.9\ht0\hss}\box0}}
{\setbox0=\hbox{$\textstyle\rm Q$}\hbox{\hbox to0pt
{\kern0.4\wd0\vrule height0.9\ht0\hss}\box0}}
{\setbox0=\hbox{$\scriptstyle\rm Q$}\hbox{\hbox to0pt
{\kern0.4\wd0\vrule height0.9\ht0\hss}\box0}}
{\setbox0=\hbox{$\scriptscriptstyle\rm Q$}\hbox{\hbox to0pt
{\kern0.4\wd0\vrule height0.9\ht0\hss}\box0}}}}
\def\Rl{{\mathchoice
{\setbox0=\hbox{$\displaystyle\rm R$}\hbox{\hbox to0pt
{\kern0.4\wd0\vrule height0.9\ht0\hss}\box0}}
{\setbox0=\hbox{$\textstyle\rm R$}\hbox{\hbox to0pt
{\kern0.4\wd0\vrule height0.9\ht0\hss}\box0}}
{\setbox0=\hbox{$\scriptstyle\rm R$}\hbox{\hbox to0pt
{\kern0.4\wd0\vrule height0.9\ht0\hss}\box0}}
{\setbox0=\hbox{$\scriptscriptstyle\rm R$}\hbox{\hbox to0pt
{\kern0.4\wd0\vrule height0.9\ht0\hss}\box0}}}}
\def\Cl{{\mathchoice
{\setbox0=\hbox{$\displaystyle\rm C$}\hbox{\hbox to0pt
{\kern0.4\wd0\vrule height0.9\ht0\hss}\box0}}
{\setbox0=\hbox{$\textstyle\rm C$}\hbox{\hbox to0pt
{\kern0.4\wd0\vrule height0.9\ht0\hss}\box0}}
{\setbox0=\hbox{$\scriptstyle\rm C$}\hbox{\hbox to0pt
{\kern0.4\wd0\vrule height0.9\ht0\hss}\box0}}
{\setbox0=\hbox{$\scriptscriptstyle\rm C$}\hbox{\hbox to0pt
{\kern0.4\wd0\vrule height0.9\ht0\hss}\box0}}}}
\def\Hl{{\mathchoice
{\setbox0=\hbox{$\displaystyle\rm H$}\hbox{\hbox to0pt
{\kern0.4\wd0\vrule height0.9\ht0\hss}\box0}}
{\setbox0=\hbox{$\textstyle\rm H$}\hbox{\hbox to0pt
{\kern0.4\wd0\vrule height0.9\ht0\hss}\box0}}
{\setbox0=\hbox{$\scriptstyle\rm H$}\hbox{\hbox to0pt
{\kern0.4\wd0\vrule height0.9\ht0\hss}\box0}}
{\setbox0=\hbox{$\scriptscriptstyle\rm H$}\hbox{\hbox to0pt
{\kern0.4\wd0\vrule height0.9\ht0\hss}\box0}}}}
\def\Ol{{\mathchoice
{\setbox0=\hbox{$\displaystyle\rm O$}\hbox{\hbox to0pt
{\kern0.4\wd0\vrule height0.9\ht0\hss}\box0}}
{\setbox0=\hbox{$\textstyle\rm O$}\hbox{\hbox to0pt
{\kern0.4\wd0\vrule height0.9\ht0\hss}\box0}}
{\setbox0=\hbox{$\scriptstyle\rm O$}\hbox{\hbox to0pt
{\kern0.4\wd0\vrule height0.9\ht0\hss}\box0}}
{\setbox0=\hbox{$\scriptscriptstyle\rm O$}\hbox{\hbox to0pt
{\kern0.4\wd0\vrule height0.9\ht0\hss}\box0}}}}
\newcommand{\cc}{\mathcal C}

\newcommand{\cg}{\mathcal G}
\newcommand{\ch}{\mathcal H}

\newcommand{\cp}{\mathcal P}
\newcommand{\cq}{\mathcal Q}

\newcommand{\fh}{\mathfrak{h}}

\newcommand{\fu}{\mathfrak{u}}

\def\nn{\nonumber}

\newcommand{\eqa}{\begin{eqnarray}}
\newcommand{\neqa}{\end{eqnarray}}

\newcommand{\p}{\partial}

\def\f{\frac}

\usepackage{bbm}

\def\q{{\quad}}

\begin{document}

{\renewcommand{\thefootnote}{\fnsymbol{footnote}}

\title{Quantization of systems with temporally varying discretization II:\\
 Local evolution moves}
\author{Philipp A H\"ohn\footnote{e-mail address: {\tt phoehn@perimeterinstitute.ca}}\\
 \small Perimeter Institute for Theoretical Physics,\\
 \small 31 Caroline Street North, Waterloo, Ontario, Canada N2L 2Y5
}

\date{}

}

\setcounter{footnote}{0}
\maketitle

{\abstract{Several quantum gravity approaches and field theory on an evolving lattice involve a discretization changing dynamics generated by evolution moves. Local evolution moves in variational discrete systems (1) are a generalization of the Pachner evolution moves of simplicial gravity models, (2) update only a small subset of the dynamical data, (3) change the number of kinematical and physical degrees of freedom, and (4) generate a dynamical (or canonical) coarse graining or refining of the underlying discretization. To systematically explore such local moves and their implications in the quantum theory, this article suitably expands the quantum formalism for global evolution moves, constructed in the companion paper \cite{Hoehn:2014fka}, by employing that global moves can be decomposed into sequences of local moves. This formalism is spelled out for systems with Euclidean configuration spaces. Various types of local moves, the different kinds of constraints generated by them, the constraint preservation and possible divergences in resulting state sums are discussed. It is shown that non-trivial local coarse graining moves entail a non-unitary projection of (physical) Hilbert spaces and `fine grained' Dirac observables defined on them. Identities for undoing a local evolution move with its (time reversed) inverse are derived. Finally, the implications of these results for a Pachner move generated dynamics in simplicial quantum gravity models are commented on. 

}}

\section{Introduction}

A discretization or graph changing dynamics is a generic feature of several quantum gravity approaches, such as Loop Quantum Gravity \cite{Thiemann:1996ay,Thiemann:1996aw,Alesci:2010gb,Bonzom:2011hm,Thiemann:2007zz}, the related spin foam models \cite{Noui:2004iy,Perez:2012wv,Dittrich:2013xwa} and Regge Calculus \cite{Dittrich:2011ke,Dittrich:2013jaa}. Likewise, field theory on an evolving lattice involves a temporally varying discretization \cite{Dittrich:2013jaa,Hoehn:2014fka,Hoehn:2014aoa,Dittrich:2013xwa,Foster:2004yc,Jacobson:1999zk}. Such a discretization changing dynamics is generated by so-called evolution moves, leads to a temporally varying number of kinematical and physical degrees of freedom \cite{Dittrich:2011ke,Dittrich:2013jaa,Hoehn:2014fka,Hoehn:2014aoa,Dittrich:2013xwa} and gives rise to the notion of evolving Hilbert spaces \cite{Hoehn:2014fka,Doldan:1994yg}. 

In order to systematically understand discretization changing dynamics in the quantum theory, at least for variational discrete systems, the companion paper \cite{Hoehn:2014fka} establishes a quantization formalism for {\it global} evolution moves. A global evolution move is characterized by the property that neighbouring discrete time steps do not overlap and thus do not share coinciding subsets of variables (except in a possible boundary). In a space-time context this corresponds to evolving an entire hypersurface at once. 

However, in discrete gravity models and in lattice field theory, such global moves can always be decomposed into sequences of {\it local evolution moves} which only locally update a small subset of the data; neighbouring time steps overlap and thus share coinciding subsets of variables. For instance, the Pachner moves \cite{pachner1,pachner2} are of fundamental importance in simplicial gravity because they establish an elementary and ergodic set of local moves by means of which one can map between any two finite triangulations of fixed topology. The Pachner moves constitute the `atoms' of discrete evolution in simplicial gravity, i.e.\ they compose the most elementary set of discretization changing local evolution moves out of which {\it any} other evolution move (also discretization preserving ones) can be built \cite{Dittrich:2011ke}. On the other hand, the tent moves \cite{Barrett:1994ks,Dittrich:2009fb,Bahr:2009ku} are an example of local discretization preserving evolution moves, however, these can be decomposed into the Pachner moves \cite{Dittrich:2011ke}. The tent and Pachner moves have been implemented in a classical canonical language in \cite{Dittrich:2011ke,Dittrich:2013jaa,Dittrich:2009fb,Hoehn:2011cm}, while local evolution moves in classical scalar lattice field theory are discussed in \cite{Dittrich:2013jaa}. 

In the present manuscript we shall extend the quantum formalism of \cite{Hoehn:2014fka} to also encompass local moves for variational discrete systems. We shall discuss four types of local moves which, at the level of the action, are generalizations of the Pachner and tent moves or local moves for lattice field theories. These moves generate a dynamical refinement or coarse graining of the underlying discretization (while others can be viewed as generating an entangling operation \cite{Dittrich:2013xwa}) and give rise to different kinds of canonical constraints. Not all of these constraints are symmetry generators. Specifically, in simplicial gravity models, constraints appear that do not correspond to Hamiltonian or diffeormosphism constraints \cite{Dittrich:2011ke,Dittrich:2013jaa}. In fact, those constraints only arise on temporally varying discretizations as non-trivial coarse graining or refining consistency conditions 
(ensuring that a state on a coarse discretization can be consistently represented on a finer discretization) \cite{Dittrich:2013xwa,Hoehn:2014fka}. Constraints of this kind will not result in any divergences in the state sum, only the symmetry generating constraints are responsible for the latter. As an aside, the present formalism offers a systematic method for tracking and regularizing such divergences and suggests a new perspective on (gauge) divergences occurring in spin foam models \cite{Bonzom:2013ofa,Bonzom:2010ar,Riello:2013bzw}.

The temporally varying discretization and number of degrees of freedom naturally leads to raise the question as regards the unitarity of such dynamics. In fact, as we shall see, non-trivial coarse graining moves in systems with propagating degrees of freedom lead to non-unitary projections of (physical) Hilbert spaces. These moves irreversibly project out Dirac observables corresponding to degrees of freedom beyond a given refinement scale; `too finely grained' Dirac observables fail to commute with non-trivial coarse graining constraints arising in such moves. 

As in \cite{Hoehn:2014fka}, we shall, for simplicity, restrict to systems with Euclidean configuration spaces $\mathbb{R}^N$, although the formalism can be suitably generalized to incorporate systems with arbitrary configuration manifolds. Such a generalization should not affect the general properties of the formalism for local moves. The results below could therefore contribute to the understanding of local evolution moves also in quantum gravity models even though Euclidean configuration spaces $\mathbb{R}^N$ are not appropriate for non-perturbative quantum gravity. We shall, in particular, comment on this in section \ref{sec_qg} and the conclusions.

The rest of this manuscript is organized as follows. In section \ref{sec_rev}, we provide a review of global evolution moves in the classical and quantum theory since the quantization of the local moves will be based on this. Subsequently, in section \ref{sec_cllocal}, we recapitulate the local evolution moves in the classical theory. Section \ref{sec_loc} proceeds by a quantization of the local moves. After the general construction, we shall consider four different types of local moves more explicitly in sections \ref{sec_I}--\ref{sec_IV} and, in particular, study (non--)unitarity of and possible divergences arising in these moves. To illustrate these moves, we provide explicit examples from scalar field theory on a space-time lattice. In section \ref{sec_comp}, we derive some identities for the composition of local moves with their respective (time reversed) inverses which help to determine the measure (updating factors) corresponding to the various moves. Section \ref{sec_dirac} investigates the status of quantum Dirac observables on temporally varying discretizations and their behaviour under the various local moves. Section \ref{sec_qg} comments on the special features of the (local) dynamics in simplicial gravity that distinguish it from, e.g., lattice field theory. Finally, section \ref{sec_conc} finishes with a conclusion and an outlook on the application of the present formalism to Pachner evolution moves in quantum gravity models. Technical details have been moved to the appendices.

\section{Review of global evolution moves}\label{sec_rev}

In order to understand the local evolution moves, it is necessary to provide a synopsis of global evolution moves in both the classical and quantum theory. We shall be brief on this; the details of the classical formalism are introduced in \cite{Dittrich:2013jaa,Dittrich:2011ke}, while the quantum formalism is developed in \cite{Hoehn:2014fka}. An explicit application of the classical and quantum formalism to systems with quadratic discrete actions, including a comprehensive classification of constraints and degrees of freedom, appears in \cite{Hoehn:2014aoa}.

\subsection{Global evolution moves in the classical formalism}\label{sec_clglobal}

\begin{figure}[hbt!]
\begin{center}
\psfrag{0}{$0$}
\psfrag{1}{$1$}
\psfrag{2}{$2$}
\psfrag{s1}{$S_1$}
\psfrag{s2}{$S_2$}
\psfrag{p0}{\small${}^-p^0,{}^-C^0$}
\psfrag{p1}{\small ${}^+p^1,{}^+C^1$}
\psfrag{p12}{\small${}^-p^1,{}^-C^1$}
\psfrag{p2}{\small${}^+p^2,{}^+C^2$}
\includegraphics[scale=.5]{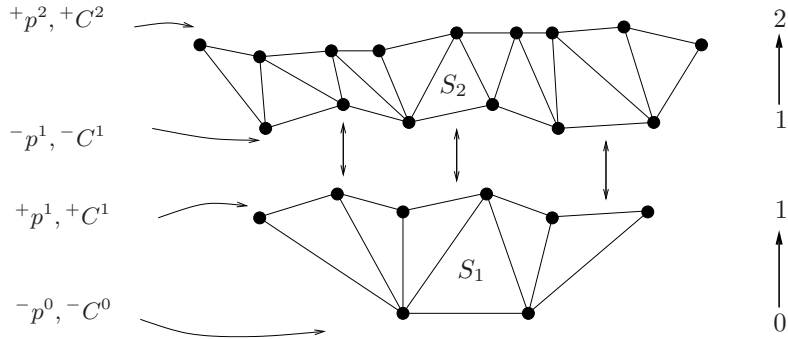}
\caption{\small Schematic illustration of two global evolution moves $0\rightarrow1$ and $1\rightarrow2$. In discrete gravity models, an evolution move corresponds to a region of triangulated space-time. Composing the moves is equivalent to gluing the corresponding regions together at time $1$. This process requires a momentum matching ${}^+p^1={}^-p^1$ and an implementation of both pre-- and post--constraints at step $1$.}\label{fig1}
\end{center}
\end{figure}

We consider variational discrete systems \cite{Dittrich:2013jaa,marsdenwest,DiBartolo:2004cg} in which the discrete time evolution is generated by so-called evolution moves. Such systems do {\it not} feature a Hamiltonian which generates the dynamics (it would be continuous!) and appear in discrete gravity models \cite{Dittrich:2011ke,Dittrich:2009fb,Gambini:2002wn,Gambini:2005vn,DiBartolo:2002fu,DiBartolo:2004dn,Campiglia:2006vy}, field theory on a space-time lattice \cite{Sorkin:1975jz,Dittrich:2013jaa,Hoehn:2014aoa} and discrete mechanics \cite{marsdenwest,Jaroszkiewicz:1996gr}. The discrete time steps shall be labeled by $n\in\mathbb{Z}$ and $\cq_n$ denotes the configuration manifold of the system at time step $n$ which can be quite arbitrary. $\cq_n$ is coordinatized by $x^i_n$, $i=1,\ldots,\dim\cq_n$, but for notational convenience the index $i$ shall often be dropped. A {\it global time evolution move} $n\rightarrow n+1$ maps the system from time step $n$ to time step $n+1$ in such a way that $\cq_n$ and $\cq_{n+1}$ do not share any subsets of coinciding variables. That is, neighbouring time steps of a global move $n\rightarrow n+1$ do not overlap (except in a possible boundary). For an illustration, see figure \ref{fig1}. By contrast, for {\it local evolution moves} neighbouring time steps can overlap. These will be the main focus of the subsequent sections. 

To every global evolution move $n\rightarrow n+1$ there is associated a discrete action or Hamilton's principal function $S_{n+1}(x_n,x_{n+1})$ which depends on `old' variables $x_n$ and `new' variables $x_{n+1}$. Consequently, $S_{n+1}$ constitutes a generating function of the first kind for the canonical evolution,
\ba
{}^-p^n:=-\f{\p S_{n+1}(x_n,x_{n+1})}{\p x_n},\q\q\q{}^+p^{n+1}:=\f{\p S_{n+1}(x_n,x_{n+1})}{\p x_{n+1}}.\label{ham}
\ea
The {\it pre--momenta} ${}^-p^n$ and the $x_n$ form a canonical Darboux coordinate system on $\cp_n:=T^*\cq_n$, the phase space at step $n$ \cite{Dittrich:2013jaa,Dittrich:2011ke,marsdenwest}. Similarly, the {\it post--momenta} ${}^+p^{n+1}$ together with the $x_{n+1}$ establish a canonical coordinate system on $\cp_{n+1}:=T^*\cq_{n+1}$, i.e.\ the phase space of step $n+1$. We emphasize that this formalism works, in particular, if $\cq_n\ncong\cq_{n+1}$ which corresponds to a temporally varying discretization with varying numbers of degrees of freedom \cite{Dittrich:2013jaa,Dittrich:2011ke}.

Equations (\ref{ham}) define an implicit global Hamiltonian time evolution map $\mathfrak{H}_n:\cp_n\rightarrow\cp_{n+1}$. However, if $\det\left(\f{\p^2S_{n+1}}{\p x^i_n\p x_{n+1}^j}\right)=0$, $\mathfrak{H}_n$ is not uniquely defined. In this case, the pre--momentum equations in (\ref{ham}) cannot be independent and non-trivial relations ${}^-C^n(x_n,{}^-p^n)=0$ at $n$ exist which are called {\it pre--constraints}. Similarly, the post--momentum equations in (\ref{ham}) cannot all be independent and there exist non-trivial relations ${}^+C^{n+1}(x_{n+1},{}^+p^{n+1})=0$ at $n+1$ which are called {\it post--constraints} \cite{Dittrich:2013jaa,Dittrich:2011ke}. In this case, $\mathfrak{H}_n$ defines a pre--symplectic transformation from $\cc^-_n$, the pre--constraint surface in $\cp_n$, to $\cc^+_{n+1}$, the post--constraint surface in $\cp_{n+1}$ \cite{Dittrich:2011ke}.

The pre--constraints ${}^-C^n$ are first class among themselves \cite{Dittrich:2013jaa} and thus define a {\it pre--orbit} $\cg^-_n$ in $\cc^-_n$ which is parametrized by free parameters. Analogously, the post--constraints ${}^+C^{n+1}$ are first class among themselves and generate a {\it post--orbit} $\cg^+_{n+1}$ in $\cc^+_{n+1}$ which is parametrized by free parameters associated to the constraints.

Consider the composition of two global moves $n-1\rightarrow n$ and $n\rightarrow n+1$ to an `effective' move $n-1\rightarrow n+1$. The prerequisite for such a composition is that the actions associated to the different moves are {\it additive} such that solving the equations of motion at $n$ and inserting the solutions into $S_n+S_{n+1}$ yields Hamilton's principal function for the move $n-1\rightarrow n+1$. The equations of motion at $n$ are equivalent to a {\it momentum matching} 
\ba
{}^-p^n={}^+p^n\nn
\ea
 at step $n$ \cite{marsdenwest,Dittrich:2011ke,Dittrich:2013jaa,Hoehn:2014aoa,Jaroszkiewicz:1996gr,Gambini:2002wn,DiBartolo:2004cg,Gambini:2005vn,DiBartolo:2002fu}. For constrained systems and, in particular, systems with temporally varying discretization this momentum matching leads to all kinds of non-trivialities: many canonical concepts such as the pre-- and post--constraints, propagating degrees of freedom, the reduced phase space, symmetries and the constraint classification become evolution move dependent (for a detailed discussion, see \cite{Dittrich:2013jaa,Hoehn:2014aoa}). If a constraint at a step $n$ is both a pre-- and post--constraint, $C^n={}^-C^n={}^+C^n$, it is necessarily a gauge symmetry generator \cite{Dittrich:2013jaa,Hoehn:2014aoa}.\footnote{Such gauge symmetries are always associated to degeneracies of the Hessian of the action \cite{Dittrich:2013jaa,Hoehn:2014aoa}. However, the converse is not quite true: degenerate directions of the Hessian of the action associated to a certain set of moves can occur which become non-degenerate directions of the Hessian of the action associated to a larger set of moves (containing the previous ones) \cite{Hoehn:2014aoa}. That is, these directions correspond to variable transformations which are symmetries for a set of moves, but not for a larger set of moves containing the previous ones. These move dependent symmetries, however, are not associated to constraints which are {\it simultaneously} pre-- and post--constraints \cite{Hoehn:2014aoa}. On the other hand, genuine (move independent) gauge symmetries of the action are always generated by constraints which are both pre-- and post--constraints \cite{Dittrich:2013jaa}.} This will become relevant below.

\subsection{Global evolution moves in the quantum theory}\label{sec_qglobal}

Since the quantization of local moves below will directly connect to the quantum formalism for global moves \cite{Hoehn:2014fka}, we shall briefly recall its basic properties. We use the Dirac algorithm \cite{Dirac,Henneaux:1992ig} for quantizing constrained systems. For simplicity, we shall restrict ourselves to systems with Euclidean configuration spaces $\cq_n\simeq\mathbb{R}^{N_n}$ (and $x_n,p^n\in\mathbb{R}$) to avoid global or topological non-trivialities \cite{isham2}, although the formalism can be adapted to general configuration manifolds. The quantum pre-- and post--constraints at $n$ are assumed to be self-adjoint operators on the kinematical Hilbert space $\ch^{\rm kin}_n=L^2(\mathbb{R}^{N_n},dx_n)$, where $N_n$ can vary with $n$. We assume the quantization to be consistent and anomaly free such that the set of quantum pre--constraints form a set of commuting operators and, likewise, all quantum post--constraints commute as in the classical theory \cite{Dittrich:2013jaa}. Furthermore, we assume the spectra of the constraints to be absolutely continuous and the pre--orbits $\cg^-_n$ of the pre--constraints and the post--orbits $\cg^+_n$ of the post--constraints at any step $n$ to be non-compact.

The {\it post--physical Hilbert space} ${}^+\ch^{\rm phys}_n$ at $n$ consists of {\it post--physical states} ${}^+\psi^{\rm phys}_n$ which are annihilated by all quantum post--constraints ${}^+\hat{C}^n_I$ at $n$. Similarly, the {\it pre--physical Hilbert space} ${}^-\ch^{\rm phys}_n$ at $n$ consists of {\it pre--physical states} ${}^-\phi^{\rm phys}_n$ which are annihilated by the quantum pre--constraints ${}^-\hat{C}^n_I$. Using group averaging \cite{Marolf:1995cn,Marolf:2000iq,Thiemann:2007zz,Rovelli:2004tv}, the post-- and pre--physical states can be constructed from kinematical states $\psi^{\rm kin}_n\in\ch^{\rm kin}_n$ via an improper projection \cite{Hoehn:2014fka}
\ba
{}^+\psi^{\rm phys}_n={}^+\mathbb{P}_n\,\psi^{\rm kin}_n,\q\q\q\q{}^-\phi^{\rm phys}_n={}^-\mathbb{P}_n\,\phi^{\rm kin}_n,\label{proj}
\ea
with {\it post--} and {\it pre--projector}
\ba\label{1234}
{}^+\mathbb{P}_n:=\prod_{I=1}^k\delta({}^+\hat{C}^n_I),\q\q\q\q {}^-\mathbb{P}_n:=\prod_{J=1}^l\delta({}^-\hat{C}^n_J),
\ea
where 
\ba
\delta(\hat{C})=\f{1}{2\pi\hbar}\int_\mathbb{R}ds\,e^{is\hat{C}/\hbar}\nn.
\ea

\begin{R}
\emph{The analysis of compositions of evolution moves below will require a formal decomposition of the projectors into sub-projectors, projecting onto only a subset of the constraints. In this regard, the formal construction (\ref{1234}) of the pre-- and post--projectors which uses the factorization into the projectors onto the individual constraints is notationally advantageous and we shall therefore henceforth work with this definition. However, alternatively, one could construct the post--projector as
\ba
{}^+\mathbb{P}_n':=\f{1}{(2\pi\hbar)^k}\int_{\mathbb{R}^k}\prod_{I=1}^kds^I_n\,e^{i\hbar\sum_{I=1}^ks^I_n{}^+\hat{C}^n_I}\label{alternative}
\ea
(and analogously for the pre--projector). ${}^+\mathbb{P}_n$ and ${}^+\mathbb{P}_n'$ are generally inequivalent because the constraints do not always commute; for Abelian constraints the two definitions coincide. The advantage of ${}^+\mathbb{P}_n'$ is that, in analogy to the classical theory, it is invariant under linear transformations of the constraints. The disadvantage of ${}^+\mathbb{P}_n'$ is that the necessary formal decomposition of the projectors below would become notationally more complicated. This is ultimately the reason why we choose to work with the projectors in the form (\ref{1234}).  }

\emph{This difference notwithstanding, we emphasize that many of the formal calculations below are independent on which of the two constructions are employed because only the fact that the constraints are imposed will matter. On the other hand, the more detailed calculations which employ the formal factorization of the constraints as in (\ref{1234}) could also be carried out with minor modifications with the alternative construction (\ref{alternative}). Nevertheless, the general conclusions drawn from this formalism would be unaltered. We refer the reader to \cite{Hoehn:2014fka} for further discussion of the two alternatives.}
\end{R}

The pre--projectors ${}^-\mathbb{P}_n$ (and similarly the post--projectors) are improper: the first action of ${}^-\mathbb{P}_n$ in a repeated action ${}^-\mathbb{P}_n\cdot{}^-\mathbb{P}_n\,\phi^{\rm kin}_n$ projects on the pre--physical Hilbert space, while the second action of ${}^-\mathbb{P}_n$ spuriously integrates the physical state over the non--compact pre--orbit $\cg^-_n$ and thus leads to a divergence. The {\it pre--physical inner product} in ${}^-\ch^{\rm phys}_n$ and the {\it post--physical inner product} in ${}^+\ch^{\rm phys}_n$ read
\ba
\langle{}^-\phi^{\rm phys}_n|{}^+\psi^{\rm phys}_n\rangle_{\rm phys-}=\langle\phi^{\rm kin}_n|{}^-\mathbb{P}_n|\psi^{\rm kin}_n\rangle_{\ch^{\rm kin}_n},\q\q\q\langle{}^+\chi^{\rm phys}_n|{}^+\zeta^{\rm phys}_n\rangle_{\rm phys+}=\langle\chi^{\rm kin}_n|{}^+\mathbb{P}_n|\zeta^{\rm kin}_n\rangle_{\ch^{\rm kin}_n}\nn
\ea
and are generally inequivalent \cite{Hoehn:2014fka}.

The dynamics in the systems under consideration is {\it not} generated by a Hamiltonian or a set of constraints. Related to this, one cannot simply write the action in terms of phase space variables as in the continuum such that there is no natural candidate for a phase space path integral to define the quantum dynamics. This is, however, not a problem because, as in the classical theory, one can make use of the action $S_{n+1}$ --- which contains the entire information about the dynamics --- to generate a time evolution map for a global move $n\rightarrow n+1$. Adhering to the construction in \cite{Hoehn:2014fka}, we shall use the propagator to construct a map from $\ch^{\rm kin}_n$ to ${}^+\ch^{\rm phys}_{n+1}$
\ba
{}^+\psi^{\rm phys}_{n+1}=\int dx_n\,K_{n\rightarrow n+1}(x_n,x_{n+1})\,\psi^{\rm kin}_n,\label{propmap}
\ea
where, in the spirit of the configuration space path integral expression for the continuum propagator, we make the following propagator ansatz which assigns a measure and phase factor to each move: 
\ba
K_{n\rightarrow n+1}(x_n,x_{n+1})=M_{n\rightarrow n+1}(x_n,x_{n+1})\,e^{iS_{n+1}(x_n,x_{n+1})/\hbar}.\label{form1}
\ea
The propagator associated to a move constitutes a building block of the (configuration space) path integral for the system under consideration: the composition of many such moves $0\rightarrow1\rightarrow\cdots\rightarrow n+1$ amounts to a convolution of the associated propagators and an integration over any bulk variables to form the path integral for the evolution from step $0$ to $n+1$. The path integral can then itself be written in the form (\ref{form1}) as a propagator depending on old and new configuration variables.

The factorization of the propagator for a move $n\rightarrow n+1$ into an integration measure $M_{n\rightarrow n+1}$ and a phase containing the classical action is motivated from the desire to obtain the correct expression in the semiclassical limit. Indeed, for quadratic discrete actions \cite{Hoehn:2014aoa} and semiclassical limit approximations \cite{Hoehn:2014fka,miller1,miller2,Klauderbook} in particular, propagators are precisely of the form (\ref{form1}). Furthermore, making such an ansatz for a discretized harmonic oscillator leads to the correct continuum expression upon coarse graining \cite{Bahr:2011uj}. We shall therefore write a general (not necessarily semiclassical) propagator for arbitrary actions in the form (\ref{form1}) and simply absorb the quantum corrections to the semiclassical expression in the generally complex measure $M_{n\rightarrow n+1}$ (for more details, see \cite{Hoehn:2014fka}). 

There are non-trivial consistency conditions on the measure which will help to determine it. While we shall discuss these in the context of local moves in detail below, for the moment it is important to mention one of them: the propagator has to satisfy both pre-- and post--constraints \cite{Hoehn:2014fka}:
\ba
{}^+\hat{C}^{n+1}_I\,K_{n\rightarrow n+1}=0={}^-\hat{C}^n_I(K_{n\rightarrow n+1})^*.\label{cond}
\ea
This ensures that (\ref{propmap}) (and its time reverse) map onto solutions to the constraints. In this way, the time evolution map (or path integral) acts as a projector onto the constraints \cite{Hoehn:2014fka}, as generally expected in quantum gravity models \cite{Halliwell:1990qr,Rovelli:1998dx,Noui:2004iy,Thiemann:2007zz,Dittrich:2013xwa,Thiemann:2013lka}, despite the absence of a natural phase space path integral.

In analogy to kinematical states $\psi^{\rm kin}_n\in\ch^{\rm kin}_n$ one can introduce the notion of a {\it kinematical propagator} $\kappa_{n\rightarrow n+1}(x_n,x_{n+1})$ as a function on $\cq_n\times\cq_{n+1}$ which is square integrable with respect to both the kinematical inner products in $\ch^{\rm kin}_n$ and $\ch^{\rm kin}_{n+1}$. The kinematical propagator itself does not satisfy any constraints. But the {\it physical propagator} can now be written as \cite{Hoehn:2014fka}
\ba
K_{n\rightarrow n+1}={}^+\mathbb{P}_{n+1}\kappa_{n\rightarrow n+1}{}^-\mathbb{P}_n.\label{kinprop}
\ea
Using (\ref{proj}) and rewriting (\ref{propmap}) yields a map
\ba
U_{n\rightarrow n+1}:=\int dx_n{}^+\mathbb{P}_n\,\kappa_{n\rightarrow n+1}\label{unitary}
\ea
from ${}^-\ch^{\rm phys}_n$ to ${}^+\ch^{\rm phys}_{n+1}$ which is a unitary isomorphism \cite{Hoehn:2014fka} and the quantum analogue of the classical global Hamiltonian time evolution map $\mathfrak{H}_n$ \cite{Dittrich:2013jaa}.

The non-trivial aspects of the formalism appear when two moves $n-1\rightarrow n$ and $n\rightarrow n+1$ are composed to an `effective' move $n-1\rightarrow n+1$ \cite{Hoehn:2014fka}. This can lead to new `effective' quantum constraints at $n-1$ and $n+1$ such that the pre--physical Hilbert space ${}^-\ch^{\rm phys}_{n-1}$ at $n-1$ and the post--physical Hilbert space ${}^+\ch^{\rm phys}_{n+1}$ at $n+1$ may change. That is, in general, the pre-- and post--physical Hilbert spaces at a given step $n$ depend on the particular evolution move. As a consequence, the Dirac observables as physical degrees of freedom become evolution move dependent too. All of this parallels the classical formalism \cite{Dittrich:2013jaa}.

The introduction of kinematical propagators and the use of projectors provide a novel construction method for the path integral of constrained variational discrete systems which, in particular, handles systems with temporally varying discretization. The path integral is given by the composition of a sequence of physical propagators. This will lead to divergences in the presence of gauge symmetries, but fortunately, the formalism easily keeps track of them \cite{Hoehn:2014fka}: gauge symmetries are generated by constraints $\hat{C}^n$ at step $n$ that are both pre-- and post--constraints \cite{Dittrich:2013jaa}. Consequently, the projection onto solutions to $\hat{C}^n$ is contained in both projectors ${}^+\mathbb{P}_n$ and ${}^-\mathbb{P}_n$ which are both implemented in a composition. This amounts to a spurious integration over the non-compact gauge orbit at $n$ and thus a divergence. The latter can be regularized by dropping one instance of the doubly occurring projector onto solutions to $\hat{C}^n$ \cite{Hoehn:2014fka}.

\section{Summary of local evolution moves in the classical theory}\label{sec_cllocal}

Neighbouring time steps of a {\it local evolution move} overlap and share a coinciding set of variables which are updated in the course of the move. The classical implementation of such local moves is discussed in detail in \cite{Dittrich:2011ke,Dittrich:2013jaa,Hoehn:2011cm}. As in these references, let us now label the discrete time steps corresponding to local evolution moves by $k\in\mathbb{Z}$ to distinguish them from $n\in\mathbb{Z}$ which labels the global moves.  

Just as in the case of global moves, to every local evolution move $k\rightarrow k+1$ there is associated an action contribution $S_{k+1}$. However, in contrast to the global moves the new piece of action $S_{k+1}$ is {\it not} a generating function for the time evolution equations of the move $k\rightarrow k+1$ because it only depends on the local variables involved in the move (for details see \cite{Dittrich:2013jaa}). To understand the local time evolution equations, we firstly need to distinguish among four different types of configuration variables which a local move may involve and which are illustrated in figure \ref{local}: 
\begin{figure}[hbt!]
\begin{center}
\psfrag{fn}{\small $\phi_{k+1}^n$}
\psfrag{fo}{\small  $\phi_k^o$}
\psfrag{fe1}{\small $\phi^{e_1}_k$}
\psfrag{fe2}{\small  $\phi^{e_2}_k$}
\psfrag{fb1}{\small $\phi^{b_1}_k$}
\psfrag{fb2}{\small  $\phi^{b_2}_k$}
\psfrag{fb3}{\small $\phi^{b_3}_k$}
\psfrag{fb4}{\small  $\phi^{b_4}_k$}
\psfrag{Sk0}{\small $\Sigma_k$}
\psfrag{fe12}{\small $\phi^{e_1}_{k+1}$}
\psfrag{fe22}{\small  $\phi^{e_2}_{k+1}$}
\psfrag{fb12}{\small $\phi^{b_1}_{k+1}$}
\psfrag{fb22}{\small  $\phi^{b_2}_{k+1}$}
\psfrag{fb32}{\small $\phi^{b_3}_{k+1}$}
\psfrag{fb42}{\small  $\phi^{b_4}_{k+1}$}
\psfrag{Sk1}{\small $\Sigma_{k+1}$}
\psfrag{s}{\small $S_{k+1}$}
\hspace*{-4cm}\begin{subfigure}[b]{.22\textwidth}
\centering
\includegraphics[scale=.45]{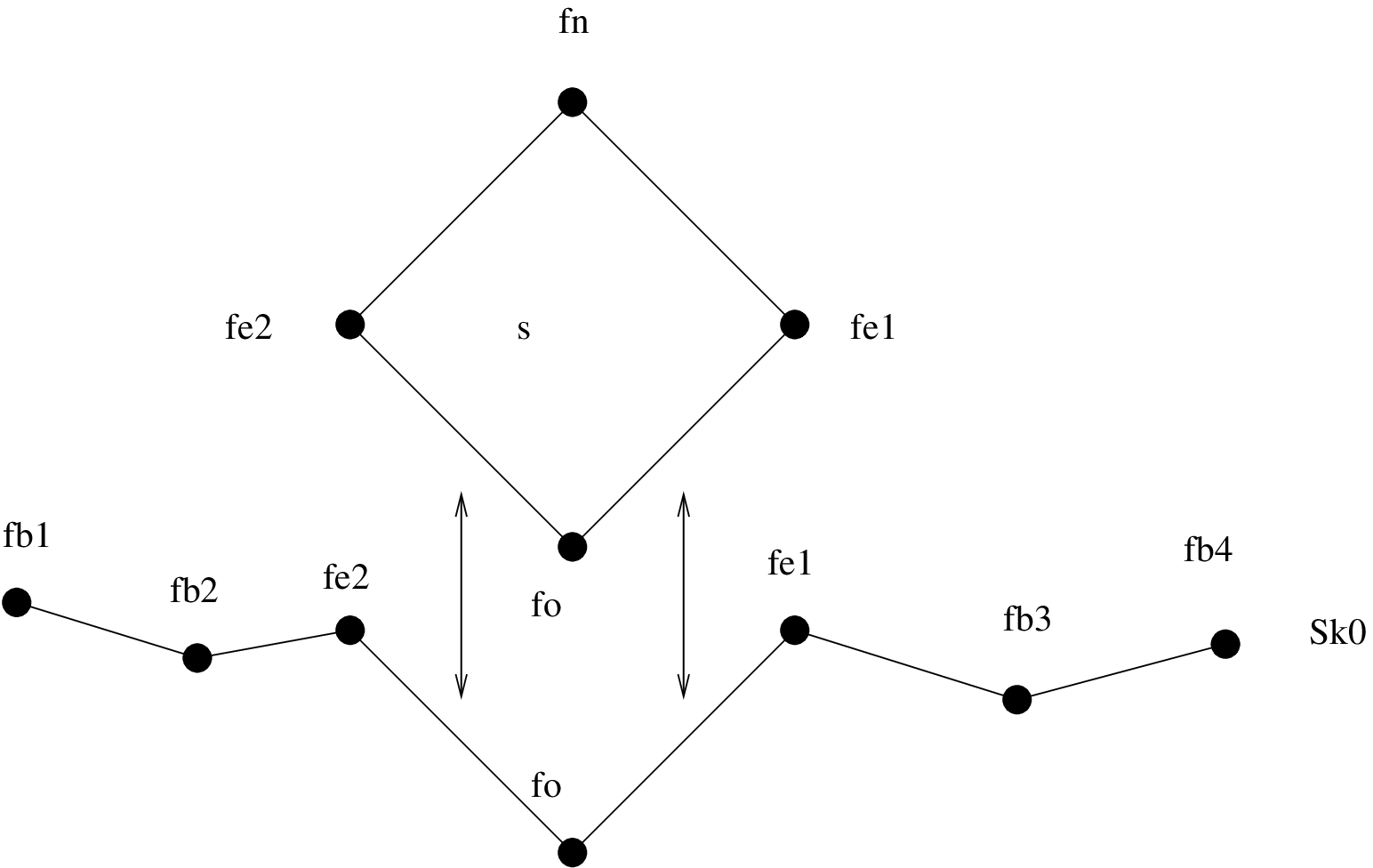}
\centering
\caption{\small }
\end{subfigure}
\hspace*{4.8cm}
\begin{subfigure}[b]{.22\textwidth}
\centering
\includegraphics[scale=.45]{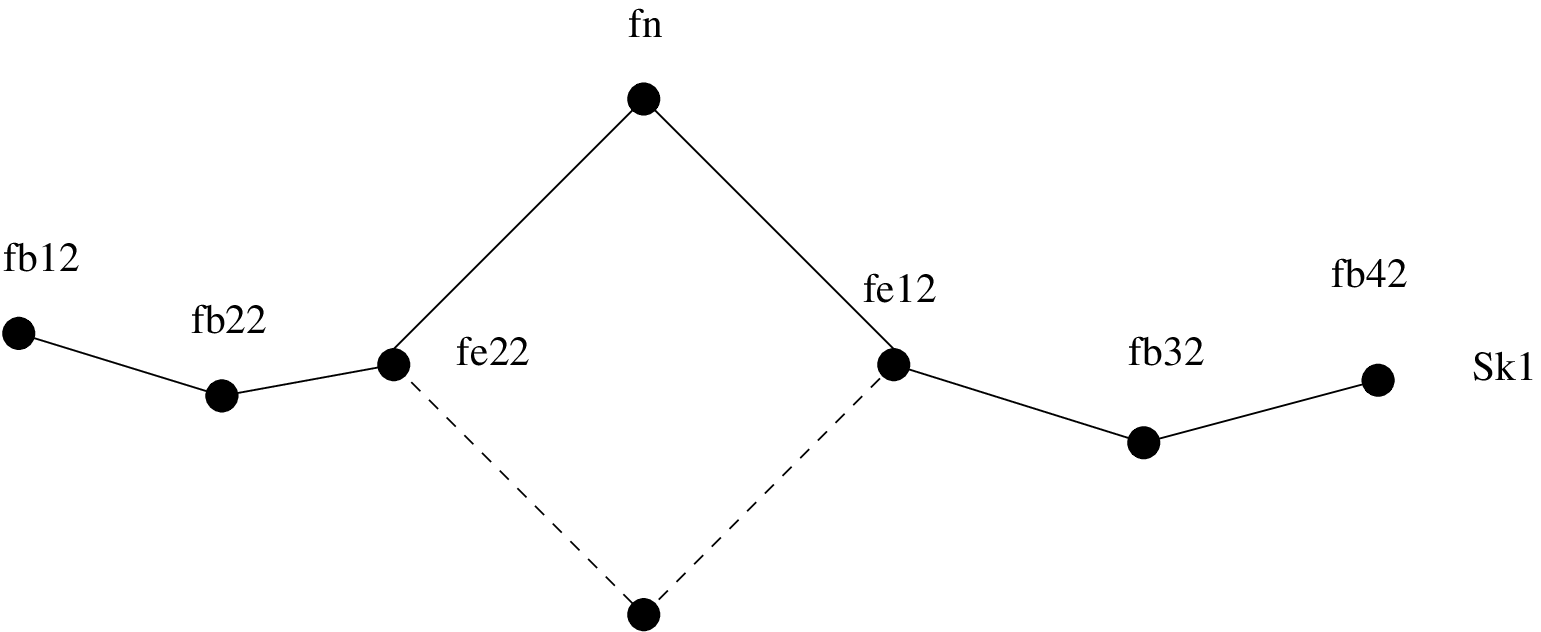}
\caption{\small }
\end{subfigure}
\caption{\small Illustration of the four types of variables labeled by $e,b,o,n$ appearing in local evolution moves. The picture shows a local evolution move for a scalar field living on the vertices of a 2D discretized spacetime. The local evolution move corresponds to gluing a new piece of discrete 2D spacetime with corresponding scalar field action $S_{k+1}$ to the 1D hypersurface $\Sigma_k$ which constitutes the time step $k$. The `old' field variable $\phi^o_k$ disappears from $\Sigma$ and becomes bulk in this move, while a new vertex with a `new' field variable $\phi^n_{k+1}$ is introduced. Clearly, the neighbouring time steps overlap, $\Sigma_k\cap\Sigma_{k+1}\neq\emptyset$. The two field variables $\phi^{e_1}_k,\phi^{e_2}_k$ occur before and after the move and contribute to $S_{k+1}$. The remaining variables $\phi^{b_i}_k$ occur at both time steps, however, do not contribute to $S_{k+1}$. }\label{local}
\end{center}
\end{figure}
\begin{itemize}
\item $x^o_k$ denotes `old variables' in $S_{k+1}$ that occur at step $k$ but which disappear at step $k+1$ because they become `bulk',
\item $x^n_{k+1}$ denotes `new variables' in $S_{k+1}$ which do not occur at step $k$ but appear at step $k+1$,
\item $x^e_k$ denotes variables which occur at both steps $k,k+1$, but are involved in the move because they contribute to $S_{k+1}$, and
\item $x^b_k$ denotes variables which occur at both steps $k,k+1$, however, do not participate in the move and thereby do not contribute to $S_{k+1}$.
\end{itemize}
In general, the number of `new variables' does not coincide with the number of `old variables'. Classically, in order to account for this changing number of variables in the move $k\rightarrow k+1$, it is convenient (but not necessary) to extend the configuration manifolds at $k$ and $k+1$ by introducing the `missing' variables $x^n_k$ at $k$ and $x^o_{k+1}$ at $k+1$ as auxiliary variables. The extended configuration manifolds $\overline{\cq}_k:=\cq_k\times\cq^{ext}_k$ and $\overline{\cq}_{k+1}:=\cq_{k+1}\times\cq^{ext}_{k+1}$ are then of equal dimension, where $\cq^{ext}_k$ and $\cq^{ext}_{k+1}$ are coordinatized by $x^n_k$ and $x^o_{k+1}$, respectively. The dynamics of the move $k\rightarrow k+1$ can now be formulated on the corresponding extended phase spaces $\overline{\cp}_k:=T^*\overline{\cq}_k$ and $\overline{\cp}_{k+1}:=T^*\overline{\cq}_{k+1}$.

Given that subsets of the data in $\cq_k$ and $\cq_{k+1}$ coincide, the canonical data in the move $k\rightarrow k+1$ only needs to be appropriately updated. The corresponding local time evolution map is denoted by $\fh_k:\overline{\cp}_k\rightarrow \overline{\cp}_{k+1}$ and is called {\it momentum updating} \cite{Dittrich:2013jaa,Dittrich:2011ke}. Specifically, for the variables $x^e,x^b$ appearing at both times $k,k+1$ it reads
\ba
x^e_{k+1}&=&x^e_k,\q\q\q\q p^{k+1}_e=p^k_e+\f{\p S_{k+1}}{\p x^e_k},\nn\\
x^b_{k+1}&=&x^b_k,\q\q\q\q p^{k+1}_b=p^k_b.\nn
\ea
$p^k_b$ does not need to be updated in the move $k\rightarrow k+1$, because $x^b$ is not involved in the move such that $S_{k+1}$ does not depend on it. On the other hand, the `old variables' $x^o_k$ generally lead to non-trivial pre--constraints at step $k$, while the `new variables' $x^n_{k+1}$ are generally accompanied by post--constraints at step $k+1$ \cite{Dittrich:2013jaa,Dittrich:2011ke}. These constraints and their effect on the (non-)preservation of the symplectic structure will be amply discussed when studying different types of local moves below. We shall therefore not go into further detail here.

\section{Quantization of local evolution moves}\label{sec_loc}

We shall now extend the quantization formalism for global evolution moves of \cite{Hoehn:2014fka}, summarized in section \ref{sec_qglobal}, to local evolution moves. Again, for simplicity, we shall restrict ourselves to systems with Euclidean configuration spaces $\cq_k\simeq\mathbb{R}^{N_k}$ (and $x_k,p^k\in\mathbb{R}$) to avoid global or topological non-trivialities \cite{isham2}, although the formalism can be adapted to general configuration manifolds. We begin with a general construction, followed by a detailed analysis of four different types of local moves below. The latter implement a dynamical coarse graining, refining or entangling of the discrete degrees of freedom (see also \cite{Dittrich:2013xwa} for a conceptual discussion).

\subsection{General construction}\label{sec_locgen}

A local move $k\rightarrow k+1$ updates a global move $0\rightarrow k$ from some initial step $n=0$ to $k$ to a new global move $0\rightarrow k+1$. Classically, the momentum updating map $\fh_k$ updates the canonical data such that after the move (\ref{ham}) holds again. Similarly, in the quantum theory the new global move must also satisfy (\ref{propmap}). In order for this to be possible, the local move $k\rightarrow k+1$ must update the measure and propagator 
\ba
K_{0\rightarrow k+1}(x_0,x_{k+1})=\int \prod_odx^o_k\,M_{k\rightarrow k+1}(x^o_k,x^e_k,x^n_{k+1})\,e^{iS_{k+1}(x^o_k,x^e_k,x^n_{k+1})/\hbar}\,K_{0\rightarrow k}(x_0,x_k)\label{pup}
\ea
and, hence, the physical post--states
\ba
{}^+\psi^{\rm phys}_{k+1}(x_{k+1})=\int  \prod_odx^o_k\,M_{k\rightarrow k+1}(x^o_k,x^e_k,x^n_{k+1})\,e^{iS_{k+1}(x^o_k,x^e_k,x^n_{k+1})/\hbar}\,{}^+\psi^{\rm phys}_k(x_k).\label{sup}
\ea
The `old variables' $x^o_k$ are integrated out because they become `bulk'. We choose the measure updating factor $M_{k\rightarrow k+1}$ independend of $x^b_k$ as these variables are not involved in the move. We emphasize that the updating factor $M_{k\rightarrow k+1}\,e^{iS_{k+1}/\hbar}$ is itself {\it not} a propagator because it only contains the information about the local data involved in the move, but not about the remaining degrees of freedom.

Consistency of such an updating requires:
\begin{itemize}
\item[(a)] The move $k\rightarrow k+1$ must preserve the quantum post--constraints from step $k$, i.e.\ ${}^+\psi^{\rm phys}_{k+1}$ must satisfy (the time evolved version of) the post--constraints that ${}^+\psi^{\rm phys}_k$ satisfied.
\item[(b)] New pre--constraints ${}^-\hat{C}^k_\nu$ at $k$ and/or new post--constraints ${}^+\hat{C}^{k+1}_{\nu'}$ at $k+1$ arising in the move must be implemented.
\end{itemize}
Equivalently, (a) means that $K_{0\rightarrow k+1}$ satisfies the time evolved version of the post--constraints that annihilated $K_{0\rightarrow k}$ in (\ref{cond}). ($K_{0\rightarrow k+1}$ will automatically satisfy all pre--constraints at $n=0$ that annihilate $K_{0\rightarrow k}$.) The new post--constraints ${}^+\hat{C}^{k+1}_{\nu'}$ in (b) are imposed on ${}^+\psi^{\rm phys}_{k+1}$ or, equivalently, on $K_{0\rightarrow k+1}$. On the other hand, the new pre--constraints of (b) lead to some subtleties. Namely, for any new quantum pre--constraint ${}^-\hat{C}^k_\nu$ at $k$ there are three possibilities for each of which we use a different index $\alpha,\beta$ or $s$ (see also \cite{Hoehn:2014fka,Dittrich:2013jaa}):
\begin{itemize}
\item[(i)] ${}^-\hat{C}^k_\alpha\,{}^+\psi^{\rm phys}_k=0$ is automatically satisfied because ${}^-\hat{C}^k_\alpha$ coincides with some post--constraint ${}^+\hat{C}^k_\alpha$, $\hat{C}^k_\alpha:={}^-\hat{C}^k_\alpha={}^+\hat{C}^k_\alpha$. These simultaneous pre-- and post--constraints are henceforth denoted by $\hat{C}^k_\alpha$.  
\item[(ii)] The constraint is non-trivial, ${}^-\hat{C}^k_\beta\,{}^+\psi^{\rm phys}_k\neq0$, and first class. 
\item[(iii)] The constraint is non-trivial, ${}^-\hat{C}^k_s\,{}^+\psi^{\rm phys}_k\neq0$, and second class. 
\end{itemize}
Constraints that are both pre-- and post--constraints (case (i)) are gauge symmetry generators \cite{Dittrich:2013jaa}. Pre--constraints of case (ii) are not symmetry generators \cite{Dittrich:2013jaa} but impose non-trivial restrictions on the dynamics that, as we shall see shortly, result in a non-unitarity of time evolution. Such pre--constraints can be understood as non-trivial coarse graining conditions that must be satisfied in order to map a state on a finer discretization to a coarser one. 
We shall elaborate further on this below and in section \ref{sec_dirac} in the context of Dirac observables (see also \cite{Dittrich:2013xwa} for a conceptual discussion and \cite{Hoehn:2014fka} for a concrete analysis for global moves). Lastly, second class constraints are usually solved in the classical theory prior to quantization \cite{Henneaux:1992ig}. We shall therefore henceforth assume that in case (iii) the constraints have been solved classically.

The new pre--constraints ${}^-\hat{C}^k_\beta$ at step $k$ of case (ii) do {\it not} need to be satisfied by the physical post--states ${}^+\psi^{\rm phys}_k$ at step $k$ in order to obtain a consistent dynamics (see also the discussion in \cite{Hoehn:2014fka}). Instead, it is sufficient to implement them on the propagator $K_{k\rightarrow f}$ of a global evolution move $k\rightarrow f$ to some future final step $n=f$. This is because of the projection property (\ref{propmap}) of the propagator which maps states at $k$, not satisfying the pre--constraints, to the physical post--states at $f$. Using the arguments above in reversed time direction, this propagator must be decomposable in analogy to (\ref{pup}) as
\ba
K_{k\rightarrow f}(x_k,x_f)=\int \prod_ndx^n_{k+1}\,M_{k\rightarrow k+1}(x^o_k,x^e_k,x^n_{k+1})\,e^{iS_{k+1}(x^o_k,x^e_k,x^n_{k+1})/\hbar}\,K_{k+1\rightarrow f}(x_{k+1},x_f)\label{revpup}
\ea
and satisfy the pre--constraints $\hat{C}^k_\alpha,{}^-\hat{C}^k_\beta$ as in (\ref{cond}). This imposes in general non-trivial restrictions on the measure $M_{k\rightarrow k+1}$ of the local move $k\rightarrow k+1$. It does not impose any restrictions on $K_{k+1\rightarrow f}$ because, as we shall see below, the new pre--constraints only act on $M_{k\rightarrow k+1}\,e^{iS_{k+1}/\hbar}$.

In analogy to (\ref{kinprop}), we can thus choose a square integrable function $\upsilon_{k\rightarrow k+1}(x^o_k,x^e_k,x^n_{k+1})$ such that
\ba
M_{k\rightarrow k+1}\,e^{iS_{k+1}/\hbar}={}^+\mathbb{P}_{k+1}^\nu\,(\mathbb{P}^A_k)^*({}^-\mathbb{P}_k^B)^*\,\upsilon_{k\rightarrow k+1},\label{kinup}
\ea
where $^*$ denotes complex conjugation and\footnote{For the time being, we ignore possible ordering ambiguities in the product of projectors.}
\ba
{}^+\mathbb{P}^\nu_{k+1}:=\prod_{\nu'}\delta({}^+\hat{C}^{k+1}_{\nu'}),\q\q\q\mathbb{P}^A_k:=\prod_\alpha\delta(\hat{C}^k_\alpha),\q\q\q{}^-\mathbb{P}^B_k:=\prod_\beta\delta({}^-\hat{C}^k_\beta),
\ea
are the improper projectors on solutions of only the new pre-- and post--constraints of (b) above. In this way, the updating of the propagator (\ref{pup}) and states (\ref{sup}) contains a projection onto solutions to the new pre--constraints at $k$.

The gauge symmetry generators $\hat{C}^k_\alpha$ of case (i) immediately lead to trouble: since the propagator $K_{0\rightarrow k}$ and, equivalently, ${}^+\psi^{\rm phys}_k$ must satisfy all post--constraints they must, in particular, already be annihilated by the $\hat{C}^k_\alpha$ because these are also post--constraints at $k$. This implies that both $K_{0\rightarrow k}$ and ${}^+\psi^{\rm phys}_k$ are already projected with $\mathbb{P}^A_k$. Since this projector is also contained in the updating factor (\ref{kinup}), the propagator and state updating (\ref{pup}, \ref{sup}) involve a double action of the improper projector $\mathbb{P}^A_k$ and thus a spurious integration over a non-compact gauge orbit which results in a divergence. This divergence can easily be regularized by dropping one instance of $\mathbb{P}^A_k$ in (\ref{pup}, \ref{sup}). Equivalently, one can often (not necessarily always) introduce Faddeev-Popov gauge fixing conditions $\prod_\alpha\delta(G_\alpha)$ in (\ref{pup}, \ref{sup}) in order to break the gauge orbit integration \cite{Hoehn:2014fka}. We shall exhibit this in detail for the various moves below. This is also the reason why physical post--states ${}^+\psi^{\rm phys}_k$ at $k$ need not also be projected on solutions to the pre--constraints ${}^-\hat{C}^k_\beta$ of case (ii), for otherwise one obtains even more divergences from the double action of ${}^-\mathbb{P}^B_k$ in (\ref{pup}, \ref{sup}). For a discussion of these issues in the context of global moves, see \cite{Hoehn:2014fka}.

If requirements (a) and (b) are fulfilled and the divergences of case (i) regularized, we can define via (\ref{sup}) a map which we shall call {\it physical state updating}
\ba
\fu_{k\rightarrow k+1}:{}^+\ch^{\rm phys}_k\rightarrow {}^+\ch^{\rm phys}_{k+1},\q\q\q{}^+\psi^{\rm phys}_{k+1}=\fu_{k\rightarrow k+1}({}^+\psi^{\rm phys}_k),\label{uk1}
\ea
where, for the time being, we write formally
\ba
\fu_{k\rightarrow k+1}:=\int\prod_odx^o_k\,M_{k\rightarrow k+1}\,e^{iS_{k+1}/\hbar}\prod_\alpha\delta(G_\alpha)=\int\prod_odx^o_k\,{}^+\mathbb{P}_{k+1}^\nu\,({}^-\mathbb{P}_k^B)^*\,\upsilon_{k\rightarrow k+1}\label{uk}
\ea
and $\delta(G_\alpha)$ are some suitable gauge fixing conditions for case (i) (including normalization). We shall make the construction (\ref{uk}) explicit for the various moves below. As we shall see below, $\fu_{k\rightarrow k+1}$ is {\it non-unitary} in the presence of coarse graining pre--constraints ${}^-\hat{C}^k_\beta$ of case (ii) above. In this case, $\fu_{k\rightarrow k+1}$ involves an improper projection ${}^-\mathbb{P}^B_k({}^+\ch^{\rm phys}_k)$ which is no longer contained in ${}^+\ch^{\rm phys}_k$.

Moreover, $\fu_{k\rightarrow k+1}$ can also be viewed as a map of unitary maps. Recall that the map $U_{0\rightarrow k}$ (\ref{unitary}) defines a unitary isomorphism from the pre--physical Hilbert space ${}^-\ch^{\rm phys}_0$ at $n=0$ to the post--physical Hilbert space ${}^+\ch^{\rm phys}_k$ at $n=k$. We now also have
\ba
U_{0\rightarrow k+1}=\fu_{k\rightarrow k+1}(U_{0\rightarrow k})
\ea
which again must be a unitary map as for any global move \cite{Hoehn:2014fka}, however, now from the pre--physical Hilbert space ${}^-\tilde{\ch}^{\rm phys}_0$ of the global move $0\rightarrow k+1$ to the post--physical Hilbert space ${}^+\ch^{\rm phys}_{k+1}$ at $k+1$. If $\fu_{k\rightarrow k+1}$ is non-unitary due to an improper projection ${}^-\mathbb{P}^B_k$, ${}^-\tilde{\ch}^{\rm phys}_0$ can no longer be contained in ${}^-\ch^{\rm phys}_0$ because $U_{0\rightarrow k+1}$ must be unitary. 

Indeed, in \cite{Hoehn:2014fka} it was discussed (for global moves) how non-trivial pre--constraints at $k$ effectively `propagate' backward to the initial step $n=0$ to project the states there to a set that evolves under $0\rightarrow k$ to post--states at $k$ that also satisfy the pre--constraints at $k$. These non-trivial pre--constraints are coarse graining conditions and the move that implements them leads to a coarse graining of the discretization at both $0$ and $k+1$. This is the reason why the pre-- and post--physical Hilbert spaces at a given time step $k$ depend on the particular evolution move one is considering. Evolving further into the future or into the past can effectively coarse grain the discretization at $k$: 
once a state at $k$ has been coarse grained and information about the finer discretization has been integrated out, it can never be regained. We shall comment on this further in section \ref{sec_dirac} below. For a related discussion, see also \cite{Hoehn:2014fka,Dittrich:2013xwa}.

We briefly mentioned in section \ref{sec_cllocal} that in the classical formalism \cite{Dittrich:2011ke,Dittrich:2013jaa} it is convenient to handle temporally varying discretizations by suitable configuration and phase space extensions. Although this is not strictly necessary it gives the momentum updating equations an intuitive form, as we shall see shortly for the explicit local moves below. In particular, the auxiliary variables $x^n_k$ and $x^o_{k+1}$ that have been introduced for dimensional reasons are immediately accompanied by constraints $p^k_n=0$ and $p^{k+1}_o=0$ which can be understood as equations of motion \cite{Dittrich:2013jaa,Dittrich:2011ke}. These constraints are (auxiliary) gauge symmetry generators whose orbits are parametrized by the free parameters $x^n_k,x^o_{k+1}$. These auxiliary variables can, of course, also be easily incorporated into the quantum theory. As in \cite{Hoehn:2014fka}, the configuration space extension translates into a Hilbert space language as
\ba
\ch^{\rm kin}_k&=&L^2(\mathbb{R}^{N_k},dx_k)\q\q\q\,\,\,\,\,\underset{\text{\tiny `extend'}}{\longmapsto}\q\q\overline{\ch}^{\rm kin}_k\,\,=L^2(\mathbb{R}^{N_k}\times\mathbb{R}^{N^{ext}_k},dx_k\prod_ndx^n_k)\label{extension}\\
\ch^{\rm kin}_{k+1}&=&L^2(\mathbb{R}^{N_{k+1}},dx_{k+1})\q\q\underset{\text{\tiny `extend'}}{\longmapsto}\q\q\overline{\ch}^{\rm kin}_{k+1}=L^2(\mathbb{R}^{N_{k+1}}\times\mathbb{R}^{N^{ext}_{k+1}},dx_{k+1}\prod_odx^o_{k+1})\nn
\ea
where $\mathbb{R}^{N^{ext}_k}$ is coordinatized by $x^n_k$ and $\mathbb{R}^{N^{ext}_{k+1}}$ is coordinatized by $x^o_{k+1}$. This is, however, not a proper Hilbert space extension because all infinite dimensional separable Hilbert spaces are isomorphic. Rather, the `extension' refers to the number of variables associated to the underlying discretization which are necessary to describe a quantum state (and thereby to the dimension of the configuration space over which the square integrable functions are defined). The constraints $p^k_n=0$ and $p^{k+1}_o=0$ can be promoted to constraint operators on $\overline{\ch}^{\rm kin}_k$ and $\overline{\ch}^{\rm kin}_{k+1}$, respectively. As derivative operators they obviously imply nothing else than independence of physical states of the auxiliary variables $x^n_k,x^o_{k+1}$. If one considers an expanding or shrinking lattice, one would extend the configuration space at a given time step $k$ more and more in the course of evolving further into the `past' or `future' in which case new auxiliary variables and new constraints at $k$ arise. This amounts to adding gauge degrees of freedom to the system at $k$. From the discussion in \cite{Hoehn:2014fka} it follows that post--physical states at $k$ and $k+1$ are cylindrical functions with respect to the unextended configuration spaces $\mathbb{R}^{N_k}$ and $\mathbb{R}^{N_{k+1}}$, respectively, and cylindrical consistency\footnote{By cylindrical consistency we mean that the integration of such cylindrical functions over a space with respect to which they are cylindrical is independent of the choice of this space.} (specifically of the physical inner product) is ensured by the cylindrical Faddeev-Popov measures for the auxiliary variables at $k$ and $k+1$
\ba
d\xi_k=\prod_ndx^n_{k+1}\,\delta(x^n_k-x'^n_k),\q\q\q\q d\xi_{k+1}=\prod_odx^o_k\,\delta(x^o_{k+1}-x'^o_{k+1}).\nn
\ea
We shall thus call the `extension' (\ref{extension}) {\it cylindrical extension}. 

The relation between the different Hilbert spaces can be summarized in a diagram 
\begin{diagram}
\ch^{\rm kin}_k&&&&&&\ch^{\rm kin}_{k+1}\\
\dTo^{\text{cyl.\ ext.}\,\,\,}&&&&&&\dTo_{\,\,\,\text{cyl.\ ext.}}\\
\overline{\ch}^{\rm kin}_k&&&&&&\overline{\ch}^{\rm kin}_{k+1}\\
\dTo^{{}^+\mathbb{P}_k\,\,\,}&&&&&&\dTo_{\,\,\,{}^+\mathbb{P}_{k+1}}\\
{}^+\ch^{\rm phys}_k&&&\rTo^{\fu_{k\rightarrow k+1}}&&&{}^+\ch^{\rm phys}_{k+1}
\end{diagram}
where ${}^+\mathbb{P}_k$ and ${}^+\mathbb{P}_{k+1}$ denote the improper projectors onto {\it all} post--constraints at $k$ (incl.\ $\hat{p}^k_n$) and {\it all} post--constraints at $k+1$ (incl.\ $\hat{p}^{k+1}_o$ and ${}^+\hat{C}^{k+1}_{\nu'}$), respectively. We emphasize that there is no direct map from $\ch^{\rm kin}_k$ to ${}^+\ch^{\rm phys}_{k+1}$ because $M_{k\rightarrow k+1}\,e^{iS_{k+1}/\hbar}$ is not a propagator.

In the sequel we shall now study in detail four different types of local evolution moves in sections \ref{sec_I}--\ref{sec_IV} which have classically been studied in \cite{Dittrich:2013jaa,Dittrich:2011ke}. The moves of type I--IV include the various types of Pachner moves \cite{pachner1,pachner2,Dittrich:2011ke} appearing in the dynamics of simplicial gravity and field theory on a triangulation. Any other conceivable local move can be treated in complete analogy. We shall henceforth not explicitly worry about cylindrical extensions and ignore the auxiliary variables $x^n_k,x^o_{k+1}$ at $k,k+1$ in the quantum theory. It is implicitly understood that physical states can be treated as cylindrical functions as sketched above. Finally, in sections \ref{sec_comp} and \ref{sec_momup} we shall discuss invertible compositions of the four types of local moves and exhibit the quantum version of momentum updating.

\subsection{Quantum moves of type I}\label{sec_I} 

Moves of type I introduce $K$ `new variables' but do not remove `old variables.' Classically, one can extend the phase space at $k$ by the $K$ auxiliary pairs $(x_k^n,p^k_n)$ corresponding to the $K$ `new' canonical pairs $(x_{k+1}^n,p^{k+1}_n)$ at $k+1$. The momentum updating map $\fh_k$ reads \cite{Dittrich:2013jaa,Dittrich:2011ke}
\ba\label{anh1}
x^b_k&=&x^b_{k+1}  \, ,\q\q\q p^{k+1}_b\,=\,p^k_b \; , \\
x^e_k&=&x^e_{k+1}\,,\q\q\q p^{k+1}_e\,=\,p^k_e+\frac{\partial S_{k+1}(x^e_{k+1}, x^n_{k+1})}{\partial x^e_{k+1}}  
\; ,    \label{anh1b}  \\
p^{k}_n&=&0 \,,\q\q\q\q \,\,\,p^{k+1}_n\,=\, \frac{\partial S_{k+1}(x^e_{k+1},x^n_{k+1})  }{\partial x^n_{k+1}}\,. \label{anh1c} 
\ea
$S_{k+1}$ is chosen as a function of only the variables at time $k+1$. Because of the $K$ constraints $C_n^k=p^k_n$ which are simultaneously pre-- and post--constraints and the $K$ post--constraints ${}^+C_n^{k+1}=p^{k+1}_n- \frac{\partial S_{k+1}(x^e_{k+1},x^n_{k+1})  }{\partial x^n_{k+1}}$ the variables $x^n_k,x^n_{k+1}$ remain undetermined.

A type I move is thus a dynamical refining move which maps a (classical or quantum) state on a coarser discretization to a state on a finer discretization without adding any new physical degrees of freedom (see also section \ref{sec_dirac}). For instance, the 1--3 Pachner evolution move in 3D Regge Calculus and the 1--4 and 2--3 Pachner evolution moves in 4D Regge Calculus are of type I \cite{Dittrich:2011ke,Dittrich:2013jaa}. In Regge Calculus one has $\cq_k\simeq\mathbb{R}^{N_k}_+$, rather than $\cq_k\simeq\mathbb{R}^{N_k}$, such that the present formalism does not immediately apply to this simplicial gravity model.\footnote{Furthermore, in 3D Regge Calculus, the canonical momenta are angles (thus taking value in a compact interval), while in 4D Regge Calculus the momenta are related to angles \cite{Dittrich:2011ke}.} But, using the techniques in \cite{isham2}, it can be suitably adapted. However, the present formalism can, in particular, be applied to linearized Regge Calculus based on a perturbation around a flat background solution \cite{Dittrich:2009fb,dh4}.

In the quantum theory the propagator and state updating (\ref{pup}, \ref{sup}) for a type I move read
\ba
K_{0\rightarrow k+1}(x_0,x_{k+1})&=&M_{k\rightarrow k+1}(x_{k+1}^e,x_{k+1}^n)\,e^{iS_{k+1}(x_{k+1}^e,x_{k+1}^n)/\hbar}\,K_{0\rightarrow k}(x_0,x_k).\label{kupI}\\
{}^+\psi^{\rm phys}_{k+1}(x_{k+1})&=&M_{k\rightarrow k+1}\,e^{iS_{k+1}(x_{k+1})/\hbar}\,{}^+\psi^{\rm phys}_k(x_k).\label{supI}\ea
(We recall the two left equations in (\ref{anh1}, \ref{anh1b}).) There is no integral involved because moves of type I do not lead to any equations of motion classically (i.e.\ no variables become `bulk'). 

In the classical formalism type I moves preserve all post--constraints and the symplectic structure restricted to the post--constraint surfaces at $k$ and $k+1$ \cite{Dittrich:2013jaa}. The analogous result holds in the quantum theory:

\begin{Theorem}\label{thm_I}{\bf{ (Type I)}}
The physical state updating map
\ba
\fu^I_{k\rightarrow k+1}=M^I_{k\rightarrow k+1}\,e^{iS_{k+1}(x^e_{k+1},x^n_{k+1})/\hbar}\nn
\ea
with constant measure $M^I_{k\rightarrow k+1}=const$ is a unitary map ${}^+\ch^{\rm phys}_k\rightarrow {}^+\ch^{\rm phys}_{k+1}$ which preserves all post--constraints from step $k$ which admit a power series expansion.
\end{Theorem}

\begin{proof}
The proof is given in appendix \ref{app_1}.
\end{proof}

We shall fix the measure to $M_{k\rightarrow k+1}^I=\left(\f{1}{2\pi\hbar}\right)^{K/2}$ below in lemmas \ref{lema} and \ref{lemb} in section \ref{sec_comp}. The state updating map $\fu^I_{k\rightarrow k+1}$ can also be viewed as a dynamical embedding of ${}^+\ch^{\rm phys}_k$ into ${}^+\ch^{\rm phys}_{k+1}$ in the sense of \cite{Dittrich:2012jq,Dittrich:2013xwa}.
 
\begin{Example}\label{ex_1}
\emph{We recall the example of a scalar field living on the vertices of a 2D triangulated space-time from \cite{Dittrich:2013jaa}. The 1--2 Pachner move within a 1D hypersurface $\Sigma$ corresponds to gluing a triangle onto one edge of $\Sigma_k$ and is of type I. It generates a new vertex $v$ at $k+1$ with one `new' field variable $\phi_{k+1}^v$ and preserves the remaining vertices of $\Sigma_k$. The move is depicted in figure \ref{fig_12}. The classical momentum updating map is given by \cite{Dittrich:2013jaa}
\ba
\phi^b_k&=&\phi^b_{k+1}  \, ,\q\q\q \pi^{k+1}_b\,=\,\pi^k_b\,,\q\q\q\q\q\q\q\q\q\q\q\q\q\q\q b=1,4,5 \; ,\nn \\
\phi^e_k&=&\phi^e_{k+1}\,,\q\q\q \pi^{k+1}_e\,=\,\pi^k_e+\phi^e_{k+1}-\f{1}{2}\left(\phi^v_{k+1}+\phi^{e+1}_{k+1}\right)\,,\q\q\q\, e=2,3\; ,    \nn  \\
\pi^{k}_v&=&0 \,,\q\q\q\q \,\,\,\pi^{k+1}_v\,=\, \phi^v_{k+1}-\f{1}{2}\left(\phi^2_{k+1}+\phi^3_{k+1}\right)\,. \nn
\ea
Note that $e+1=3$ if $e=2$ and $e+1=2$ if $e=3$. The last equation contains the single post--constraint ${}^+C^{k+1}_v=\pi^{k+1}_v- \phi^v_{k+1}+\f{1}{2}\left(\phi^2_{k+1}+\phi^3_{k+1}\right)$.}

\emph{The action of a scalar field on an equilateral triangle with vertices $v_1,v_2,v_3$ is \cite{Sorkin:1975jz}
\ba
S_\Delta=\f{1}{4}\left((\phi^{v_1}_{k+1}-\phi^{v_2}_{k+1})^2+(\phi^{v_1}_{k+1}-\phi^{v_3}_{k+1})^2+(\phi^{v_2}_{k+1}-\phi^{v_3}_{k+1})^2\right).\nn
\ea
Thus, the physical state updating map in the quantum theory corresponding to this move reads
\ba
{}^+\psi^{\rm phys}_{k+1}(\phi_{k+1})=\fu^{\tiny 1-2}_{k\rightarrow k+1}({}^+\psi^{\rm phys}_k)=\f{1}{\sqrt{2\pi\hbar}}\,e^{\f{i}{4\hbar}((\phi^v_{k+1}-\phi^2_{k+1})^2+(\phi^v_{k+1}-\phi^3_{k+1})^2+(\phi^2_{k+1}-\phi^3_{k+1})^2)}\,{}^+\psi^{\rm phys}_k(\phi_k),\nn
\ea
where we have already made use of lemma \ref{lema} below for the measure factor. Indeed, as one can check, 
\ba
{}^+\hat{C}^{k+1}_v\,\fu^{\tiny 1-2}_{k\rightarrow k+1}=\left(\hat{\pi}^{k+1}_v- \phi^v_{k+1}+\f{1}{2}\left(\phi^2_{k+1}+\phi^3_{k+1}\right)\right)\,\fu^{\tiny 1-2}_{k\rightarrow k+1}=0.\nn
\ea
}
\begin{center}
\begin{figure}[htbp!]
\psfrag{s}{\small$\Sigma_k$}
\psfrag{sk}{\small$\Sigma_{k+1}$}
\psfrag{v}{\small$v$}
\psfrag{fv}{\footnotesize$\phi^v_{k+1}$}
\psfrag{1}{\footnotesize $\phi^1_{k+1}$}
\psfrag{2}{\footnotesize$\phi^2_{k+1}$}
\psfrag{3}{\footnotesize$\phi^3_{k+1}$}
\psfrag{4}{\footnotesize$\phi^4_{k+1}$}
\psfrag{5}{\footnotesize$\phi^5_{k+1}$}
\psfrag{a}{\footnotesize$\phi^1_k$}
\psfrag{b}{\footnotesize$\phi^2_k$}
\psfrag{c}{\footnotesize$\phi^3_k$}
\psfrag{d}{\footnotesize$\phi^4_k$}
\psfrag{e}{\footnotesize$\phi^5_k$}
\begin{subfigure}[b]{.22\textwidth}
\centering
\includegraphics[scale=.4]{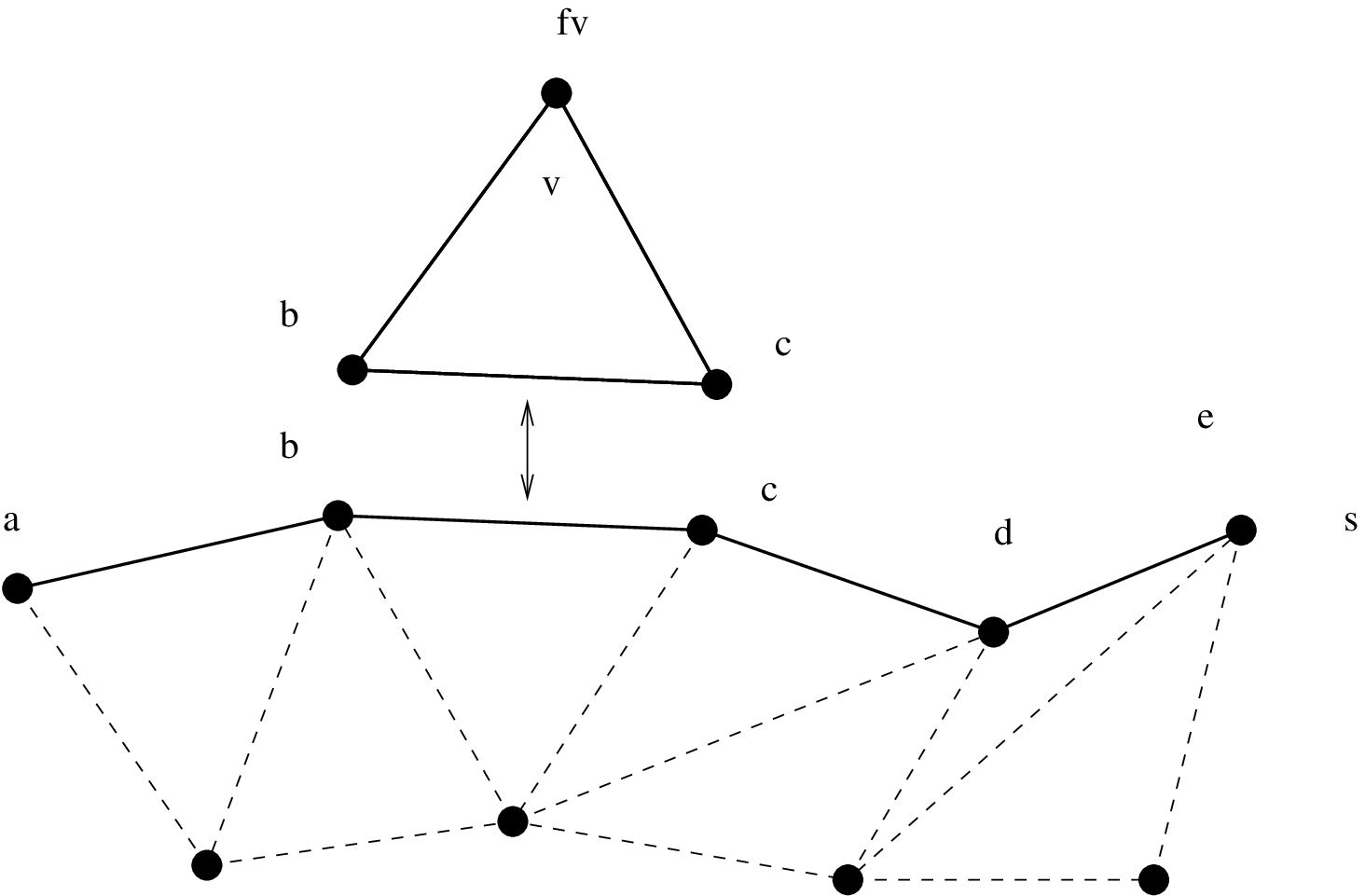}
\centering
\caption{\small }
\end{subfigure}
\hspace*{4.3cm}
\begin{subfigure}[b]{.22\textwidth}
\centering
\includegraphics[scale=.5]{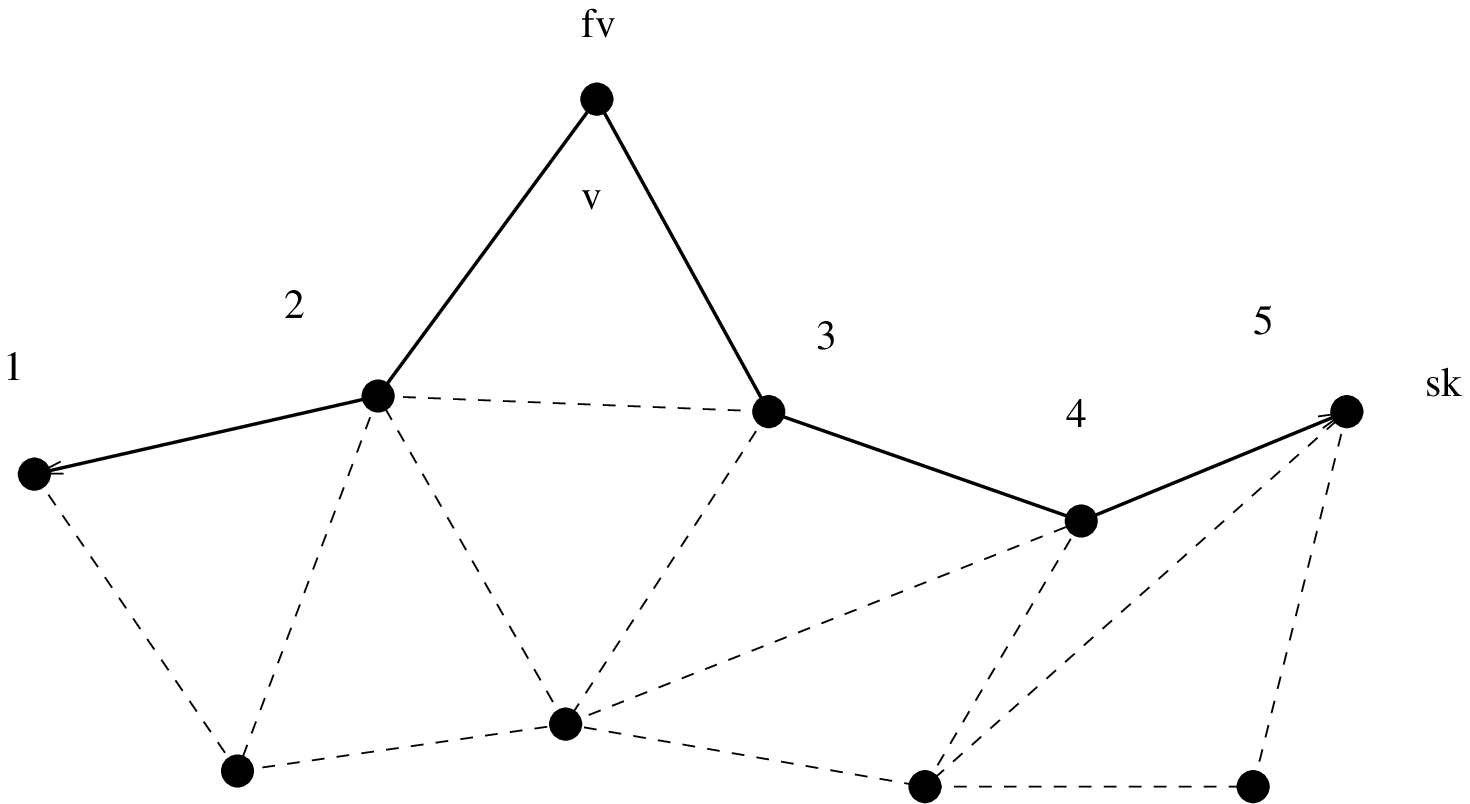}
\caption{\small }
\end{subfigure}
\caption{\small The 1--2 Pachner move for a scalar field on a 2D triangulated space-time is of type I. It introduces one new field variable $\phi^v_{k+1}$ at step $k+1$.}\label{fig_12}
\end{figure}
\end{center}
\end{Example}

\subsection{Quantum moves of type II}\label{sec_II}

Type II moves are the time reverse of type I, i.e.\ they annihilate $K$ `old variables' without introducing `new variables'. Classically, one may extend the phase space at step $k+1$ by $K$ auxiliary pairs $(x^o_{k+1},p^{k+1}_o)$. The momentum updating map $\fh_k$ takes the form \cite{Dittrich:2013jaa,Dittrich:2011ke}
\ba\label{case2}
x^b_k&=&x^b_{k+1}  \, ,\q\q\q p^{k+1}_b\,=\,p^k_b \; , \\
x^e_k&=&x^e_{k+1}\,,\q\q\q\q p^{k}_e\,=\,p^{k+1}_e-\frac{\partial S_{k+1}(x^e_{k}, x^o_{k})}{\partial x^e_{k}}  
\; ,    \label{case2b}  \\
p^{k+1}_o&=&0 \,,\q\q\q\q\q \,\,\,p^{k}_o\,=\, -\frac{\partial S_{k+1}(x^e_{k},x^o_{k})  }{\partial x^o_{k}}\,. \label{case2c} 
\ea
($S_{k+1}$ is a function of the variables of $k$.) There are $K$ pre--constraints ${}^-C^k_o=p^{k}_o+\frac{\partial S_{k+1}(x^e_{k},x^o_{k})  }{\partial x^o_{k}}$ and $K$ coinciding pre-- and post--constraints $C^{k+1}_o=p^{k+1}_o$. The $x^o_{k+1}$ are left undetermined. 

A type II move is therefore a dynamical coarse graining move which maps a (classical or quantum) state on a finer discretization to a state on a coarser discretization. For example, the 3--1 Pachner evolution move in 3D Regge Calculus and the 4--1 and 3--2 Pachner evolution moves in 4D Regge Calculus are of type II \cite{Dittrich:2011ke,Dittrich:2013jaa} (although, as previously mentioned, in Regge Calculus $\cq_k\simeq\mathbb{R}^{N_k}_+$).

For type II moves we have to recall the distinction of pre--constraints discussed in section \ref{sec_locgen}. We thus split the constraint set ${}^-\hat{C}^k_o$, $o=1,\ldots,K$, into a set of $C^k_\alpha$, $\alpha=1,\ldots,K_\alpha$, of case (i) and into a set ${}^-C^k_\beta$, $\beta=1,\ldots,K_\beta$ of case (ii), where $K_\alpha+K_\beta=K$. If case (ii) occurs, a type II move is a non-trivial coarse graining move. 
While case (ii) can, in general, occur for the 4--1 and 3--2 Pachner moves in 4D Regge Calculus, it does {\it not} arise for the 3--1 Pachner move in 3D Regge Calculus or the 4--1 Pachner move in 4D linearized Regge Calculus where the constraints $C^k_\alpha$ correspond to Hamiltonian and diffeomorphism constraints \cite{Dittrich:2011ke,dh4}.

In the presence of case (i) pre--constraints $\hat{C}^k_\alpha$ we have to regularize the physical state updating map (\ref{uk}). To this end, we make use of (\ref{kinup}) to write for the propagator updating\footnote{Since all new pre--constraints are linear in the moment, one can write for any projector $\delta({}^-\hat{C}^k_o)=e^{-iS_{k+1}/\hbar}\,\delta(\hat{p}^k_o)\,e^{iS_{k+1}/\hbar}$ such that no ordering ambiguities arise and the projectors $\mathbb{P}^A_k$ and ${}^-\mathbb{P}^B_k$ commute.}
\ba
K_{0\rightarrow k+1}(x_0,x_{k+1})&=&\int\prod_o dx^o_k\,(\mathbb{P}^A_k)^*({}^-\mathbb{P}^B_k)^*\,\upsilon_{k\rightarrow k+1}(x^o_k,x^e_k)\,K_{0\rightarrow k}(x_0,x_k)\nn\\
&=&\int\prod_o dx^o_k\,({}^-\mathbb{P}^B_k)^*\,\upsilon_{k\rightarrow k+1}(x^o_k,x^e_k)\,\mathbb{P}^A_k\,K_{0\rightarrow k}(x_0,x_k).\label{lalala}
\ea
The last step is possible because of $\upsilon_{k\rightarrow k+1}$ and $\kappa_{0\rightarrow k}$ in $K_{0\rightarrow k}$ being square integrable at $k$ and because the $\hat{C}^k_\alpha$ in $\mathbb{P}^A_k$ only contain $\hat{p}^k_o$ as derivative operators and these are self-adjoint with respect to integration over the $x^o_k$ (see also \cite{Hoehn:2014fka} on this).\footnote{In particular, we require that $\upsilon_{k\rightarrow k+1}\,\kappa_{0\rightarrow k}\rightarrow0$ for $x^o_k\rightarrow\pm\infty$.} Recall that the projector $\mathbb{P}^A_k$ is already contained in $K_{0\rightarrow k}$. One instance of it must thus be dropped or regularized in the above expression. Given that the new pre--constraints are linear in $\hat{p}^k_\alpha$ and conjugate to $x^\alpha_k$, one can apply lemma 4.1 of \cite{Hoehn:2014fka} which in this case implies
\ba
(2\pi\hbar)^{K_\alpha}\mathbb{P}^A_k\prod_\alpha\delta(x'^\alpha_k-x^\alpha_k)\,\mathbb{P}^A_k\,\psi^{\rm kin}_k=\mathbb{P}^A_k\,\psi^{\rm kin}_k.\label{lem1}
\ea
That is, for constraints linear in the momenta, dropping one instance of $\mathbb{P}^A_k$ in (\ref{lalala}) is equivalent to inserting a Faddeev-Popov gauge fixing condition. This allows us to finally write the Faddeev-Popov regularized propagator and state updating (\ref{pup}, \ref{sup}) for type II as
\ba
K^{\rm reg}_{0\rightarrow k+1}(x_0,x_{k+1})&=&(2\pi\hbar)^{K_\alpha}\int\prod_o dx^o_k\,\prod_\alpha\delta(x'^\alpha_k-x^\alpha_k)\,M^{II}_{k\rightarrow k+1}(x_{k}^e,x_{k}^o)\,e^{iS_{k+1}(x_{k}^e,x_{k}^o)/\hbar}\nn\\
&&\q\q\q\q\q\q\times K_{0\rightarrow k}(x_0,x_k),\label{kupII}\\
{}^+\psi^{\rm phys}_{k+1}(x_{k+1})&=&(2\pi\hbar)^{K_\alpha}\int\prod_o dx^o_k\,\prod_\alpha\delta(x'^\alpha_k-x^\alpha_k)\,M^{II}_{k\rightarrow k+1}(x_{k}^e,x_{k}^o)\,e^{iS_{k+1}(x_{k}^e,x_{k}^o)/\hbar}\nn\\
&&\q\q\q\q\q\q\times{}^+\psi^{\rm phys}_k(x_k).\label{supII}
\ea

Although the present formalism must be generalized in order to be applicable to Regge Calculus (with the exception of the linearized theory \cite{Dittrich:2009fb,dh4}), its qualitative features can be expected to survive. In particular, the above discussion suggests that the 3--1 move in 3D Regge Calculus and the 4--1 move in 4D linearized Regge Calculus lead to divergences in the quantum theory that must be regularized. 

Classically, type II moves preserve all post--constraints from $k$ and only preserve the symplectic form restricted to the post--constraint surface, provided no pre--constraints of case (ii) ${}^-C^k_\beta$ occur \cite{Dittrich:2013jaa}. If, on the other hand, at least one ${}^-C^k_\beta$ arises at $k$, the classical type II move reduces the rank of the symplectic form from $k$ to $k+1$ \cite{Dittrich:2013jaa}. These pre--constraints are non-trivial coarse graining conditions that reduce the number of physical degrees of freedom. 
The situation is analogous in the quantum theory.

\begin{Theorem}\label{thm_II}{\bf{ (Type II)}}
The physical state updating map
\ba
\fu^{II}_{k\rightarrow k+1}=(2\pi\hbar)^{K_\alpha}\int\prod_odx^o_k\,\prod_\alpha\delta(x'^\alpha_k-x^\alpha_k)\,M^{II}_{k\rightarrow k+1}\,e^{iS_{k+1}(x^e_{k},x^o_{k})/\hbar}\nn
\ea
with constant measure $M^{II}_{k\rightarrow k+1}=const$ defines a map ${}^+\ch^{\rm phys}_k\rightarrow {}^+\ch^{\rm phys}_{k+1}$ which preserves all post--constraints from step $k$ which admit a power series expansion. $\fu^{II}_{k\rightarrow k+1}$ is
\begin{enumerate}
\item unitary if no pre--constraints ${}^-\hat{C}^k_\beta$ of case (ii) in section \ref{sec_locgen} occur,
\item non-unitary if at least one pre--constraint ${}^-\hat{C}^k_\beta$ of case (ii) in section \ref{sec_locgen} occurs.
\end{enumerate}
\end{Theorem}

\begin{proof}
The proof is given in appendix \ref{app_1}.
\end{proof}

The measure will be fixed to $M_{k\rightarrow k+1}^{II}=\left(\f{1}{2\pi\hbar}\right)^{K/2}$ in lemmas \ref{lema} and \ref{lemb} in section \ref{sec_comp}.

\begin{Example}
\emph{This example is, again, taken from \cite{Dittrich:2013jaa}. We consider the reverse of example \ref{ex_1}. The 2--1 Pachner move for the scalar field is depicted in figure \ref{fig_21} and of type II. It pushes a vertex $v^*$ with an `old' field variable $\phi^{v^*}_k$ into the bulk. Momentum updating is given by \cite{Dittrich:2013jaa}
\ba
\phi^b_k&=&\phi^b_{k+1}  \, ,\q\q\q \pi^{k+1}_b\,=\,\pi^k_b\,,\q\q\q\q\q\q\q\q\q\q\q\q\q\q\,\,\, b=1,4 \; ,\nn \\
\phi^e_k&=&\phi^e_{k+1}\,,\q\q\q\q \pi^{k}_e\,=\,\pi^{k+1}_e-\phi^e_{k}+\f{1}{2}\left(\phi^{v^*}_{k}+\phi^{e+1}_{k}\right)\,,\q\q\q e=2,3\; ,   \nn  \\
\pi^{k+1}_{v^*}&=&0 \,,\q\q\q\q\q \,\,\,\pi^{k}_{v^*}\,=\, -\phi^{v^*}_{k}+\f{1}{2}\left(\phi^2_{k}+\phi^3_{k}\right)\,.\nn
\ea
Again, $e+1=3$ if $e=2$ and $e+1=2$ if $e=3$. The last equation contains the new pre--constraint ${}^-C^k_{v^*}=\pi^{k}_{v^*}+\phi^{v^*}_{k}-\f{1}{2}\left(\phi^2_{k}+\phi^3_{k}\right)$.}

\emph{In analogy to example \ref{ex_1}, the physical state updating map corresponding to this move reads
\ba
{}^+\psi^{\rm phys}_{k+1}(\phi_{k+1})=\fu^{\tiny 2-1}_{k\rightarrow k+1}({}^+\psi^{\rm phys}_k)=\f{1}{\sqrt{2\pi\hbar}}\int_{-\infty}^{+\infty} d\phi^{v^*}_k\,e^{\f{i}{4\hbar}((\phi^{v^*}_k-\phi^2_{k})^2+(\phi^{v^*}_k-\phi^3_{k})^2+(\phi^2_{k}-\phi^3_{k})^2)}\,{}^+\psi^{\rm phys}_k(\phi_k),\nn
\ea
where use of lemma \ref{lema} of section \ref{sec_comp} for the measure factor has been made. Clearly, (as in (\ref{cond}))
\ba
&&\!\!\!\!\!{}^-\hat{C}^{k}_{v^*}\,e^{-\f{i}{4\hbar}((\phi^{v^*}_k-\phi^2_{k})^2+(\phi^{v^*}_k-\phi^3_{k})^2+(\phi^2_{k}-\phi^3_{k})^2)}\label{2Dprecon}\\
&&\q\q\q\q=\left(\hat{\pi}^{k}_{v^*}+\phi^{v^*}_{k}-\f{1}{2}\left(\phi^2_{k}+\phi^3_{k}\right)\right)\,e^{-\f{i}{4\hbar}((\phi^{v^*}_k-\phi^2_{k})^2+(\phi^{v^*}_k-\phi^3_{k})^2+(\phi^2_{k}-\phi^3_{k})^2)}=0.\nn
\ea
If the post--physical state ${}^+\psi^{\rm phys}_k(\phi_k)$ at $k$ satisfies this pre--constraint too, a gauge fixing factor $(2\pi\hbar)\,\delta(\phi^{v^*}_k-\phi'^{v^*}_k)$ must be introduced for regularization. }
\begin{center}
\begin{figure}[htbp!]
\psfrag{s}{\small$\Sigma_k$}
\psfrag{sk}{\small$\Sigma_{k+1}$}
\psfrag{v}{\small$v^*$}
\psfrag{fv}{\footnotesize$\phi^{v^*}_{k}$}
\psfrag{1}{\footnotesize $\phi^1_{k+1}$}
\psfrag{2}{\footnotesize$\phi^2_{k+1}$}
\psfrag{3}{\footnotesize$\phi^3_{k+1}$}
\psfrag{4}{\footnotesize$\phi^4_{k+1}$}
\psfrag{5}{\footnotesize$\phi^5_{k+1}$}
\psfrag{a}{\footnotesize$\phi^1_k$}
\psfrag{b}{\footnotesize$\phi^2_k$}
\psfrag{c}{\footnotesize$\phi^3_k$}
\psfrag{d}{\footnotesize$\phi^4_k$}
\psfrag{e}{\footnotesize$\phi^5_k$}
\begin{subfigure}[b]{.22\textwidth}
\centering
\includegraphics[scale=.5]{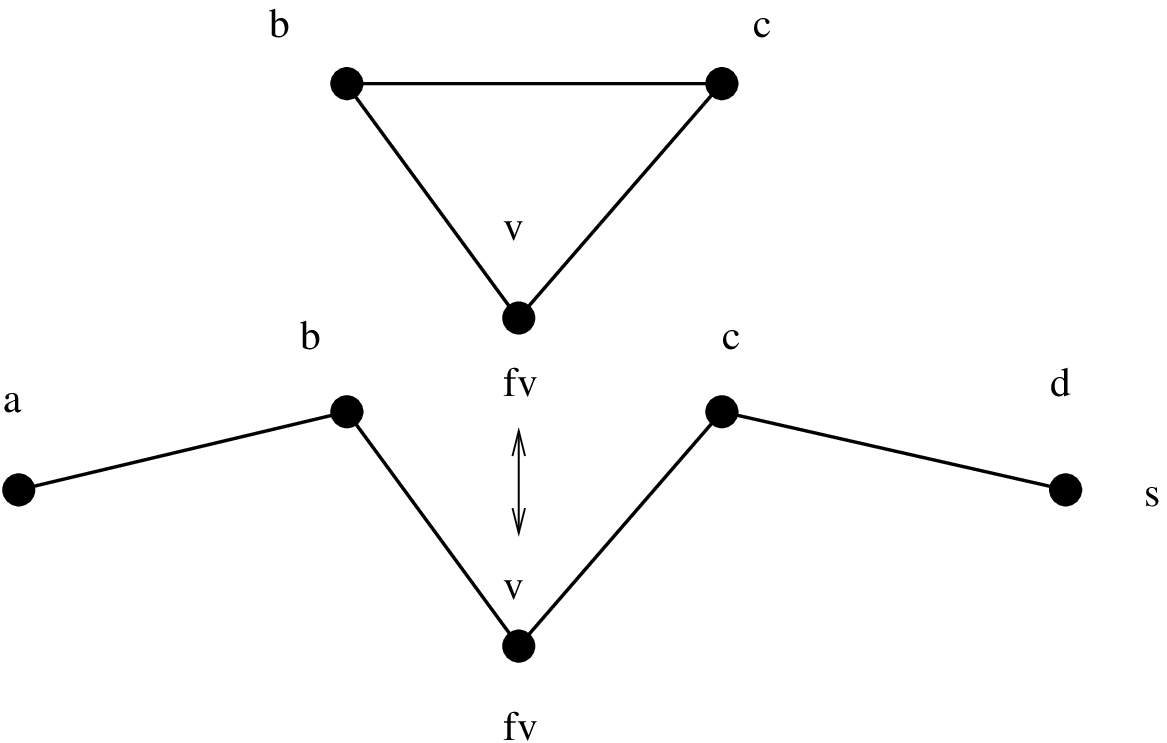}
\centering
\caption{\small }
\end{subfigure}
\hspace*{4.4cm}
\begin{subfigure}[b]{.22\textwidth}
\centering
\includegraphics[scale=.5]{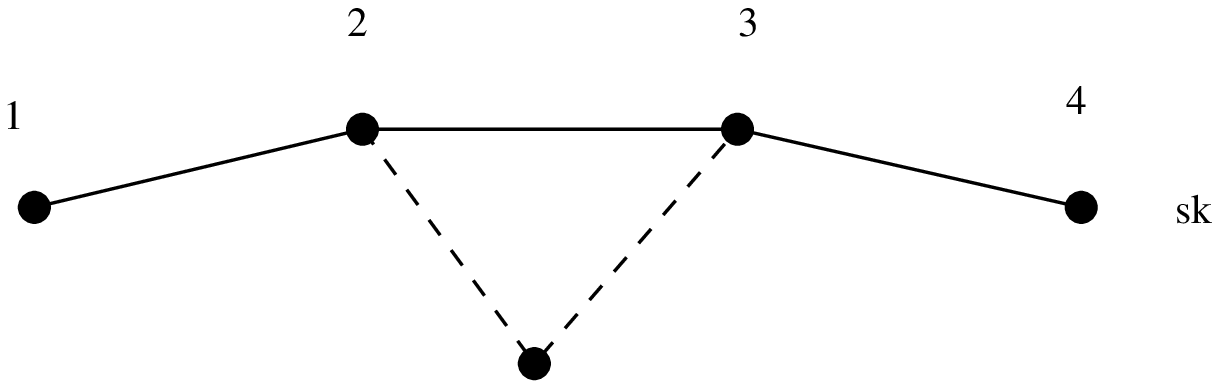}
\vspace*{.005cm}
\caption{\small }
\end{subfigure}
\caption{\small The 2--1 Pachner move for a scalar field on a 2D space-time triangulation is of type II. It corresponds to gluing a triangle onto two edges in $\Sigma_k$ which annihilates a vertex $v^*$ with the variable $\phi^{v^*}_{k}$.}\label{fig_21}
\end{figure}
\end{center}
\end{Example}

\subsection{Quantum moves of type III}\label{sec_III}

A type III move both eliminates $K$ `old variables' and creates $K$ `new variables'. Classically, the phase space at $k$ is extended by $K$ auxiliary pairs $(x^n_k,p^k_n)$ and the phase space at $k+1$ is extended by $K$ auxiliary pairs $(x^o_{k+1},p^{k+1}_o)$. The momentum updating map $\fh_k$ reads \cite{Dittrich:2013jaa,Dittrich:2011ke}
\ba\label{anh21}
x^b_k&=&x^b_{k+1}  \, ,\q\q\q p^{k+1}_b\,=\,p^k_b \, , \\
x^e_k&=&x^e_{k+1}\,,\q\q\q p^{k+1}_e\,=\,p^k_e+\frac{\partial S_{k+1}(  x^e_{k+1},x^o_k,x^n_{k+1})  }{\partial x^e_{k+1}} 
\, ,\label{anh21b}\\
p^{k}_n&=&0 \,,\q\q\q\q\,\, \,p^{k+1}_n\,=\, \frac{\partial S_{k+1}(  x^e_{k+1},x^o_k,x^n_{k+1}) }{\partial x^n_{k+1}}\,\,
 , \label{anh21c}\\
p^{k+1}_o &=& 0 \,,\q\q\q\q\,\, \,p^{k}_o\q\,=\, -\frac{\partial S_{k+1} (  x^e_{k+1},x^o_k,x^n_{k+1}) }{\partial x^o_{k}}\,
 . \label{anh21d}
\ea
($S_{k+1}$ could similarly be chosen to depend on $x^e_k$, rather than $x^e_{k+1}$.) There are $2K$ simultaneous pre-- and post--constraints $p^k_n=0$ and $p^{k+1}_o=0$. If the $K\times K$ matrix $\f{\p^2S_{k+1}}{\p x^o_k\p x^n_{k+1}}$ is of rank $K-\kappa$, then the right equation in (\ref{anh21d}) involves $\kappa$ additional pre--constraints ${}^-C^k_\nu(x^e_{k},x^o_k,p^k_o)$ at $k$ and the right equation in (\ref{anh21c}) involve $\kappa$ additional post--constraints ${}^+C^{k+1}_\nu(x^e_{k+1},x^n_{k+1},p^{k+1}_n)$ at $k+1$, $\nu=1,\ldots,\kappa$. $K-\kappa$ of the $x^n_{k+1}$ are determined via the second equation in (\ref{anh21d}), while the remaining $\kappa$ variables are left undetermined. Similarly, the auxiliary $x_k^n$ and $x_{k+1}^o$ are gauge parameters that can be fixed to arbitrary values. This move is its own inverse.

The interpretation of a type III move depends on the situation. For instance, the 2--2 Pachner evolution move in 3D Regge Calculus is of type III with $\kappa=0$ such that no non-trivial pre--constraints arise \cite{Dittrich:2011ke,Dittrich:2013jaa}. The 2--2 move neither coarse grains nor refines the triangulation. Instead, in the quantum theory, it can be viewed as an entangling move \cite{Dittrich:2013xwa}. Moreover, the tent moves \cite{Dittrich:2009fb,Dittrich:2011ke,Bahr:2009ku,Barrett:1994ks} in Regge Calculus of any dimension are of type III once the equation of the `tent pole' is solved. The tent moves are spatial triangulation preserving moves which can be decomposed into sequences of Pachner moves \cite{Dittrich:2011ke}.

However, for general type III moves new pre--constraints can arise, such that similarly to the type II moves we have to distinguish the $\kappa$ pre--constraints ${}^-\hat{C}^k_\nu$ and split them into the two sets $\hat{C}^k_\alpha$, $\alpha=1,\ldots,\kappa_\alpha$, of case (i) and ${}^-\hat{C}^k_\beta$, $\beta=1,\ldots,\kappa-\kappa_\alpha$, of case (ii) of section \ref{sec_locgen}. If case (ii) arises, a type III move includes a non-trivial dynamical coarse graining. 

Again, for the pre--constraints $\hat{C}^k_\alpha$ we have to regularize the physical state updating map (\ref{uk}). For simplicity, let us henceforth assume that the $\hat{C}^k_\alpha$ are linear in the $\hat{p}^k_o$ as for type II moves. In this case the Faddeev-Popov gauge fixing conditions for these constraints can be chosen as functions of the $x^o_k$ only, $G^k_\alpha(x^o_k)=0$. Lemma 4.1 of \cite{Hoehn:2014fka} implies in this case
\ba
(2\pi)^{\kappa_\alpha}\mathbb{P}^A_k\,|\det[G^k_\alpha(x^o_k),\hat{C}^k_{\alpha'}]|\,\prod_\alpha\delta(G^k_\alpha(x^o_k))\,\mathbb{P}^A_k\,\psi^{\rm kin}_k=\mathbb{P}^A_k\,\psi^{\rm kin}_k.\label{lem1III}
\ea

By virtue of similar arguments to the type I and II moves, the (regularized) propagator and physical state updating (\ref{pup}, \ref{sup}, \ref{revpup}) for a quantum type III move read
\ba
K^{\rm reg}_{0\rightarrow k+1}(x_0,x_{k+1}&=&(2\pi)^{\kappa_\alpha}\int\prod_o dx^o_k\,M^{III}_{k\rightarrow k+1}(x^e_{k+1},x^n_{k+1},x^o_k)\,e^{iS_{k+1}( x^e_{k+1},x^o_k,x^n_{k+1})/\hbar}\nn\\
&&\q\q\q\q\q\times|\det[G^k_\alpha(x^o_k),\hat{C}^k_{\alpha'}]|\,\prod_\alpha\delta(G^k_\alpha(x^o_k))\,K_{0\rightarrow k}(x_0,x_k),\nn\\
K_{k\rightarrow f}(x_k,x_f)&=&\int\prod_ndx^n_{k+1}\,K_{k+1\rightarrow f}(x_{k+1},x_f)\,M^{III}_{k\rightarrow k+1}(x^e_{k+1},x^n_{k+1},x^o_k)\,e^{iS_{k+1}/\hbar},\nn\\
{}^+\psi^{\rm phys}_{k+1}(x_{k+1})&=&(2\pi)^{\kappa_\alpha}\int \prod_odx^o_k \,M^{III}_{k\rightarrow k+1}(x^e_{k+1},x^n_{k+1},x^o_k)\,e^{iS_{k+1}( x^e_{k+1},x^o_k,x^n_{k+1})/\hbar}\nn\\
&&\q\q\q\q\q\times|\det[G^k_\alpha(x^o_k),\hat{C}^k_{\alpha'}]|\,\prod_\alpha\delta(G^k_\alpha(x^o_k))\,{}^+\psi^{\rm phys}_k(x_k).\nn
\ea

Similar to the type II move, the type III move preserves all post--constraints at $k$ in the classical formalism and preserves the symplectic structure restricted to the post--constraint surfaces, provided no constraints of type ${}^-C^k_\beta$ occur \cite{Dittrich:2013jaa}. If at least one pre--constraint ${}^-C^k_\beta$ occurs, a type III move decreases the rank of the symplectic form from $k$ to $k+1$. The analogous situation holds in the quantum theory.

\begin{Theorem}\label{thm_III}{\bf{ (Type III)}}
The physical state updating map
\ba
\fu^{III}_{k\rightarrow k+1}=(2\pi)^{\kappa_\alpha}\int\prod_odx^o_k\,|\det[G^k_\alpha,\hat{C}^k_{\alpha'}]|\,\prod_\alpha\delta(G^k_\alpha)\,M^{III}_{k\rightarrow k+1}\,e^{iS_{k+1}(x^e_{k},x^o_{k},x^n_{k+1})/\hbar}\nn
\ea
defines a map ${}^+\ch^{\rm phys}_k\rightarrow {}^+\ch^{\rm phys}_{k+1}$ which preserves all post--constraints from step $k$ if the measure satisfies
\ba
{}^+\hat{C}^{k+1}_\nu\left(x^e_{k+1},x^n_{k+1},\hat{p}^{k+1}_n+\f{\p S_{k+1}}{\p x^n_{k+1}}\right)\,M^{III}_{k\rightarrow k+1}&=&0,\nn\\
{}^-\hat{C}^k_\nu\left(x^e_k,x^o_k,\hat{p}^k_o-\f{\p S_{k+1}}{\p x^o_k}\right)\,(M^{III}_{k\rightarrow k+1})^*&=&0,\q\forall\,\nu=1,\ldots,\kappa\nn
\ea
and commutes with all other post--constraints at $k+1$ and if all constraints are analytical. $\fu^{III}_{k\rightarrow k+1}$ is
\begin{enumerate}
\item unitary if no pre--constraints ${}^-\hat{C}^k_\beta$ of case (ii) in section \ref{sec_locgen} occur,
\item non-unitary if at least one pre--constraint ${}^-\hat{C}^k_\beta$ of case (ii) in section \ref{sec_locgen} occurs.
\end{enumerate}
\end{Theorem}

\begin{proof}
The proof is given in appendix \ref{app_1}.
\end{proof}

In the condition that $M^{III}_{k\rightarrow k+1}$ commutes with (almost) all constraints the measure is to be understood as a multiplication operator (in the position representation) on $\ch^{\rm kin}$. The conditions on the measure updating factor will, in general, not fix it to be constant but leave a non-trivial dependence on $x^o_k,x^n_{k+1}$. In lemma \ref{lemc} of section \ref{sec_comp} we shall discuss a further non-trivial condition on $M^{III}_{k\rightarrow k+1}$. As one can easily check, for quadratic discrete actions \cite{Hoehn:2014aoa} all these conditions on the measure fix it to a unique constant value. Since $M^{III}_{k\rightarrow k+1}$ has to commute with (almost) all constraints one can view the measure essentially as a Dirac observable; it should be invariant under the corresponding symmetry transformations.

\begin{Example}\label{ex_3}
\emph{We take the example of a scalar field on a 2D quadrangulation from \cite{Dittrich:2013jaa}. The 1D hypersurface $\Sigma_k$ shall form a `zig-zag line'. An example of a type III move is to add a square as depicted in figure \ref{fig_typeiii}. This removes one vertex $v^*$ with `old' field variable $\phi_{k}^{v^*}$ and introduces a vertex $v$ with `new'  field variable $\phi_{k+1}^{v}$, while preserving the remaining vertices. The associated momentum updating equations are \cite{Dittrich:2013jaa}
\ba
\phi^b_k&=&x^b_{k+1}  \, ,\q\q\q \pi^{k+1}_b\,=\,\pi^k_b\,,\q\q\q\q\q\q\q\q\q\q\q\q\q b=1,4,5 \, , \nn\\
\phi^e_k&=&x^e_{k+1}\,,\q\q\q \pi^{k+1}_e\,=\,\pi^k_e+2\phi^e_{k+1}-\phi^{v^*}_k-\phi^v_{k+1}\,,\q\q\q\, e=2,3\, ,\nn\\
\pi^{k}_v&=&0 \,,\q\q\q\q\,\, \,\pi^{k+1}_v\,=\, 2\phi^v_{k+1}-\phi^2_{k+1}-\phi^3_{k+1}\,\,
 ,\nn \\
\pi^{k+1}_{v^*} &=& 0 \,,\q\q\q\q\q\,\pi^{k}_{v^*}\,=\, -2\phi^{v^*}_k+\phi^2_k+\phi^3_k.\nn
\ea
A new post--constraint ${}^+C^{k+1}_v:=\pi^{k+1}_v-2\phi^v_{k+1}+\phi^2_{k+1}+\phi^3_{k+1}$ and a new pre--constraint ${}^-C^k_{v^*}:=\pi^{k}_{v^*} +2\phi^{v^*}_k-\phi^2_k-\phi^3_k$ are generated.}

\emph{The action corresponding to the added (rectangular) square is \cite{Dittrich:2011ke}
\ba
S_\Box=(\phi^{v^*}_k)^2-\phi^{v^*}_k\phi^{3}_k+(\phi^{3}_k)^2-\phi^{3}_k\phi^{v}_{k+1}+(\phi^{v}_{k+1})^2-\phi^{v}_{k+1}\phi^{2}_k+(\phi^{2}_k)^2-\phi^{2}_k\phi^{v^*}_{k}.\nn
\ea
The physical state updating map corresponding to the present move is
\ba
\fu^{III}_{k\rightarrow k+1}=\f{1}{\sqrt{2\pi\hbar}}\int_{-\infty}^{+\infty}d\phi^{v^*}_k\,e^{\f{i}{\hbar}\left((\phi^{v^*}_k)^2-\phi^{v^*}_k\phi^{3}_k+(\phi^{3}_k)^2-\phi^{3}_k\phi^{v}_{k+1}+(\phi^{v}_{k+1})^2-\phi^{v}_{k+1}\phi^{2}_k+(\phi^{2}_k)^2-\phi^{2}_k\phi^{v^*}_{k}\right)}.\nn
\ea
Here we have employed that the measure for quadratic discrete actions is constant \cite{Hoehn:2014aoa}. It is straightforward to check that
\ba
{}^+\hat{C}^{k+1}_{v}\,\fu^{III}_{k\rightarrow k+1}&=&\f{1}{\sqrt{2\pi\hbar}}\int_{-\infty}^{+\infty}d\phi^{v^*}_k\,\left(\hat{\pi}^{k+1}_v-2\phi^v_{k+1}+\phi^2_{k+1}+\phi^3_{k+1}\right)\nn\\
&&\times e^{\f{i}{\hbar}\left((\phi^{v^*}_k)^2-\phi^{v^*}_k\phi^{3}_{k+1}+(\phi^{3}_{k+1})^2-\phi^{3}_{k+1}\phi^{v}_{k+1}+(\phi^{v}_{k+1})^2-\phi^{v}_{k+1}\phi^{2}_{k+1}+(\phi^{2}_{k+1})^2-\phi^{2}_{k+1}\phi^{v^*}_{k}\right)}=0\nn
\ea
and, similarly (as in (\ref{cond})),
\ba
&&\!\!\!\!\!\!\!\!\!\!{}^-\hat{C}^{k}_{v^*}\,e^{-\f{i}{\hbar}\left((\phi^{v^*}_k)^2-\phi^{v^*}_k\phi^{3}_k+(\phi^{3}_k)^2-\phi^{3}_k\phi^{v}_{k+1}+(\phi^{v}_{k+1})^2-\phi^{v}_{k+1}\phi^{2}_k+(\phi^{2}_k)^2-\phi^{2}_k\phi^{v^*}_{k}\right)}\label{2Dprecon2}\\
&&=\left(\hat{\pi}^{k}_{v^*} +2\phi^{v^*}_k-\phi^2_k-\phi^3_k\right)\, e^{-\f{i}{\hbar}\left((\phi^{v^*}_k)^2-\phi^{v^*}_k\phi^{3}_k+(\phi^{3}_k)^2-\phi^{3}_k\phi^{v}_{k+1}+(\phi^{v}_{k+1})^2-\phi^{v}_{k+1}\phi^{2}_k+(\phi^{2}_k)^2-\phi^{2}_k\phi^{v^*}_{k}\right)}=0.\nn
\ea
(We have made use of the fact that it is irrelevant whether $S_\Box$ depends on $\phi^2_k,\phi^3_k$ or $\phi^2_{k+1},\phi^3_{k+1}$.) If the post--physical state ${}^+\psi^{\rm phys}_k$ at $k$ satisfies ${}^-\hat{C}^k_{v^*}$ too, a gauge fixing factor $(2\pi\hbar)\,\delta(\phi^{v^*}_k-\phi'^{v^*}_k)$ must be inserted in $\fu^{III}_{k\rightarrow k+1}$ for regularization.}
\begin{center}
\begin{figure}[htbp!]
\psfrag{s}{\small$\Sigma_k$}
\psfrag{sk}{\small$\Sigma_{k+1}$}
\psfrag{v}{\small$v$}
\psfrag{vs}{\small$v^*$}
\psfrag{fv}{\footnotesize$\phi^v_{k+1}$}
\psfrag{fvs}{\footnotesize$\phi^{v^*}_k$}
\psfrag{1}{\footnotesize $\phi^1_{k+1}$}
\psfrag{2}{\footnotesize$\phi^2_{k+1}$}
\psfrag{3}{\footnotesize$\phi^3_{k+1}$}
\psfrag{4}{\footnotesize$\phi^4_{k+1}$}
\psfrag{5}{\footnotesize$\phi^5_{k+1}$}
\psfrag{a}{\footnotesize$\phi^1_k$}
\psfrag{b}{\footnotesize$\phi^2_k$}
\psfrag{c}{\footnotesize$\phi^3_k$}
\psfrag{d}{\footnotesize$\phi^4_k$}
\psfrag{e}{\footnotesize$\phi^5_k$}
\begin{subfigure}[b]{.22\textwidth}
\centering
\includegraphics[scale=.5]{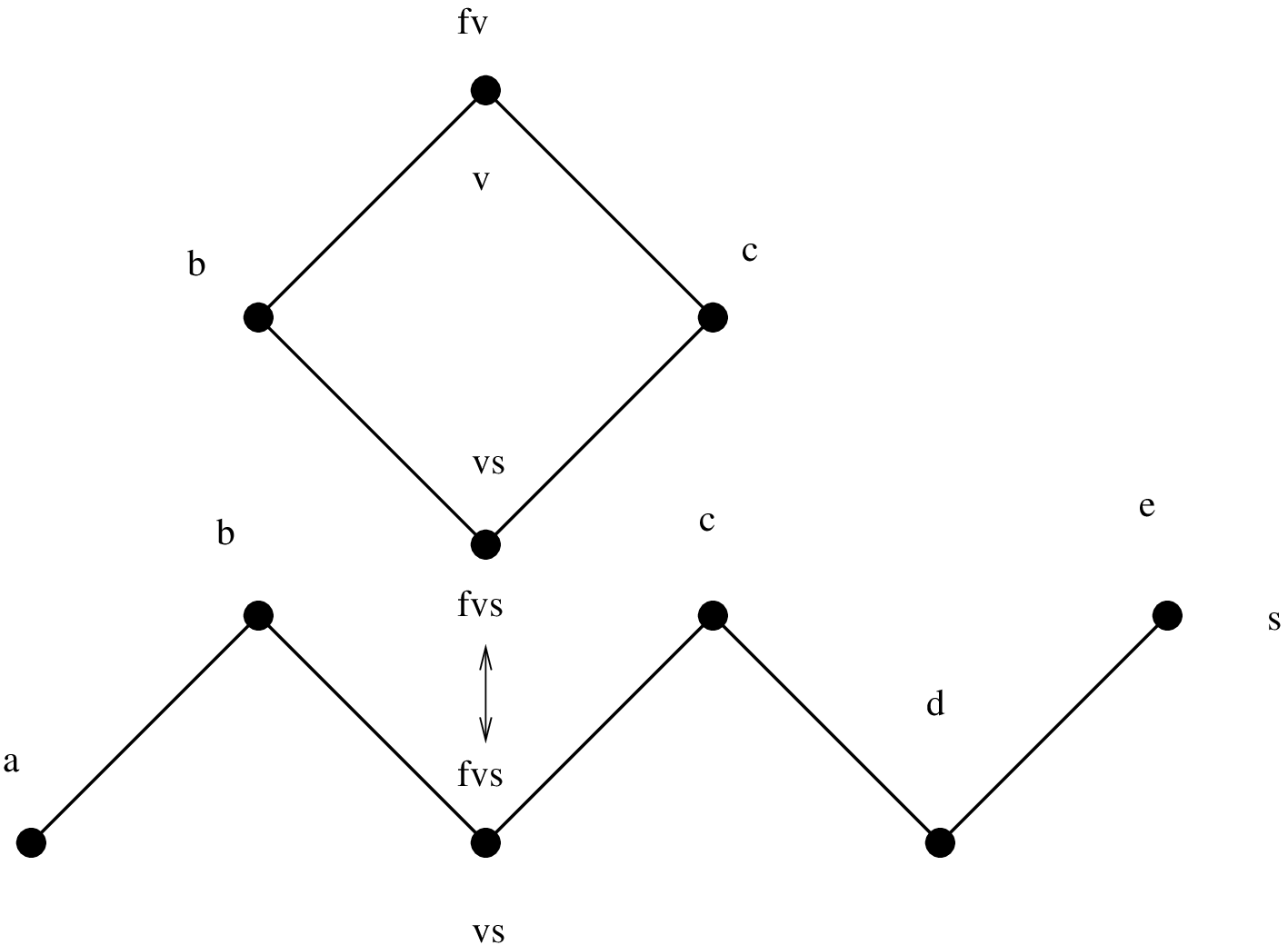}
\centering
\caption{\small }
\end{subfigure}
\hspace*{4.8cm}
\begin{subfigure}[b]{.22\textwidth}
\centering
\includegraphics[scale=.5]{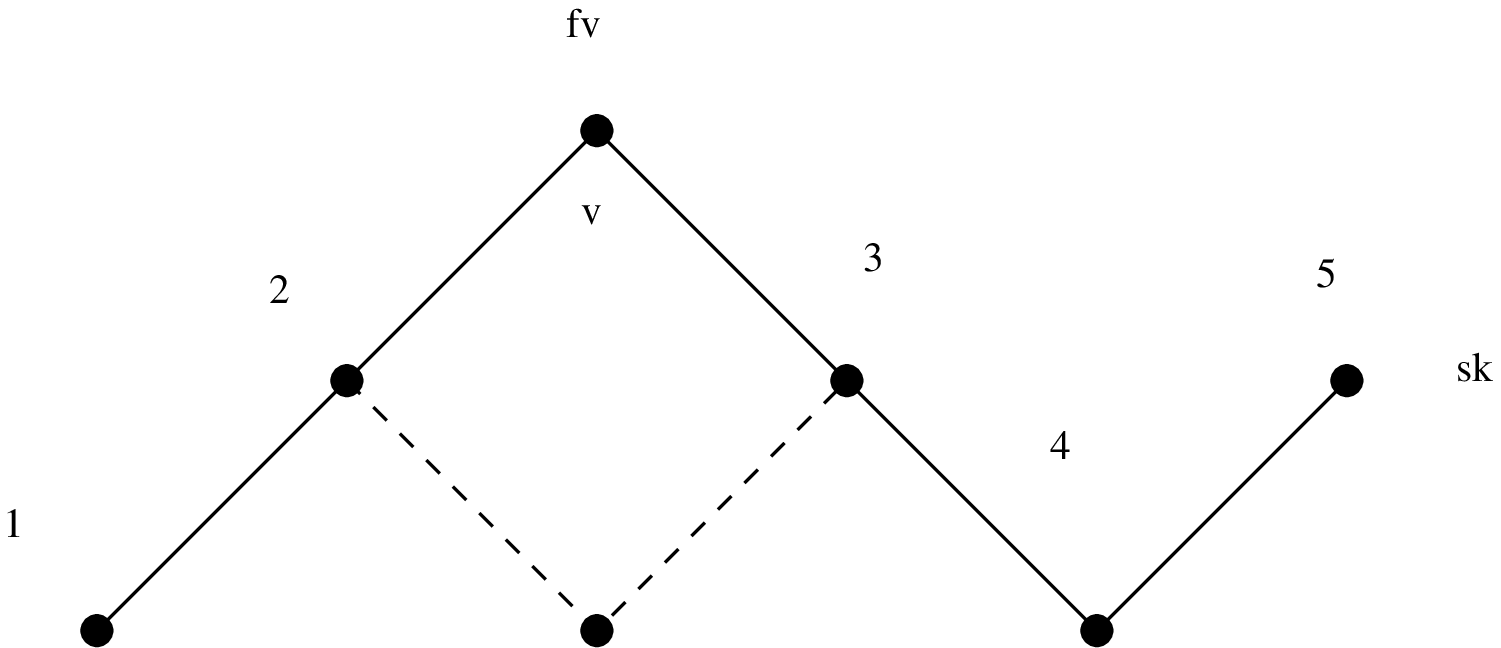}
\caption{\small }
\end{subfigure}
\caption{\small An example of a type III move for a scalar field on a 2D quadrangulation. It corresponds to gluing a square onto a 1D zig-zag line $\Sigma_k$ and removes the field variable $\phi^{v^*}_k$ and introduces $\phi^v_{k+1}$. }\label{fig_typeiii}
\end{figure}
\end{center}
\end{Example}

\subsection{Quantum moves of type IV}\label{sec_IV}

A type IV move neither generates a `new variable' nor annihilates an `old variable'. The momentum updating map $\fh_k$ is given by \cite{Dittrich:2013jaa}
\ba
x^b_k&=&x^b_{k+1}  \, ,\q\q\q p^{k+1}_b\,=\,p^k_b \, , \label{typeiv1}\\
x^e_k&=&x^e_{k+1}\,,\q\q\q p^{k+1}_e\,=\,p^k_e+\frac{\partial S_{k+1}(  x^e_{k+1})  }{\partial x^e_{k+1}} .\label{typeiv2}
\ea
($S_{k+1}$ can equally well be chosen as a function of $x^e_k$, instead of $x^e_{k+1}$.) The move does not introduce any new constraints and is its own inverse. Classically, this move trivially preserves all post--constraints and the symplectic structure restricted to the post--constraint surface \cite{Dittrich:2013jaa}. 

In the quantum theory, the {\it propagator} and {\it physical state updating} corresponding to the quantum type IV move are simply
\ba
K_{0\rightarrow k+1}(x_0,x_{k+1})&=&M^{IV}_{k\rightarrow k+1}\,e^{iS_{k+1}(x_{k+1}^e)/\hbar}\,K_{0\rightarrow k}(x_0,x_k),\nn\\
{}^+\psi^{\rm phys}_{k+1}(x_{k+1})&=&M^{IV}_{k\rightarrow k+1}\,e^{iS_{k+1}(x_{k+1}^e)/\hbar}\,{}^+\psi^{\rm phys}_k(x_k).\nn
\ea
The measure $M^{IV}_{k\rightarrow k+1}=const$ will be fixed in section \ref{sec_comp}. 

Using the methods of the quantum type I--III moves above, it is trivial to show that the quantum type IV move preserves all quantum post--constraints and is always unitary. We thus abstain from any further details.

\begin{Example}
\emph{Consider a scalar field living on the vertices of a 3D triangulation. Performing a 2--2 Pachner move neither annihilates nor introduces any vertex or field variable, as shown in figure \ref{fig_typeiv}. This move is therefore of type IV. This example is, once more, taken from \cite{Dittrich:2013jaa}. The corresponding equations are somewhat convoluted such that we shall not present them here. }
\begin{center}
\begin{figure}[htbp!]
\psfrag{s}{$\Sigma_k$}
\psfrag{sk}{$\Sigma_{k+1}$}
\psfrag{fv}{\footnotesize$\phi^v_{k+1}$}
\psfrag{fvs}{\footnotesize$\phi^{v^*}_k$}
\psfrag{1}{\footnotesize $\phi^1_{k+1}$}
\psfrag{2}{\footnotesize$\phi^2_{k+1}$}
\psfrag{3}{\footnotesize$\phi^3_{k+1}$}
\psfrag{4}{\footnotesize$\phi^4_{k+1}$}
\psfrag{5}{\footnotesize$\phi^5_{k+1}$}
\psfrag{6}{\footnotesize$\phi^6_{k+1}$}
\psfrag{7}{\footnotesize$\phi^7_{k+1}$}
\psfrag{8}{\footnotesize$\phi^8_{k+1}$}
\psfrag{9}{\footnotesize$\phi^9_{k+1}$}
\psfrag{a}{\footnotesize$\phi^1_k$}
\psfrag{b}{\footnotesize$\phi^2_k$}
\psfrag{c}{\footnotesize$\phi^3_k$}
\psfrag{d}{\footnotesize$\phi^4_k$}
\psfrag{e}{\footnotesize$\phi^5_k$}
\psfrag{f}{\footnotesize$\phi^6_k$}
\psfrag{g}{\footnotesize$\phi^7_k$}
\psfrag{h}{\footnotesize$\phi^8_k$}
\psfrag{i}{\footnotesize$\phi^9_k$}
\begin{subfigure}[b]{.21\textwidth}
\centering
\includegraphics[scale=.5]{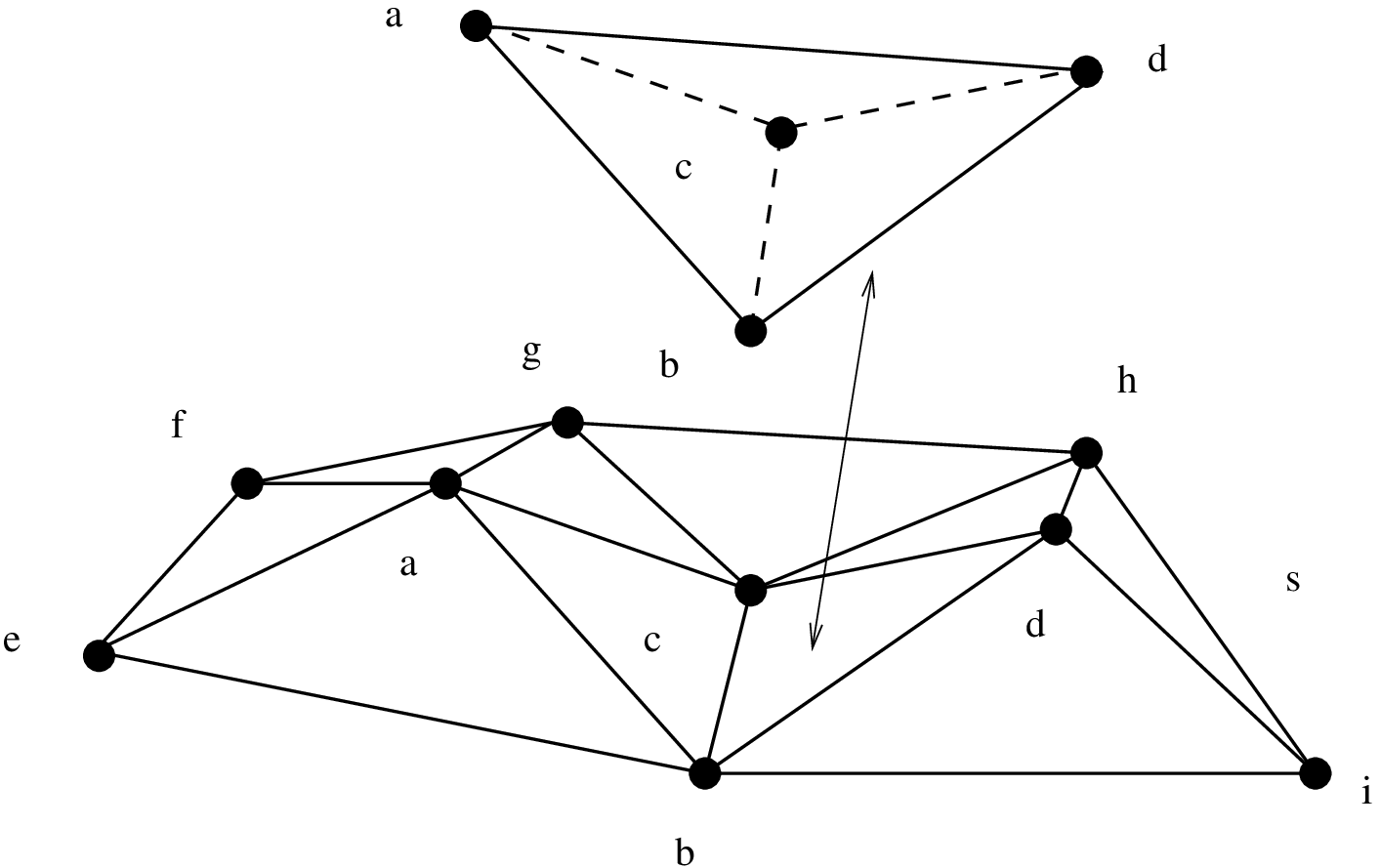}
\centering
\caption{\small }
\end{subfigure}
\hspace*{4.5cm}
\begin{subfigure}[b]{.21\textwidth}
\centering
\includegraphics[scale=.5]{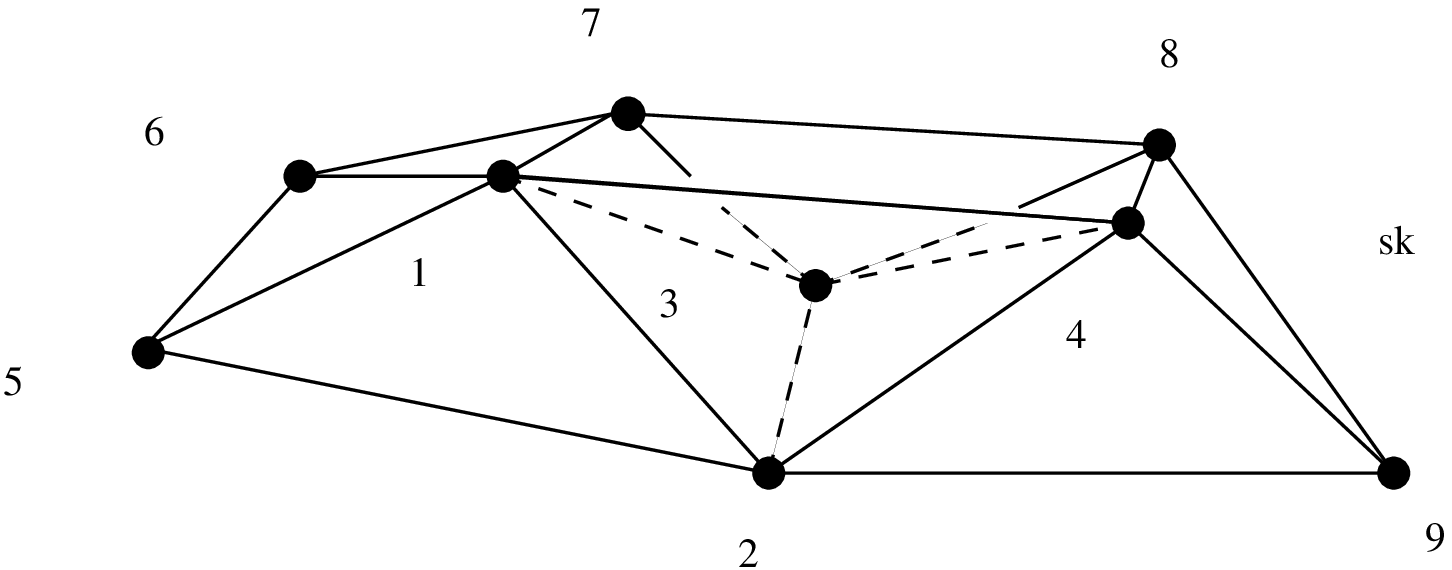}
\caption{\small }
\end{subfigure}
\caption{\small The 2--2 Pachner move for a scalar field on a 3D triangulation is an example of a type IV move. This move neither generates `new' variables nor annihilates `old' variables. }\label{fig_typeiv}
\end{figure}
\end{center}
\end{Example}

\subsection{Invertible compositions of local moves and measure updating factors}\label{sec_comp}

In this section we shall compose local moves with their time reversed inverses to (partially) fix the measure updating factors. To this end, we recall from \cite{Hoehn:2014fka} that the propagator of a time reversed global move $n+1\rightarrow n$ is given by $K_{n+1\rightarrow n}=(K_{n\rightarrow n+1})^*$. This is a condition to ensure unitarity of global moves. Since a local move updates the propagator of a global move, consistency requires that the time reversed physical state updating map (\ref{uk}) of a local move
\ba
\fu_{k+1\rightarrow k}=\int\prod_ndx^n_{k+1}\,(M_{k\rightarrow k+1})^*\,e^{-iS_{k+1}/\hbar}\prod_\nu(G^{k+1}_\nu)\nn
\ea
also comes with complex conjugation.

\begin{center}
\begin{figure}[htbp!]
\psfrag{sk}{$\Sigma_k$}
\psfrag{sk1}{$\Sigma_{k+1}$}
\psfrag{fv}{\scriptsize $\phi^v_{k+1}$}
\psfrag{f12}{\scriptsize $\phi^1_{k+1}$}
\psfrag{f22}{\scriptsize$\phi^2_{k+1}$}
\psfrag{f1}{\scriptsize$\phi^1_k$}
\psfrag{f2}{\scriptsize$\phi^2_k$}
\begin{subfigure}[b]{.17\textwidth}
\centering
\includegraphics[scale=.4]{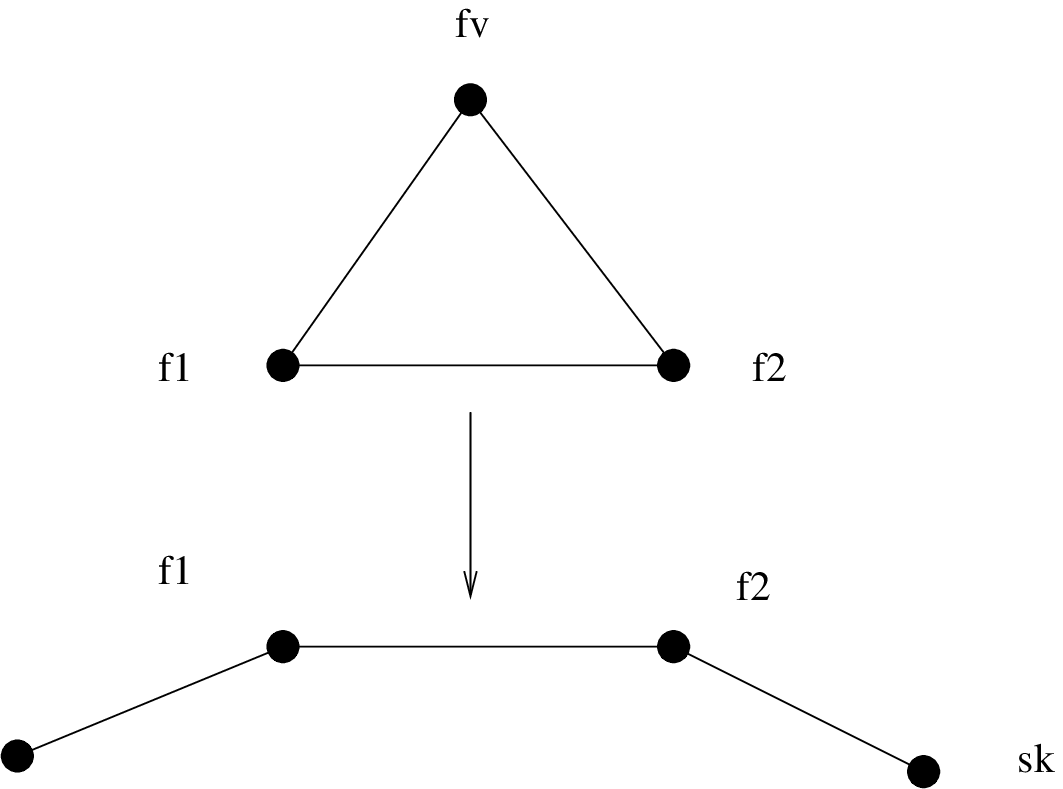}
\centering
\caption{\small }
\end{subfigure}
\hspace*{2.5cm}
\begin{subfigure}[b]{.17\textwidth}
\centering
\includegraphics[scale=.4]{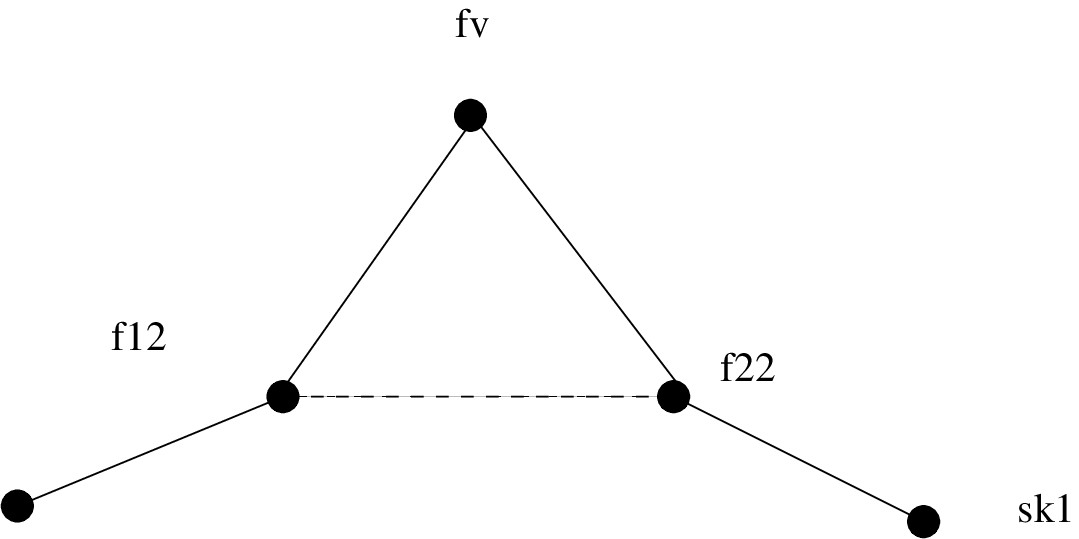}
\caption{\small }
\end{subfigure}
\hspace*{2.5cm}
\begin{subfigure}[b]{.17\textwidth}
\centering
\includegraphics[scale=.4]{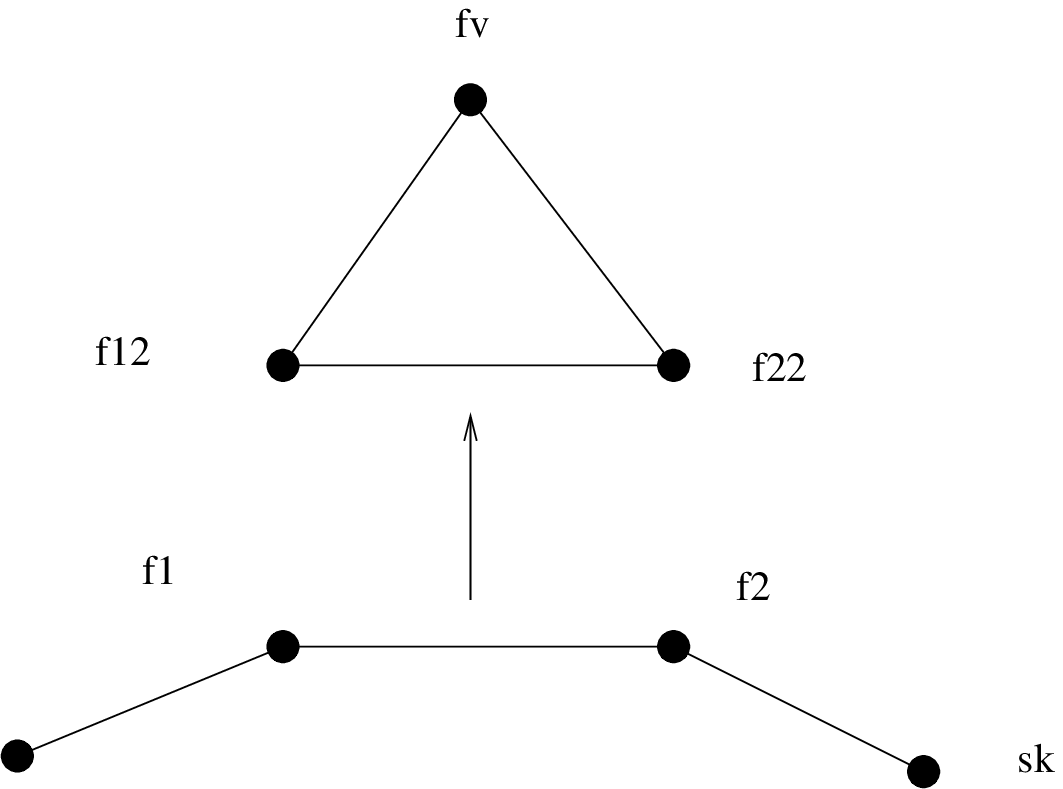}
\caption{\small }
\end{subfigure}
\caption{\small Composition of a `forward' 1--2 Pachner move with a `time reversed' 2--1 Pachner move for a scalar field on a 2D triangulation. This composition yields the identity operation on ${}^+\ch^{\rm phys}_k$. }\label{fig_inv1}
\end{figure}
\end{center}

We have seen that the type II move is the inverse move of the type I move and vice versa. We can make use of this to fix the two constant measures $M^I_{k\rightarrow k+1},M^{II}_{k\rightarrow k+1}$ from sections \ref{sec_I} and \ref{sec_II}. In particular, we can consider performing a future directed type I move and then immediately performing a time reverse type II move. Clearly, this must give the identity. For example, for a scalar field living on the vertices of a 2D space-time triangulation this corresponds to gluing a single triangle onto a single edge in a 1D hypersurface $\Sigma_k$ (a 1--2 Pachner move) and subsequently immediately removing the triangle again (a 2--1 Pachner move). This composition is depicted in figure \ref{fig_inv1}.

\begin{lem}\label{lema}
Undoing a type I move $k\rightarrow k+1$ with a reverse type II move $k+1\rightarrow k$ yields the identity at step $k$
\ba
\fu^{II}_{k+1\rightarrow k}\cdot\fu^I_{k\rightarrow k+1}=1\nn
\ea
 if (up to phase)
\ba
M^I_{k\rightarrow k+1}=M^{II}_{k\rightarrow k+1}=\left(\f{1}{2\pi\hbar}\right)^{K/2}.\nn
\ea
\end{lem}

\begin{proof}
The proof is given in appendix \ref{app_2}.
\end{proof}

We have to check that this is consistent with the reverse operation, i.e.\ with firstly carrying out a forward type II move and then immediately performing a time reversed type I move to undo the previous type II move. This is the more interesting case. For a scalar field on a 2D triangulation such a composition corresponds to performing a forward 2--1 Pachner move followed by a time reversed 1--2 Pachner move that rips out the triangle of the previous 2--1 move (see figure \ref{fig_inv2}). We have seen in theorem \ref{thm_I} that a type I move automatically projects onto the new post--constraints it generates. Given that the post--constraints of the reverse type I move are the pre--constraints of the forward type II move, we expect this composition to project onto the pre--constraints of the type II move. However, it should only project onto the non-trivial pre--constraints of case (ii) of section \ref{sec_locgen} (the coinciding pre-- and post--constraints of case (i) are regularized). Indeed, this is the case: 

\begin{center}
\begin{figure}[htbp!]
\psfrag{sk}{$\Sigma_k$}
\psfrag{sk1}{$\Sigma_{k+1}$}
\psfrag{fv}{\scriptsize $\phi^{v^*}_{k}$}
\psfrag{fv2}{\scriptsize $\phi'^{v^*}_k$}
\psfrag{f12}{\scriptsize $\phi^1_{k+1}$}
\psfrag{f22}{\scriptsize$\phi^2_{k+1}$}
\psfrag{f1}{\scriptsize$\phi^1_k$}
\psfrag{f2}{\scriptsize$\phi^2_k$}
\begin{subfigure}[b]{.17\textwidth}
\centering
\includegraphics[scale=.4]{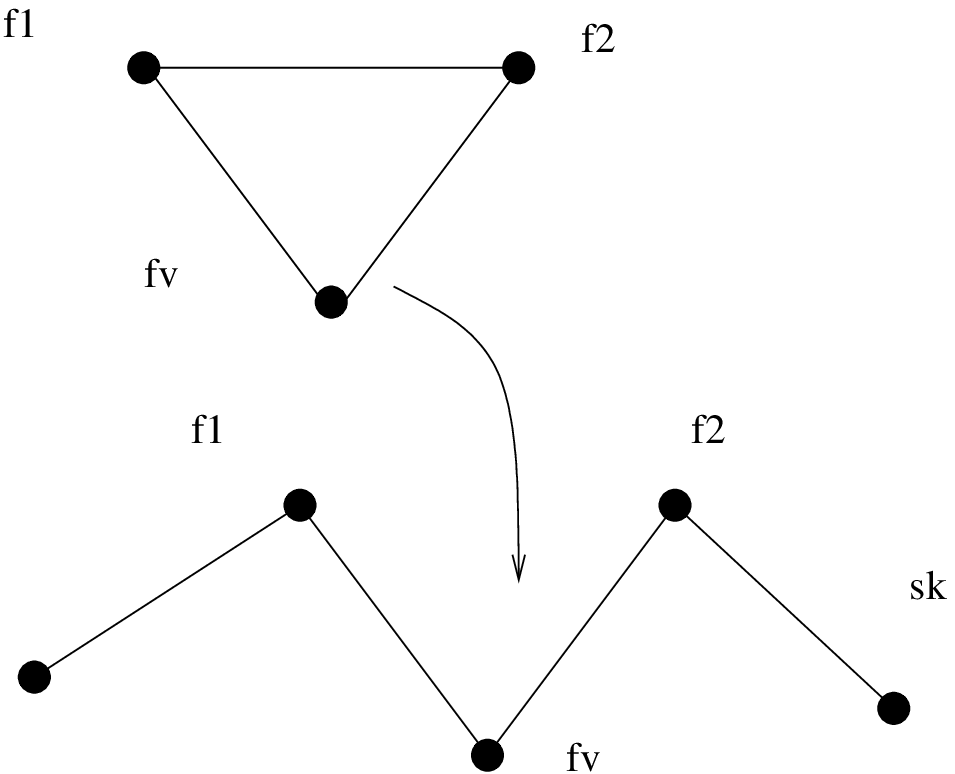}
\centering
\caption{\small }
\end{subfigure}
\hspace*{2.5cm}
\begin{subfigure}[b]{.17\textwidth}
\centering
\includegraphics[scale=.4]{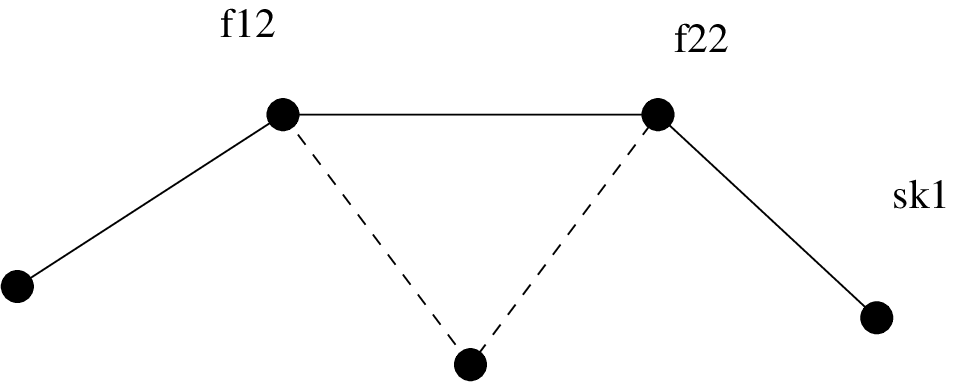}
\caption{\small }
\end{subfigure}
\hspace*{2.5cm}
\begin{subfigure}[b]{.17\textwidth}
\centering
\includegraphics[scale=.4]{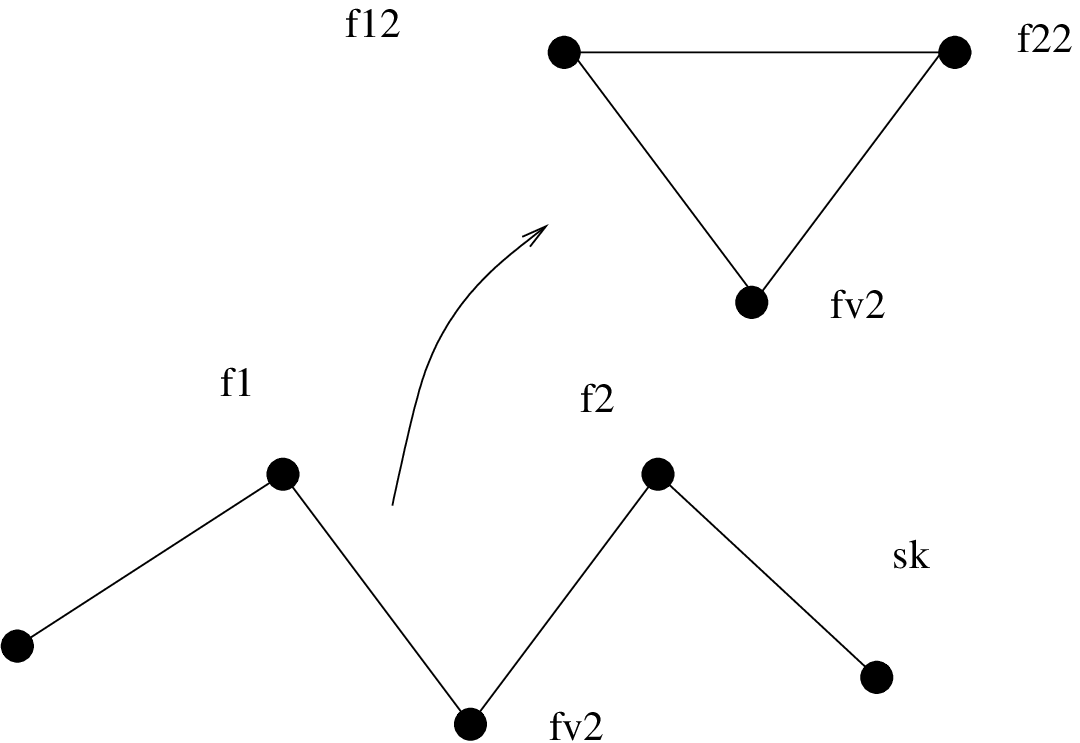}
\caption{\small }
\end{subfigure}
\caption{\small Composition of a `forward' 2--1 Pachner move with a `time reversed' 1--2 Pachner move for a scalar field on a 2D triangulation. This operation projects  ${}^+\ch^{\rm phys}_k$ onto the pre--constraint ${}^-\hat{C}^k_{v^*}$ (\ref{2Dprecon}). }\label{fig_inv2}
\end{figure}
\end{center}

\begin{lem}\label{lemb}
Undoing a type II move $k\rightarrow k+1$ with a reverse type I move $k+1\rightarrow k$ is equivalent to projecting ${}^+\psi^{\rm phys}_k\in{}^+\ch^{\rm phys}_k$ onto the non-trivial pre--constraints of the type II move
\ba
\fu^{I}_{k+1 \rightarrow k}\cdot\fu^{II}_{k\rightarrow k+1}\,{}^+\psi^{\rm phys}_k={}^-\mathbb{P}^B_k\,{}^+\psi^{\rm phys}_k\nn
\ea
at step $k$ if the measure factors are chosen as in lemma \ref{lema}.
\end{lem}

\begin{proof}
The proof is given in appendix \ref{app_2}.
\end{proof}

\begin{center}
\begin{figure}[htbp!]
\psfrag{sk}{$\Sigma_k$}
\psfrag{sk1}{$\Sigma_{k+1}$}
\psfrag{fva}{\scriptsize $\phi^{v^*}_{k}$}
\psfrag{fv}{\scriptsize $\phi^v_{k+1}$}
\psfrag{fvb}{\scriptsize $\phi'^{v^*}_k$}
\psfrag{f12}{\scriptsize $\phi^1_{k+1}$}
\psfrag{f22}{\scriptsize$\phi^2_{k+1}$}
\psfrag{f1}{\scriptsize$\phi^1_k$}
\psfrag{f2}{\scriptsize$\phi^2_k$}
\begin{subfigure}[b]{.17\textwidth}
\centering
\includegraphics[scale=.4]{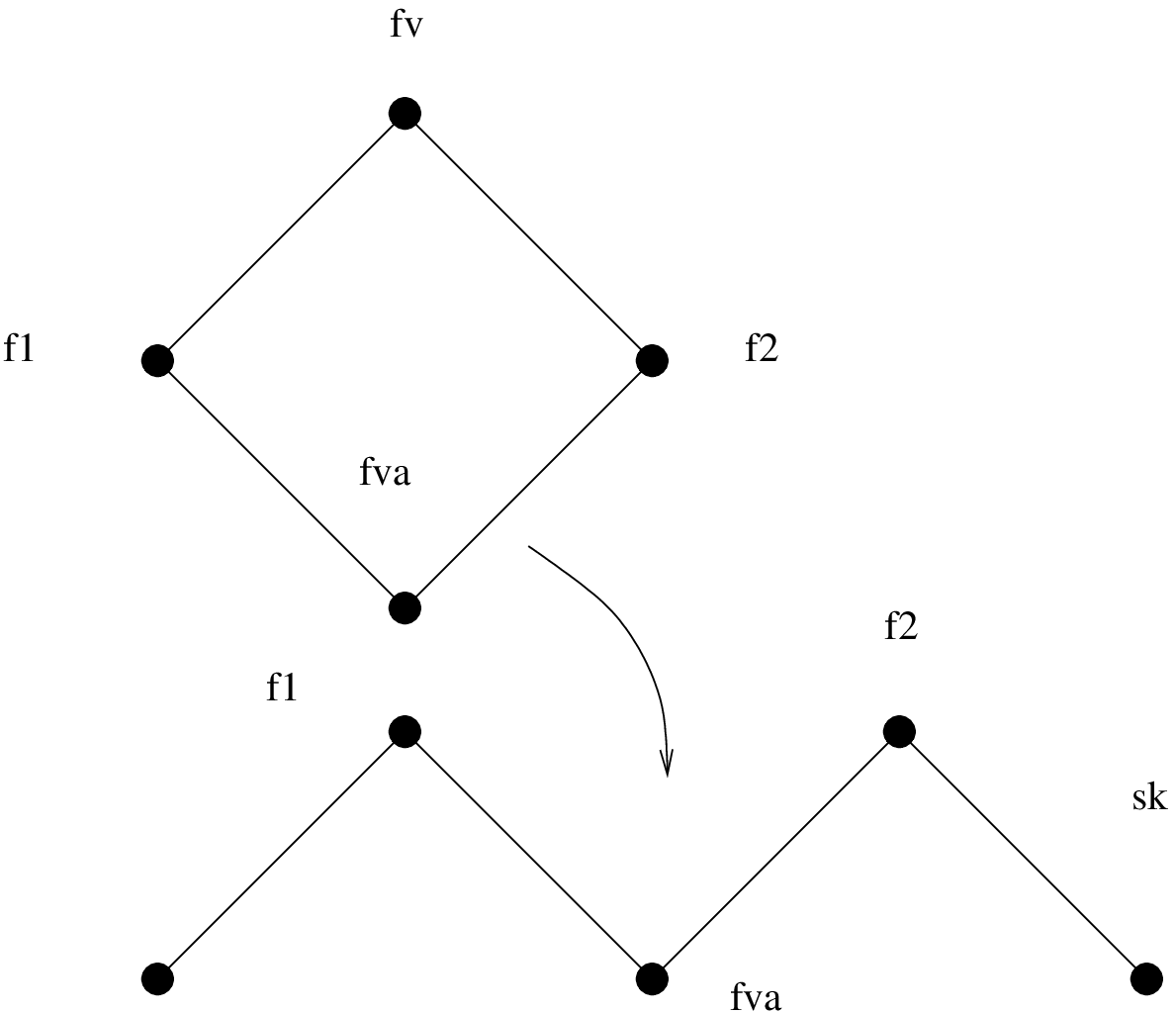}
\centering
\caption{\small }
\end{subfigure}
\hspace*{2.5cm}
\begin{subfigure}[b]{.17\textwidth}
\centering
\includegraphics[scale=.4]{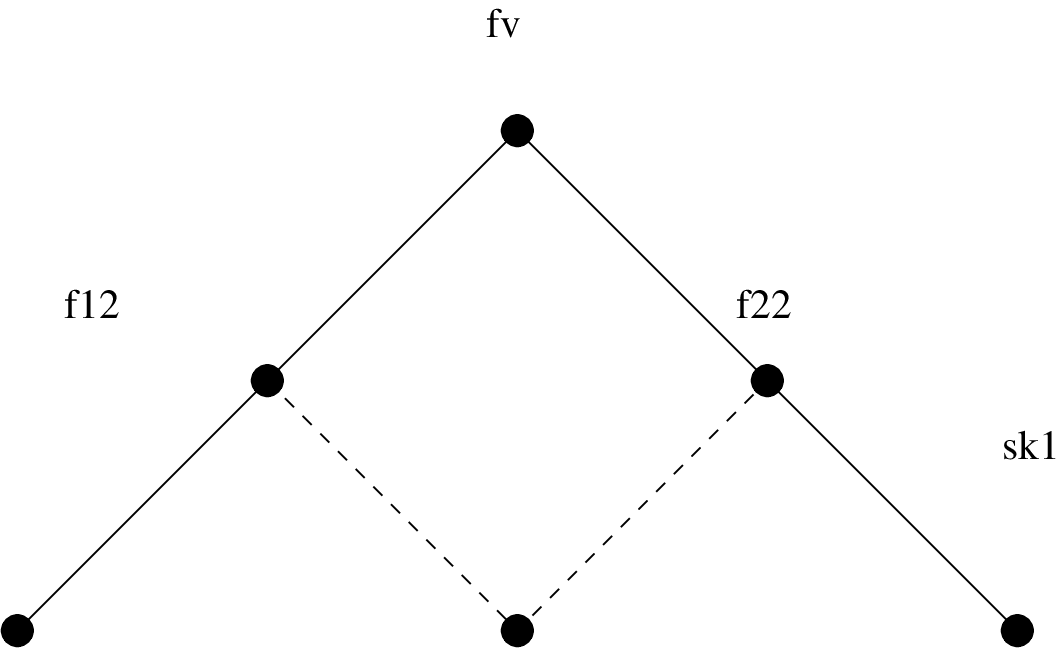}
\caption{\small }
\end{subfigure}
\hspace*{2.5cm}
\begin{subfigure}[b]{.17\textwidth}
\centering
\includegraphics[scale=.4]{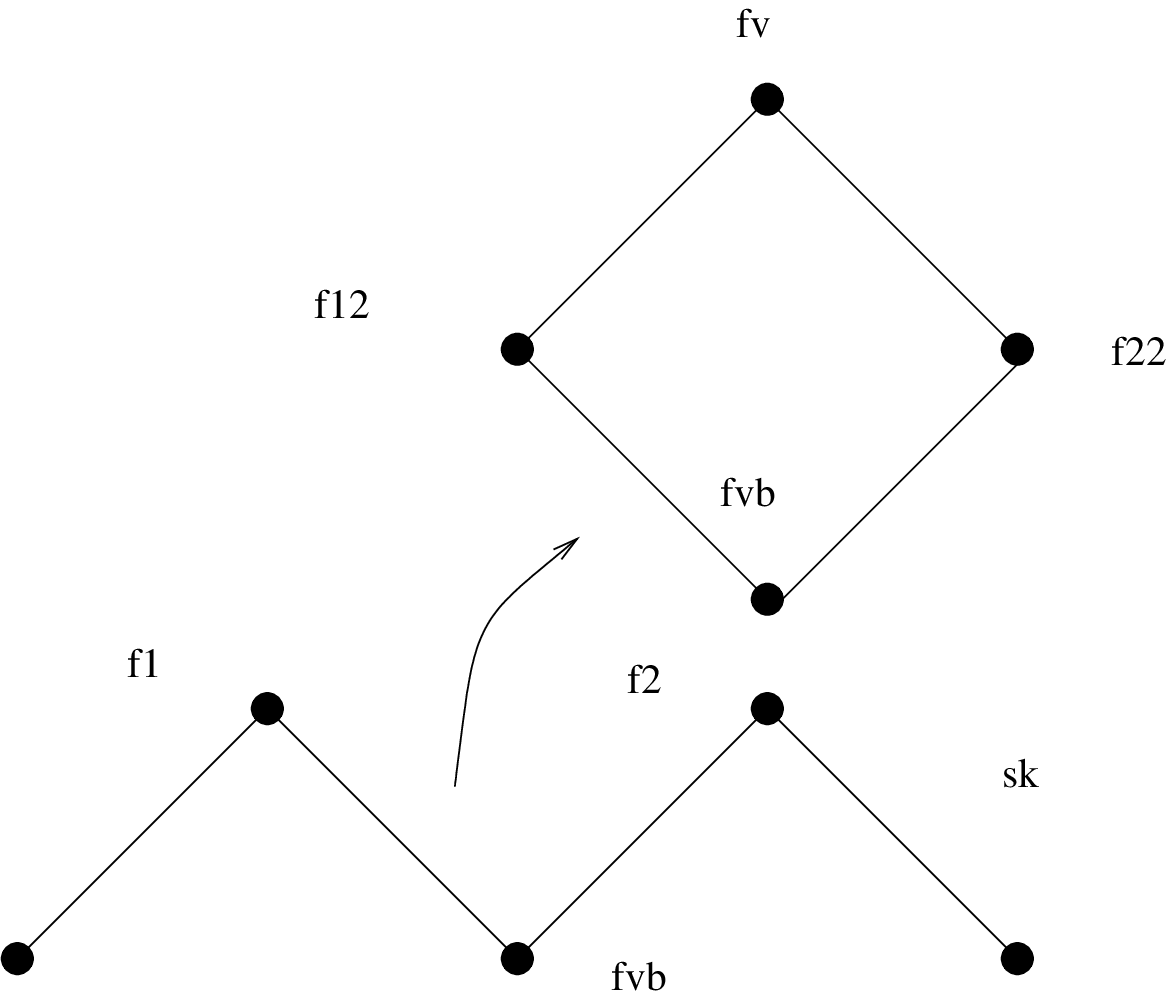}
\caption{\small }
\end{subfigure}
\caption{\small Composition of a `forward' type III move with a `time reversed' type III move for a scalar field on a 2D quadrangulation. This operation projects  ${}^+\ch^{\rm phys}_k$ onto the pre--constraint ${}^-\hat{C}^k_{v^*}$ (\ref{2Dprecon2}). }\label{fig_inv3}
\end{figure}
\end{center}

For a type III move we expect a similar result. Namely, given that a type III move is its own inverse, undoing a type III move with a reverse type III move should enforce a projection onto the non-trivial pre--constraints. In fact, while theorem \ref{thm_III} already imposes a number of non-trivial conditions on the measure factor $M^{III}_{k\rightarrow k+1}$ of the type III move, we can use this as a further restricting condition. The situation for the type III move is, unfortunately, more complicated such that for simplicity we shall restrict to new constraints that are linear in the momenta. The composition of a `forward' and `backward' type III move for the case of a scalar field on a 2D quadrangulation from example \ref{ex_3} is shown in figure \ref{fig_inv3}. 

\begin{lem}\label{lemc}
Suppose all new pre-- and post--constraints of a type III move are linear in the momenta. Then, undoing a type III move $k\rightarrow k+1$ with a reverse type III move $k+1\rightarrow k$ is equivalent to projecting ${}^+\psi^{\rm phys}_k\in{}^+\ch^{\rm phys}_k$ onto the non-trivial pre--constraints of the type III move at $k$,
\ba
\fu^{III}_{k+1 \rightarrow k}\cdot\fu^{III}_{k\rightarrow k+1}\,{}^+\psi^{\rm phys}_k={}^-\mathbb{P}^B_k\,{}^+\psi^{\rm phys}_k,\nn
\ea
if the following condition on $M^{III}_{k\rightarrow k+1}$  
\ba
\mathbb{P}^A_k\,{}^-\mathbb{P}^B_k\,\cdot\!\!\!\!&=&\!\!\!(2\pi)^\kappa\int\prod_ndx^n_{k+1}|\det[G^{k+1}_\nu,{}^+\hat{C}^{k+1}_{\nu'}]|\,\prod_\nu\delta(G^{k+1}_\nu)\,(M^{III}_{k\rightarrow k+1}(x'^o_k,x^e_k,x^n_{k+1}))^*\label{IIIcon}\\
&&\q\q\q\q\times e^{-iS_{k+1}(x'^o_k,x^e_k,x^n_{k+1})/\hbar}\,\int\prod_odx^o_k\,M_{k\rightarrow k+1}^{III}(x^o_k,x^e_k,x^n_{k+1})\,e^{iS_{k+1}(x^o_k,x^e_k,x^n_{k+1})/\hbar}\,\cdot\nn
\ea
is fulfilled together with the conditions of theorem \ref{thm_III}.
\end{lem}

\begin{proof}
The proof is given in appendix \ref{app_2}.
\end{proof}

The conditions of this lemma on $M^{III}_{k\rightarrow k+1}$ will in general allow it to non-trivially depend on $x^o_k,x^n_{k+1}$. For instance, if $\kappa=0$, as for the 2--2 Pachner evolution move in 3D Regge Calculus \cite{Dittrich:2011ke}, the condition on the measure factor in the lemma translates into
\ba
&&\!\!\!\!\!\!\!\!\!\int\prod_ndx^n_{k+1}\,M^{III}_{k\rightarrow k+1}(x^o_k,x^e_k,x^n_{k+1})\,(M^{III}_{k\rightarrow k+1}(x'^o_k,x^e_k,x^n_{k+1}))^*\,e^{i(S_{k+1}(x^o_k,x^e_k,x^n_{k+1})-S_{k+1}(x'^o_k,x^e_k,x^n_{k+1}))/\hbar}\nn\\
&&\q\q\q\q\q\q\q\q\q\q\q\q\q\q\q=\prod_o\delta(x'^o_k-x^o_k).\nn
\ea
For a general action this will not be solvable for a constant measure factor.\footnote{Of course, for an arbitrary action one could set $M^{III}_{k\rightarrow k+1}=(\f{1}{2\pi\hbar})^K\,e^{i(x^0_kx^n_{k+1}-S_{k+1})/\hbar}$ to solve this condition. But this would produce a trivial dynamics.} Furthermore, this condition may in the general case admit multiple inequivalent solutions for $M^{III}_{k\rightarrow k+1}$.

Finally, the type IV move is its own inverse. One trivially finds
\ba
\fu^{IV}_{k+1\rightarrow k}\cdot\fu^{IV}_{k\rightarrow k+1}=1\nn
\ea
if (up to phase)
\ba
M^{IV}_{k\rightarrow k+1}=1.\nn
\ea

\subsection{Momentum updating in the quantum theory}\label{sec_momup}

In sections \ref{sec_I}--\ref{sec_IV} we have given the classical momentum updating equations of the local moves of type I--IV. In this section we shall exhibit the quantum analogue of these equations, however, only for the momenta $\hat{p}^k_e$ that are actively updated in the move. 

It should be noted that $\hat{p}^k_e$ (and similarly for the other momentum operators) is, in general, not an operator on ${}^+\ch^{\rm phys}_k$ unless it commutes with all post--constraints at step $k$ and thus is a genuine Dirac post--observable (see section \ref{sec_dirac}). Nevertheless, the action of $\hat{p}^k_e$ as a derivative operator (in the position representation) on ${}^+\ch^{\rm phys}_k$ may still be well-defined although possibly $\hat{p}^k_e\,{}^+\psi^{\rm phys}_k\notin{}^+\ch^{\rm phys}_k$. In this sense it is still meaningful to consider quantum momentum updating.

Consider a type I move $k\rightarrow k+1$. To map the momentum operator from one time step to the next, we also need its inverse move, i.e.\ the time reversed type II move $k+1\rightarrow k$ as in lemma \ref{lema}. The only non-trivial quantum version of momentum updating (\ref{anh1b}) takes the form\footnote{Recall that $\hat{p}^{k+1}_e$ is the derivative operator with respect to $x^e_{k+1}$ acting on states at $k+1$, while $\hat{p}^k_e$ is the derivative operator with respect to $x^e_k$ acting on states at $k$.}
\ba
\fu^{II}_{k+1\rightarrow k}\,\hat{p}^{k+1}_e\,\fu^I_{k\rightarrow k+1}\,{}^+\psi^{\rm phys}_k(x_k)
&=&\fu^{II}_{k+1\rightarrow k}\,\left(\f{1}{2\pi\hbar}\right)^{K/2}e^{iS_{k+1}/\hbar}\,\left(\hat{p}^{k}_e+\f{\p S_{k+1}}{\p x^e_{k}}\right)\,{}^+\psi^{\rm phys}_{k}(x_{k})\nn\\
&=&\int\prod_ndx^n_{k+1}\,\delta(x'^n_{k+1}-x^n_{k+1})\,\left(\hat{p}^{k}_e+\f{\p S_{k+1}}{\p x^e_{k}}\right)\,{}^+\psi^{\rm phys}_{k}(x_{k})\nn\\
&=&\left(\hat{p}^{k}_e+\f{\p S_{k+1}}{\p x^e_{k}}\right)\,{}^+\psi^{\rm phys}_{k}(x_{k}).\nn
\ea
In complete analogy, one finds for the reverse direction
\ba
\fu^I_{k\rightarrow k+1}\,\hat{p}^k_e\,\fu^{II}_{k+1\rightarrow k}\,{}^+\psi^{\rm phys}_{k+1}(x_{k+1})=\left(\hat{p}^{k+1}_e-\f{\p S_{k+1}}{\p x^e_{k+1}}\right){}^+\psi^{\rm phys}_{k+1}(x_{k+1}).\nn
\ea

As in lemma \ref{lemb} consider now a type II move $k\rightarrow k+1$ followed by a time reversed type I move $k+1\rightarrow k$. The quantum version of the momentum updating equation (\ref{case2b}) reads
\ba
\fu^{I}_{k+1 \rightarrow k}\,\hat{p}^{k+1}_e\,\fu^{II}_{k\rightarrow k+1}{}^+\psi^{\rm phys}_k(x_k)&=&\left(\f{1}{2\pi\hbar}\right)^Ke^{-iS_{k+1}(x'^o_{k},x^e_k)/\hbar}\,(2\pi\hbar)^{K_\alpha}\int\prod_odx^o_k\,\hat{p}^{k+1}_e\nn\\
&&\q\q\q\q\q\times e^{iS_{k+1}(x^o_k,x^e_k)/\hbar}\,\prod_\alpha\delta(x'^\alpha_k-x^\alpha_k)\,{}^+\psi^{\rm phys}_k(x^o_k,x^e_k)\nn\\
&=&\left(\f{1}{2\pi\hbar}\right)^Ke^{-iS_{k+1}(x'^o_{k},x^e_k)/\hbar}\,(2\pi\hbar)^{K_\alpha}\int\prod_odx^o_k\, e^{iS_{k+1}(x^o_k,x^e_k)/\hbar}\nn\\
&&\q\q\q\q\times\prod_\alpha\delta(x'^\alpha_k-x^\alpha_k)\,\left(\hat{p}^k_e+\f{\p S_{k+1}}{\p x^e_k}\right)\,{}^+\psi^{\rm phys}_k(x^o_k,x^e_k)\nn\\
&=&{}^-\mathbb{P}^B_k\,\left(\hat{p}^k_e+\f{\p S_{k+1}}{\p x^e_k}\right)\,{}^+\psi^{\rm phys}_k(x_k)\nn.
\ea
The reasoning here is analogous to that in lemma \ref{lemb}. $\hat{p}^k_e$ commutes with the gauge fixing conditions at $k$. For the reverse direction one finds by similar arguments
\ba
\fu^{II}_{k\rightarrow k+1}\,\hat{p}^k_e\,\fu^I_{k+1\rightarrow k}\,{}^+\psi^{\rm phys}_{k+1}(x_{k+1})=\left(\hat{p}^{k+1}_e-\f{\p S_{k+1}}{\p x^e_{k+1}}\right){}^+\psi^{\rm phys}_{k+1}(x_{k+1}).\nn
\ea

Next, we consider the quantum momentum updating of $\hat{p}^k_e$ for a type III move. We only state the result as the reasoning is identical to that in lemma \ref{lemc}:
\ba
\fu^{III}_{k+1\rightarrow k}\,\hat{p}^{k+1}_e\,\fu^{III}_{k\rightarrow k+1}\,{}^+\psi^{\rm phys}_k(x_k)&=&{}^-\mathbb{P}^B_k\,\left(\hat{p}^k_e+\f{\p S_{k+1}}{\p x^e_k}-i\f{\p \ln(M^{III}_{k\rightarrow k+1})}{\p x^e_k}\right)\,{}^+\psi^{\rm phys}_k(x_k),\nn\\
\fu^{III}_{k\rightarrow k+1}\,\hat{p}^{k}_e\,\fu^{III}_{k+1\rightarrow k}\,{}^+\psi^{\rm phys}_{k+1}(x_{k+1})&=&\left(\hat{p}^{k+1}_e-\f{\p S_{k+1}}{\p x^e_{k+1}}-i\f{\p \ln((M^{III}_{k\rightarrow k+1})^*)}{\p x^e_{k+1}}\right)\,{}^+\psi^{\rm phys}_{k+1}(x_{k+1}).\nn
\ea
For the last equation to be true one needs $|M^{III}_{k\rightarrow k+1}|=(1/2\pi\hbar)^{\kappa/2}$ for all values of $x^o_k,x^e_k,x^n_{k+1}$.

Finally, for the type IV move the quantum momentum updating equations follow trivially
\ba
\fu^{IV}_{k+1\rightarrow k}\,\hat{p}_e^{k+1}\,\fu^{IV}_{k\rightarrow k+1}\,{}^+\psi^{\rm phys}_{k}(x_k)&=&\left(\hat{p}^k_e+\f{\p S_{k+1}}{\p x^e_k}\right)\,{}^+\psi^{\rm phys}_k(x_k),\nn\\
\fu^{IV}_{k\rightarrow k+1}\,\hat{p}^{k}_e\,\fu^{IV}_{k+1\rightarrow k}\,{}^+\psi^{\rm phys}_{k+1}(x_{k+1})&=&\left(\hat{p}^{k+1}_e-\f{\p S_{k+1}}{\p x^e_{k+1}}\right)\,{}^+\psi^{\rm phys}_{k+1}(x_{k+1}).\nn
\ea

This concludes our discussion of quantum momentum updating.

\section{Dirac observables and local evolution}\label{sec_dirac}

The local evolution moves refine and coarse grain (or entangle \cite{Dittrich:2013xwa} degrees of freedom in) the discretization. Such changes of the discretization result in changes of the algebra of Dirac observables, which represent the physical degrees of freedom, at a given time step. 

In the classical formalism (see \cite{Dittrich:2013jaa} for details), Dirac observables as propagating degrees of freedom are associated to a given global time evolution move $0\rightarrow1$, rather than a given time step. The set of pre--observables of the move $0\rightarrow1$ corresponds to those functions on the pre--constraint surface $\cc^-_0$ that weakly Poisson commute with all pre--constraints on $\cc^-_0$. Similarly, the set of post--observables of the move $0\rightarrow1$ corresponds to the functions on the post--constraint surface $\cc^+_1$ which weakly Poisson commute with all post--constraints at step $1$. The global Hamiltonian time evolution map then takes the set of pre--observables at $0$ isomorphically into the set of post--observables at $1$ and vice versa. However, since in the context of a temporally varying discretization the pre-- and post--constraints at a given time step depend on the time evolution moves one is considering, the pre-- and post--observables at a given time step likewise depend on the evolution move under consideration \cite{Dittrich:2013jaa}. 

In particular, if a local time evolution move which updates a global move $0\rightarrow k$ to a new global move $0\rightarrow k+1$ is of type II or III it can introduce non-trivial pre--constraints at step $k$. In this case, the move performs a coarse graining of the discretization which has two effects: first, the rank of the symplectic form (restricted to the post--constraint surface) is reduced, and, second, the number of pre--observables at $0$ propagating in the move $0\rightarrow k+1$ to post--observables at step $k+1$ is {\it smaller} than the number of pre--observables at $0$ propagating via $0\rightarrow k$ to post--observables at $k$ \cite{Dittrich:2013jaa}. 


This is, of course, a one-way business: previously propagating pre-- and post--observables which become non-propagating (and are factored out of the physical phase space) because of the emergence of new coarse graining conditions can never be regained because all evolution moves preserve the already existing constraints but can introduce additional ones \cite{Dittrich:2013jaa}. 
In this way one obtains an `arrow of discrete time' which is determined by the direction of `information loss' (see also \cite{Dittrich:2013xwa} on this). We emphasize that this notion of `information loss' refers solely to the shrinking of the physical phase space at a given step and no information is literally `lost': in the course of the coarse graining moves the equations of motion of `finer' degrees of freedom at other steps are solved. Accordingly, the `finer' degrees of freedom, although no longer dynamical or reconstructable, contribute to the coarse grained `effective' dynamics.

On the other hand, refining the discretization on the evolving `time slice' via type I moves can be viewed as only adding vacuum degrees of freedom \cite{Dittrich:2013xwa,Dittrich:2014wpa}. The degrees of freedom added in the move $k\rightarrow k+1$ cannot be predicted by the data at $k$ because they are not post--observables at step $k+1$ \cite{Dittrich:2011ke,Dittrich:2013jaa}. Such a refinement move thus corresponds to representing the same (coarse) physical state at step $k$ on a finer discretization at step $k+1$ without adding any new relevant information. However, these newly added finer degrees of freedom, although not being post--observables at $k+1$, may be pre--observables at $k+1$ of a global evolution $k+1\rightarrow X$ to some time step $X>k+1$ and thus become propagating in the `future' evolution \cite{Dittrich:2013jaa}. 

\begin{figure}[hbt!]
\begin{center}
\psfrag{0}{$0$}
\psfrag{n}{$n$}
\psfrag{nx}{$n+x$}
\psfrag{no}{\large `Nothing'}
\hspace*{-3.5cm}\begin{subfigure}[b]{.22\textwidth}
\centering
\includegraphics[scale=.3]{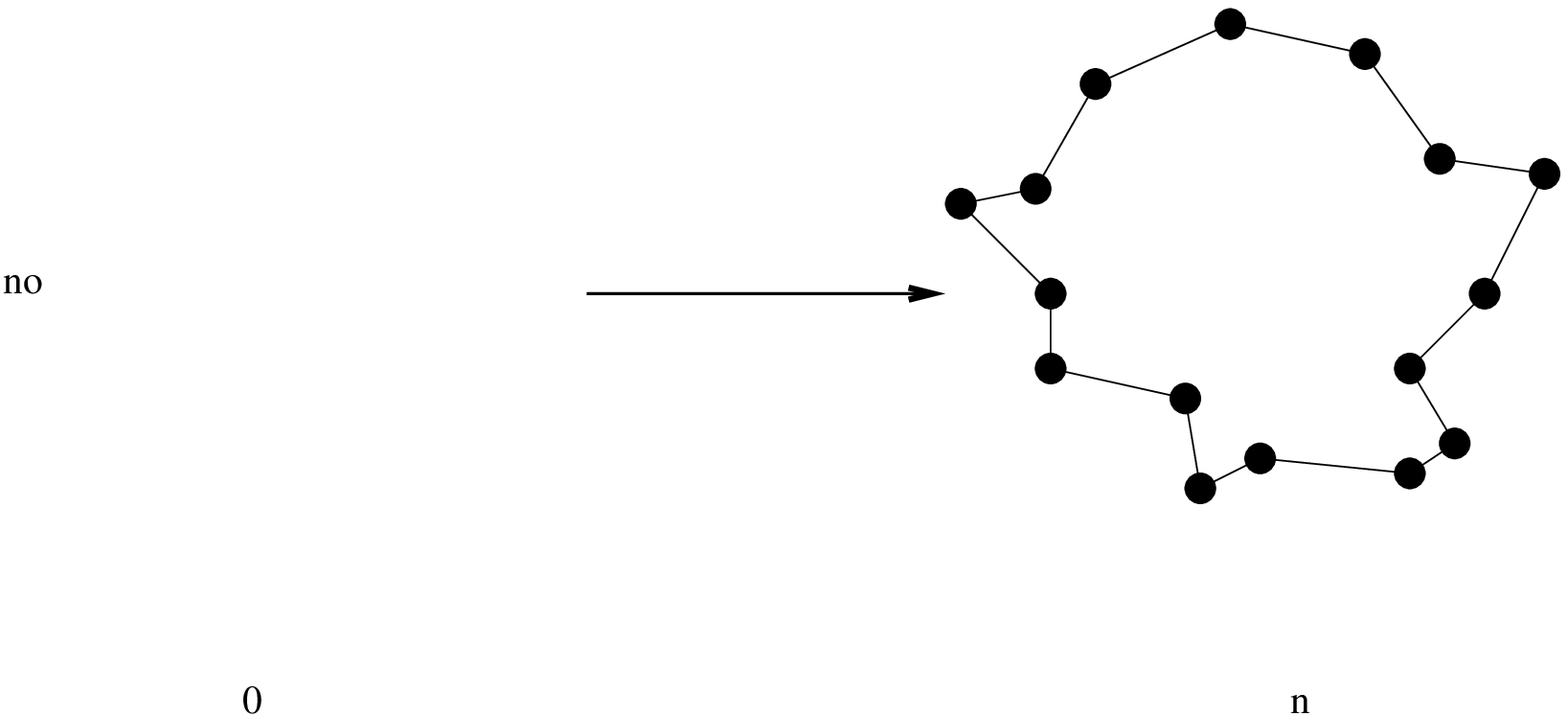}
\centering
\caption{\small }
\end{subfigure}
\hspace*{4.5cm}
\begin{subfigure}[b]{.22\textwidth}
\centering
\includegraphics[scale=.3]{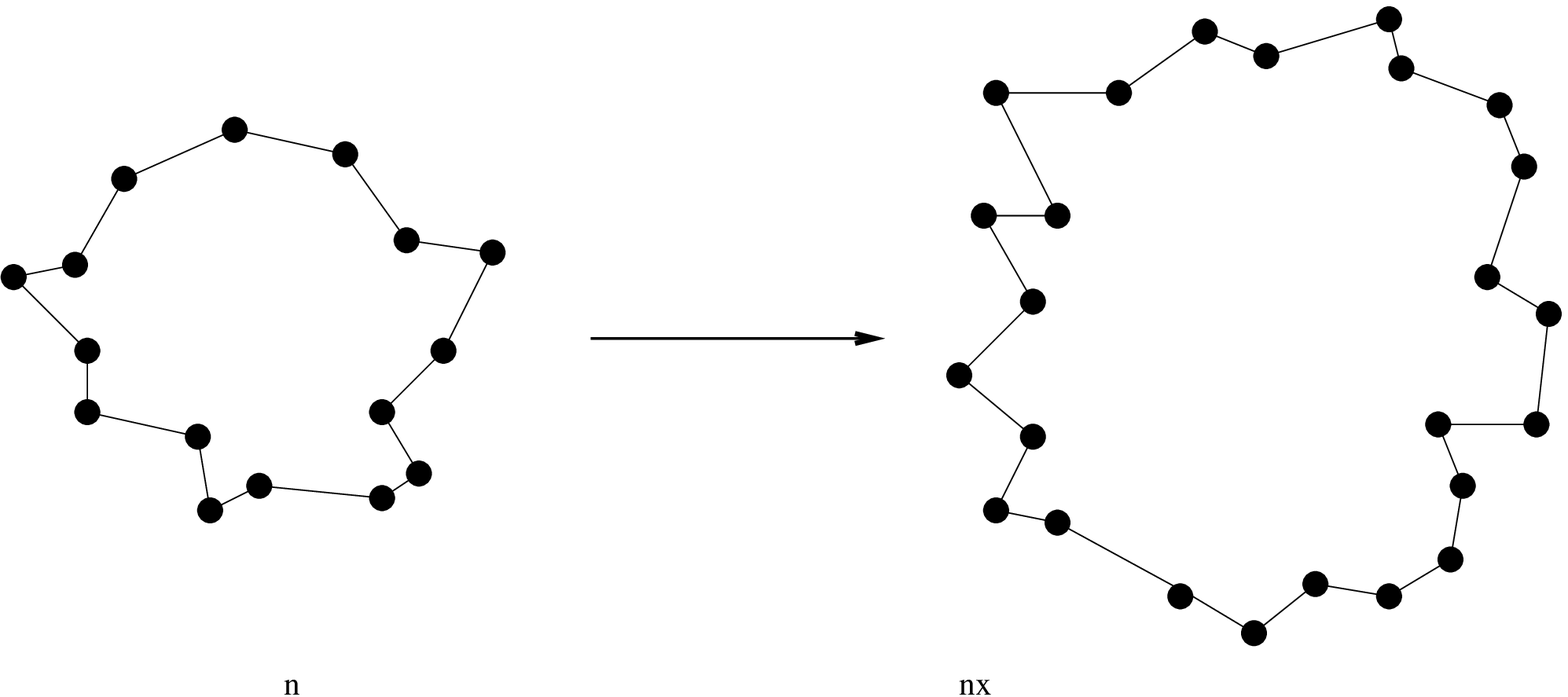}
\centering
\caption{\small }
\end{subfigure}
\caption{\small The `no boundary' proposal translated into the discrete. (a) There are no pre--observables at $0$ and no post--observables at $n$ for a move $0\rightarrow n$ from an empty triangulation to a later spherical hypersurface. (b) However, the next move $n\rightarrow n+x$ can admit pre--observables at $n$.}\label{fig_nb}
\end{center}
\end{figure}

For instance, consider a discrete version of the `no-boundary' proposal \cite{Hartle:1983ai} in gravitation. This corresponds to starting with an empty triangulation at a step $0$ and evolving to a triangulated spherical hypersurface at later time steps \cite{Dittrich:2013jaa,Hoehn:2014fka,Dittrich:2011ke} (see figure \ref{fig_nb}). For any global evolution $0\rightarrow n$ from the empty triangulation at $0$ to any subsequent spherical hypersurface at $n$ there are no non-trivial pre-- and post--observables propagating because such moves are fully constrained \cite{Dittrich:2013jaa,Hoehn:2014fka}. This, however, does not imply that spherical triangulations are devoid of `gravitons'. Quite the contrary, an evolution from some non-vanishing spherical hypersurface at $n$ to a later larger spherical hypersurface at $n+x$ will, in general, contain non-trivial propagating information because, although there are no non-trivial post--observables at $n$, there will generally be non-trivial pre--observables at $n$ \cite{Dittrich:2013jaa,Hoehn:2014fka}. 

The situation in the quantum theory is completely analogous. Quantum Dirac observables in the context of global evolution moves are discussed in \cite{Hoehn:2014fka}. Here we will only consider the situation for a local evolution move $k\rightarrow k+1$ with state updating map $\fu_{k\rightarrow k+1}:{}^+\ch^{\rm phys}_k\rightarrow{}^+\ch^{\rm phys}_{k+1}$ which updates a global move $0\rightarrow k$. Clearly, in order for an operator $\hat{O}^+_k$ to be a well-defined Dirac observable on ${}^+\ch^{\rm phys}_k$ it must commute with {\it all} quantum post--constraints at $k$,
\ba
[\hat{O}^+_k,{}^+\hat{C}^k_I]=0,\q\q\q\forall\,I.\nn
\ea
We shall call such operators $\hat{O}^+_k$ {\it quantum post--observables} on ${}^+\ch^{\rm phys}_k$ of the (global) move $0\rightarrow k$. Similarly, {\it quantum pre--observables} $\hat{O}^-_0$ on ${}^-\ch^{\rm phys}_0$ of the move $0\rightarrow k$ commute with {\it all} quantum pre--constraints at step $0$,
\ba
[\hat{O}^-_0,{}^-\hat{C}^0_I]=0,\q\q\q\forall\,I.\nn
\ea
The local move $k\rightarrow k+1$ can introduce non-trivial pre--constraints (case (ii) of section \ref{sec_locgen}) which `propagate backward' to step $0$ when updating $0\rightarrow k$ to $0\rightarrow k+1$ \cite{Hoehn:2014fka,Dittrich:2013jaa}. This will non-trivially project the pre--physical Hilbert space ${}^-\ch^{\rm phys}_0$ to a new pre--physical Hilbert space ${}^-\tilde{\ch}^{\rm phys}_0$ and thereby project out those pre--observables at $0$ which do not commute with the new effective quantum pre--constraints at $0$. The quantum pre-- and post--observables are thus associated to a given global evolution move rather than a single time step. This can be explicitly seen in the toy model of \cite{Hoehn:2014fka} which mimics a `creation from nothing' as in the `no boundary' proposal mentioned above. The quantum pre-- and post--observables associated to a move represent the physical degrees of freedom propagating in this move.

Let us now consider a local move $k\rightarrow k+1$. The quantum post--observables $\hat{O}^+_{k+1}$ on ${}^+\ch^{\rm phys}_{k+1}$ must likewise commute with {\it all} post--constraints at $k+1$, $[\hat{O}^+_{k+1},{}^+\hat{C}^{k+1}_J]=0$, $\forall\,J$. This implies different repercussions for different local moves.

Consider a type I local refinement move $k\rightarrow k+1$. As in sections \ref{sec_comp} and \ref{sec_momup} we have to consider its composition with a reverse type II move $k+1\rightarrow k$ in order to map quantum post--observables from $k$ to $k+1$ and vice versa. For instance, if $\hat{O}^+_{k+1}$ is a quantum post--observable on ${}^+\ch^{\rm phys}_{k+1}$ then
\ba
\hat{O}^+_k:=\fu^{II}_{k+1\rightarrow k}\,\hat{O}^+_{k+1}\,\fu^I_{k\rightarrow k+1}\nn
\ea
is a well-defined quantum post--observable at step $k$ since
\ba
&&\!\!\!\!\![\fu^{II}_{k+1\rightarrow k}\,\hat{O}^+_{k+1}\,\fu^I_{k\rightarrow k+1},{}^+\hat{C}^k]\,{}^+\psi^{\rm phys}_k=\nn\\
&&\fu^{II}_{k+1\rightarrow k}\,\hat{O}^+_{k+1}\,\fu^I_{k\rightarrow k+1}\,\underset{=0}{\underbrace{{}^+\hat{C}^k{}^+\psi^{\rm phys}_k}}-\underset{=0}{\underbrace{{}^+\hat{C}^k\,\fu^{II}_{k+1\rightarrow k}\,\hat{O}^+_{k+1}\,\fu^I_{k\rightarrow k+1}\,{}^+\psi^{\rm phys}_k}}=0\nn
\ea
The last equality requires $\hat{O}^+_{k+1}$ to commute with all post--constraints at $k+1$. Hence, $\hat{O}^+_{k+1}\,\fu^I_{k\rightarrow k+1}\,{}^+\psi^{\rm phys}_k$ $\in{}^+\ch^{\rm phys}_{k+1}$. The reverse map $\fu^{II}_{k+1\rightarrow k}$, as seen in section \ref{sec_II} preserves all quantum post--constraints such that the result will be annihilated by ${}^+\hat{C}^k$.

Equivalently, if $\hat{O}^+_k$ is a post--observable on ${}^+\ch^{\rm phys}_k$,
\ba
\hat{O}^+_{k+1}:=\fu^I_{k\rightarrow k+1}\,\hat{O}^+_k\,\fu^{II}_{k+1\rightarrow k}\nn
\ea
is a well-defined quantum post--observable at step $k+1$ because
\ba
[\fu^I_{k\rightarrow k+1}\,\hat{O}^+_k\,\fu^{II}_{k+1\rightarrow k},{}^+\hat{C}^{k+1}]\,{}^+\psi^{\rm phys}_{k+1}=0\nn
\ea 
for analogous reasons as above. Consequently, the post--observables at step $k$ are in one-to-one correspondence with the post--observables at step $k+1$. The new degrees of freedom added in the refining type I move thus do not contribute to the quantum post--observables at $k+1$. Indeed, as argued in \cite{Dittrich:2013xwa} these degrees of freedom can be viewed as additional vacuum degrees of freedom. In the type I move the coarser states ${}^+\psi^{\rm phys}_k\in{}^+\ch^{\rm phys}_k$ are mapped to states ${}^+\psi^{\rm phys}_{k+1}$ carrying the same amount of `coarse' information but represented on a `finer' Hilbert space ${}^+\ch^{\rm phys}_{k+1}$ corresponding to a refined discretization. We emphasize, however, that the new vacuum degrees of freedom added in the move, although not being post--observables at $k+1$, may still feature in non-trivial pre--observables $\hat{O}^-_{k+1}$ at $k+1$ such that $[\hat{O}^-_{k+1},{}^-\hat{C}^{k+1}]=0$ for some future global move $k+1\rightarrow f$.

The situation for a type II coarse graining move $k\rightarrow k+1$ is more complicated than for the type I refining move. As in sections \ref{sec_comp} and \ref{sec_momup} we must consider its composition with a reverse type I move $k+1\rightarrow k$ in order to map observables from one time step to the other. Let $\hat{O}^+_{k+1}$ be a quantum post--observable on ${}^+\ch^{\rm phys}$. Then, by similar arguments to lemma \ref{lemb},
\ba
\widehat{\tilde O}^+_k:=\fu^I_{k+1\rightarrow k}\,\hat{O}^+_{k+1}\,\fu^{II}_{k\rightarrow k+1}\nn
\ea
defines a map ${}^+\ch^{\rm phys}_k\rightarrow{}^-\mathbb{P}^B_k({}^+\ch^{\rm phys}_k)$. One can check that, in analogy to above,
\ba
&&[\fu^I_{k+1\rightarrow k}\,\hat{O}^+_{k+1}\,\fu^{II}_{k\rightarrow k+1},{}^+\hat{C}^k]\,{}^+\psi^{\rm phys}_k=0.\nn
\ea
However, $\widehat{\tilde O}^+_k$ will not commute with the pre--constraints ${}^-\hat{C}^k_\beta$ because it defines a map ${}^+\ch^{\rm phys}_k\rightarrow{}^-\mathbb{P}^B_k({}^+\ch^{\rm phys}_k)$.

The reverse direction for the type II move is more subtle. The composition
\ba
\hat{O}^+_{k+1}:=\fu^{II}_{k\rightarrow k+1}\,\widehat{\tilde O}^+_k\,\fu^I_{k+1\rightarrow k}\nn
\ea
defines a quantum post--observable on ${}^+\ch^{\rm phys}_{k+1}$, however, only if $\widehat{\tilde O}^+_k$ is a quantum post--observable at $k$ that also satisfies $[\widehat{\tilde O}^+_k,{}^-\hat{C}^k_\beta]=0$, $\forall\,\beta=1,\ldots,K_\beta$. This is because
\ba
\fu^{II}_{k\rightarrow k+1}\,\widehat{\tilde O}^+_k\,\fu^I_{k+1\rightarrow k}\,{}^+\psi^{\rm phys}_{k+1}&=&\fu^{II}_{k\rightarrow k+1}\,\widehat{\tilde O}^+_k\,\fu^I_{k+1\rightarrow k}\,\cdot\fu^{II}_{k\rightarrow k+1}\,{}^+\psi^{\rm phys}_k\nn\\
&\underset{\text{\tiny lemma \ref{lemb}}}{=}&\fu^{II}_{k\rightarrow k+1}\,\widehat{\tilde O}^+_k\,{}^-\mathbb{P}^B_k\,{}^+\psi^{\rm phys}_k.\nn
\ea
(Note that in this composition $\fu^{II}_{k\rightarrow k+1}$ must be fully regularized as all of its pre--constraints are already implemented thanks to ${}^-\mathbb{P}^B_k$.) In this case, since the moves preserve quantum post--constraints (see theorems \ref{thm_I} and \ref{thm_II}),
\ba
[\fu^{II}_{k\rightarrow k+1}\,\widehat{\tilde O}^+_k\,\fu^I_{k+1\rightarrow k},{}^+\hat{C}^{k+1}]\,{}^+\psi^{\rm phys}_{k+1}=0.\nn
\ea
We thus see that only those post--observables $\widehat{\tilde O}^+_k$ on ${}^+\ch^{\rm phys}_k$ can be mapped to post--observables at $k+1$ if they are also post--observables on ${}^-\mathbb{P}^B_k({}^+\ch^{\rm phys}_k)$ at $k$. That is, all post--observables at step $k$ which do {\it not} commute with the non-trivial pre--constraints ${}^-\hat{C}^k_\beta$ of case (ii) in section \ref{sec_locgen} are `too fine grained', {\it projected out} and cannot be regained. 
In this case, the type II move implements a coarse graining that maps a state ${}^+\psi^{\rm phys}_k\in{}^+\ch^{\rm phys}_k$, carrying `finer' information, to a new state ${}^+\psi^{\rm phys}_{k+1}$ carrying `coarser' information on a new post--Hilbert space ${}^+\ch^{\rm phys}_{k+1}$ corresponding to a coarser discretization (see also \cite{Dittrich:2013xwa,Hoehn:2014fka} on this issue). The `arrow of discrete time' alluded to above thus also appears in the quantum theory as the direction of projecting out `finer' post--observables. This goes in hand with the non-unitarity of the type II move in the presence of non-trivial pre--constraints (see theorem \ref{thm_II}). We emphasize, again, that, nevertheless, no information is literally `lost' because the `finer' degrees of freedom (at step $k$) are integrated in (\ref{kupII}, \ref{supII}) and thereby contribute to the construction of the physical state at $k+1$.

In this sense the physical Hilbert spaces `evolve' in the course of an evolution generated by discretization changing moves. This also entails the notion of `evolving' physical Hilbert spaces for global moves, discussed in \cite{Hoehn:2014fka}, since the latter can be decomposed into the local ones presented here.

The state of affairs for the type III move is completely analogous to that of the type II move. We shall thus abstain from discussing it in detail. Also in this case, only post--observables at $k$ that also commute with the non-trivial pre--constraints at $k$ can be mapped to post--observables at $k+1$. That is, the type III move will also project out Dirac observables if pre--constraints of case (ii) in section \ref{sec_locgen} occur. Note, however, that the 2--2 Pachner evolution move in 3D Regge Calculus does not generate any non-trivial pre--constraints \cite{Dittrich:2011ke}. In the light of the present discussion, this comes as no surprise, given that in 3D quantum gravity there are no local degrees of freedom that could be projected out. Instead, as argued in \cite{Dittrich:2013xwa}, the 2--2 move in spin net models constitutes an entangling move.

As a side remark, we emphasize that the above reasoning does not apply if the operators do not commute with the constraints. For instance, if the momentum operators in the quantum momentum updating \ref{sec_momup} do not commute with the constraints then the time evolved version will in general also not commute with the constraints. In this case, the action of such a kinematical operator will throw a physical state out of the physical Hilbert space and the local maps $\fu_{k\rightarrow k+1}$ do not contain the global projectors necessary to map the result back into a physical Hilbert space. For instance, if one carried out the momentum updating for the operators $\hat{p}^k_b$ these would be trivially mapped from step $k$ to step $k+1$. If they did not commute with the post--constraints at $k$ they surely will also not commute with the post--constraints at $k+1$.

\section{Distinct features of dynamics in simplicial gravity}\label{sec_qg}

\begin{figure}[hbt!]
\begin{center}
\psfrag{a}{\large +}
\psfrag{b}{\Large --}
\psfrag{c}{\large ?}
\psfrag{s}{$\Sigma$}
\includegraphics[scale=.35]{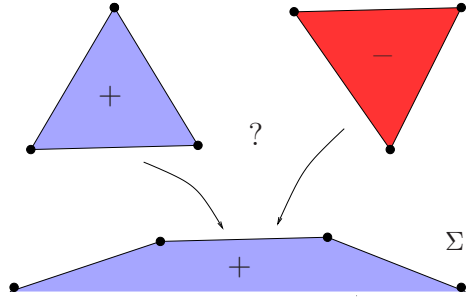}
\caption{\small 1--2 Pachner move in 2D as an analogue of the 1--3 Pachner move in 3D. In a simplicial gravity context gluings with both positive and negative orientation are possible. But from the point of view of $\Sigma$ both amount to a 1--2 Pachner move. }\label{anti}
\end{center}
\end{figure}
\begin{center}
\begin{figure}[htbp!]
\psfrag{a}{ +}
\psfrag{b}{ --}
\begin{subfigure}[b]{.17\textwidth}
\centering
\includegraphics[scale=.4]{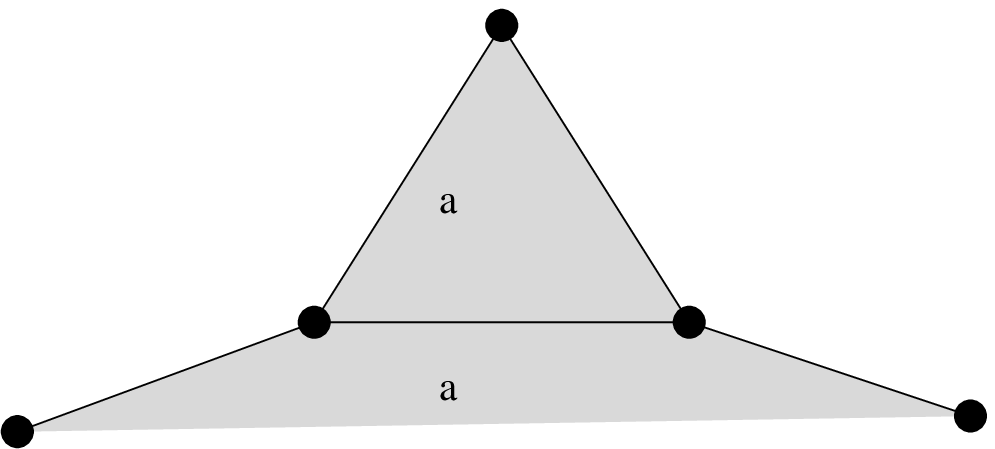}
\centering
\caption{\small }
\end{subfigure}
\hspace*{2.5cm}
\begin{subfigure}[b]{.17\textwidth}
\centering
\includegraphics[scale=.4]{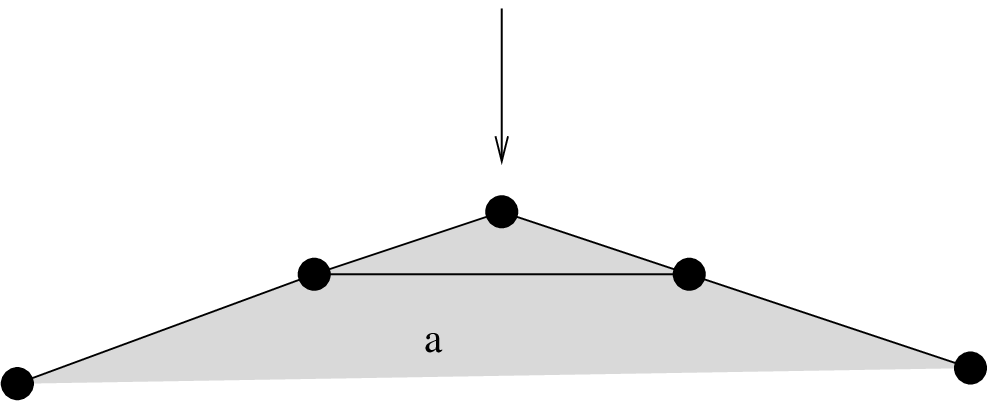}
\caption{\small }
\end{subfigure}
\hspace*{2.5cm}
\begin{subfigure}[b]{.17\textwidth}
\centering
\includegraphics[scale=.4]{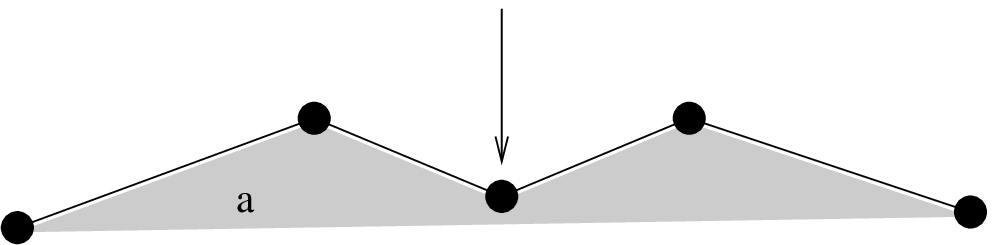}
\caption{\small }
\end{subfigure}
\caption{\small 2D analogue of the 1--3 Pachner move in 3D simplicial gravity. The new vertex of the new triangle (with positive orientation) generating the 1--2 move in $\Sigma$ can be freely displaced \cite{Dittrich:2011ke}. In particular, it can be pushed into $\Sigma$ in which case the move corresponds to a gluing of a triangle with negative orientation. }\label{fig_pos}
\end{figure}
\end{center}

In the above constructions we have always assumed that the distinction between `forward' and `backward' evolution is clear. In simplicial gravity, however, the situation is different. Canonical time evolution corresponds to gluing single $D$-dimensional simplices to a $(D-1)$ dimensional `spatial' hypersurface $\Sigma$ and thereby evolving $\Sigma$ in `time' \cite{Dittrich:2011ke,Dittrich:2009fb}. When given the simplex to be glued to $\Sigma$, the variables describing it do not carry information about the simplex' orientation. For instance, in 3D Regge Calculus, a 1--3 Pachner evolution move within $\Sigma$ corresponds to gluing a single tetrahedron $\tau$ onto a single triangle in $\Sigma$ \cite{Dittrich:2011ke}. But given $\tau$, it is not clear whether its tip points into or away from $\Sigma$ and both scenarios are possible (see figure \ref{anti} for the 2D analogue). Gluing $\tau$ with the tip pointing away from $\Sigma$ can be viewed as `forward' evolution, while gluing $\tau$ with the tip pointing into $\Sigma$ is equivalent to removing a tetrahedron from the underlying triangulation that led to $\Sigma$ and thus to `backward' evolution. In the latter case, $\tau$ can be viewed as a piece of `anti-spacetime' which has opposite orientation to spacetime (see also \cite{Christodoulou:2012af} for a related discussion). Equivalently, one can glue $\tau$ with tip pointing away from $\Sigma$ and then push the tip into $\Sigma$, see figure \ref{fig_pos}. This is possible because the `position' of the tip cannot be predicted, given only the data in $\Sigma$ at the previous step \cite{Dittrich:2011ke,Dittrich:2013jaa}. This freedom of placing the top vertex can (under certain conditions) be related to a diffeomorphism symmetry in the discrete \cite{Dittrich:2009fb,Dittrich:2011ke,Dittrich:2013jaa,Bahr:2009ku}.

However, from the perspective of $\Sigma$, the distinction between the `forward' and `backward' evolving case is not meaningful, both appear as a 1--3 Pachner move within $\Sigma$. Similarly, for any Pachner evolution move in any dimension this distinction is of little importance and only the dynamics of $\Sigma$ matters. The dynamical nature of the spacetime triangulation puts simplicial gravity apart from other discrete systems. By contrast, for a field theory on a triangulation the orientation (and thus `time direction') is fixed from the outset because only the field on the triangulation, but not the triangulation itself is dynamical.

If one now considers the Pachner moves as local evolution moves in simplicial quantum gravity, 
the distinction between `forward' and `backward' evolution, which we made use of in the lemmas of section \ref{sec_comp}, no longer holds. For the quantum dynamics of $\Sigma$ there are only the various types of Pachner moves, but {\it without} distinction of the orientation or, equivalently `forward' and `backward'. Dropping this distinction 
requires a superposition of `forward' and `backward' propagation\footnote{In addition, but independently of this, the quadratic nature of the Hamiltonian constraint in gravitational systems also generically causes a superposition of `forward' and `backward' internal time directions in the quantum theory. See \cite{Bojowald:2010xp,Bojowald:2010qw,Hohn:2011us} for a detailed exposition.} or, equivalently, a superposition of evolution in a spacetime and `anti-spacetime' (see also \cite{Dittrich:2013xwa}). This amounts to an integration over both positive and negative values of lapse and shift which helps to ensure diffeomorphism symmetry, but is in tension with causality (in a time direction sense) as argued in \cite{Teitelboim:1983fh}. 
This is the origin of the appearance of the Regge action in a cosine (rather than exponential) in semiclassical spinfoam amplitudes \cite{Conrady:2008mk,Barrett:2009gg,Perez:2012wv,Oriti:2014aka}.

But there are also other subtleties that arise when adapting the present formalism to discrete gravity models. These have been amply discussed in \cite{Hoehn:2014fka} such that we shall be brief on this topic here. In particular, thus far we have examined how the local evolution moves evolve physical states satisfying the constraints in discrete time. However, continuum quantum gravity, whose dynamics is generated by constraints, is devoid of any external time parameters; physical states satisfying all constraints are {\it a priori} timeless and do {\it not} evolve in an external time. This manifests the infamous `problem of time' in quantum gravity \cite{Kuchar:1991qf,Isham:1992ms,Anderson:2012vk}. Instead, physical quantum states of gravitational systems contain the entire information about the dynamics and a time evolution in {\it internal} clock degrees of freedom can often be extracted from them using the relational paradigm of dynamics \cite{Rovelli:2004tv,Rovelli:1989jn,Bojowald:2010xp,Bojowald:2010qw,Hohn:2011us,Tambornino:2011vg}. On the other hand, in discrete gravity time evolution is {\it not} generated by constraints but by evolution moves. As a consequence, an `external' discrete time (counting the moves) survives even upon implementation of the constraints in the quantum theory such that it appears as if physical states in discrete quantum gravity models `evolve' in this `external' time. 

However, such a dynamics {\it cannot} be interpreted as an external evolution. Especially, if the space-time discretization is perfect in the sense that it preserves the continuum symmetries and dynamics \cite{Bahr:2009qc,Bahr:2009mc,Bahr:2011uj}---as, e.g., 3D (vacuum) Regge Calculus without cosmological constant \cite{Bahr:2009ku,Bahr:2009qc,Dittrich:2011vz} and, by construction, Loop Quantum Gravity \cite{Thiemann:2007zz,Rovelli:2004tv} in general---time evolution acts as a projector onto solutions to the quantum constraints \cite{Halliwell:1990qr,Rovelli:1998dx,Noui:2004iy,Thiemann:2013lka}. In this case, the discretization changing Hamiltonian dynamics \cite{Dittrich:2011ke,Thiemann:1996ay,Thiemann:1996aw,Alesci:2010gb,Dittrich:2013xwa,Bonzom:2011hm} can clearly {\it not} be interpreted as `evolving' physical states in some external time. So how should the discretization changing Hamiltonian dynamics be interpreted?

The discretization changing Hamiltonian dynamics can be regarded as coarse graining or refining physical states and thereby sometimes changing the representation of the latter. In fact, for perfect discretizations the action of the refining local moves on physical states can be viewed as representing one and the same physical state on different discretizations without losing (or gaining) any dynamical information \cite{Dittrich:2013xwa}. The initial and final state may thus be safely identified as one and the same physical state. Accordingly, there is no evolution of the state in an external time despite the change of discretization. Rather, the `external' discrete time can be interpreted as counting the steps towards an embedding into the continuum. Refining moves generate dynamical embeddings which map states on a coarser discretization onto a finer discretization and can ultimately be used to embed states on discretizations into the continuum Hilbert space in a cylindrically consistent manner, as proposed in \cite{Dittrich:2012jq,Dittrich:2013xwa}. This requires a path independence of the dynamical refining operations from an initial state on a coarsely to a final state on a finely discretized hypersurface \cite{Dittrich:2013xwa} and yields a notion of hyperbolicity in the discrete \cite{Hoehn:2014fka}. In the recent \cite{Dittrich:2014wpa} such refining Pachner moves have also been successfully employed to construct a new (geometrically non-degenerate) vacuum for Loop Quantum Gravity using a simplicial formulation of the theory.

By contrast, coarse graining evolution moves will generically change physical states non-unitarily for systems with propagating degrees of freedom---even if the discretization is perfect (see also \cite{Hoehn:2014fka}). This is because states on finer discretizations will generically carry more dynamical information than states on coarser discretizations if there are propagating degrees of freedom. That is, in contrast to the refining moves, coarse graining moves cannot in general be used to identify states on different discretizations; mapping a state on a fine discretization to a state on a coarse discretization is generally irreversible. In this case, the change of physical states in the `external' discrete time can be interpreted as changes in a discrete renormalization flow rather than in a time evolution.


On the other hand, 4D Regge Calculus and 4D spin foam models do not constitute perfect discretizations because the continuum diffeomorphism symmetry is generically broken for simplicial geometries featuring non-trivial curvature \cite{Rocek:1982tj,Bahr:2009ku,Dittrich:2008pw,Bahr:2011xs,Bonzom:2013ofa}. Related to this, the set of pre-- and post--constraints generally does not include the Hamiltonian and diffeomorphism constraints in 4D Regge Calculus \cite{Dittrich:2009fb,Dittrich:2011ke,Dittrich:2013jaa}. The latter rather arise in the form of approximate or `pseudo'-constraints \cite{Dittrich:2009fb,Dittrich:2011ke,Gambini:2002wn,Gambini:2005vn} such that time evolution can only be expected to yield an approximate projector onto the Hamiltonian and diffeomorphism constraints for large scales \cite{Dittrich:2013xwa,Dittrich:2008pw}. In this case, {\it all} 4D Pachner evolution moves can be regarded as changing (pre-- and post--)physical states non-trivially: the refining 1--4 and 2--3 moves will generally change physical states because, given the absence of symmetries, there will be no path independence of evolution from coarser to finer discretizations which would permit an identification of states in a cylindrically consistent manner, while the non-trivial coarse graining 4--1 and 3--2 moves can be expected to change physical states by non-unitary projections as discussed above. Nonetheless, the local evolution moves in 4D simplicial gravity models with broken continuum symmetries should likewise be viewed as generating a dynamical coarse graining or refining of physical states, rather than generating a genuine `evolution' in some external time. 

In order to 
build a cylindrically consistent (continuum) quantum gravity theory also in 4D, coarse graining techniques seem a promising tool to construct effective theories which are better behaved with regards to symmetries, as examined in \cite{Bahr:2009ku,Bahr:2009qc,Bahr:2011uj,Dittrich:2012qb,Dittrich:2012jq,Dittrich:2013xwa,Dittrich:2013voa}. This would ultimately also remove the `external' discrete time  counting the time evolution moves in discrete gravity models.

 \section{Conclusions and outlook}\label{sec_conc}


Discretization (or graph) changing local evolution moves feature in several quantum gravity models \cite{Thiemann:1996ay,Thiemann:1996aw,Alesci:2010gb,Dittrich:2011ke,Dittrich:2013jaa,Dittrich:2013xwa,Bonzom:2011hm} and in field theory on an evolving lattice \cite{Dittrich:2013jaa,Hoehn:2014aoa,Foster:2004yc,Jacobson:1999zk}. While such local evolution moves have been classically studied in detail for variational discrete systems with arbitrary configuration manifolds in \cite{Dittrich:2011ke,Dittrich:2013jaa}, this article exhibits how such local evolution moves can be quantized. However, to avoid global and topological non-trivialities in the quantization \cite{isham2}, this manuscript restricts to variational discrete systems with Euclidean configuration spaces. For instance, this quantization directly applies to a scalar field theory on a triangulated space-time lattice, as illustrated in the main body and, similarly, to perturbative 4D Regge Calculus to linear order \cite{Dittrich:2009fb,dh4}. 

Notwithstanding this restriction, there is no principle obstruction which could inhibit an extension to systems with arbitrary configuration manifolds. In this case the formalism could be applied, e.g., to non-perturbative Regge Calculus and possibly spin foam models. Furthermore, the qualitative results 
of this formalism can be expected to hold analogously in systems with more general configuration manifolds. Thus, we hope, in particular, that the present work, incl.\ the companion paper on global moves \cite{Hoehn:2014fka}, sheds new light on the discretization changing dynamics in quantum gravity models. More precisely: \\~\\
(1) The (type II) 4--1 and 3--2 Pachner evolution moves in 4D simplicial quantum gravity models will, in general, be non-trivial coarse graining moves which must lead to non-unitary projections of physical Hilbert spaces because of the presence of propagating degrees of freedom. This would give rise to an irreversible evolution of the post--physical Hilbert space on the evolving hypersurface under local moves. 
The (type I) 1--4 and 2--3 Pachner evolution moves, on the other hand, are refining moves which can be employed to generate dynamical embeddings of states on coarser to states on finer triangulated hypersurfaces. If diffeomorphism symmetry is preserved, these embeddings should be cylindrically consistent and could be employed to construct a physical vacuum (or ground) state for quantum gravity \cite{Dittrich:2012jq,Dittrich:2013xwa}. This has recently been successfully implemented for Loop Quantum Gravity \cite{Dittrich:2014wpa,biancamarc2}.

We emphasize that all four of the 1--4, 2--3, 3--2 and 4--1 Pachner evolution moves are necessary in order to get any interesting dynamics in 4D Regge Calculus \cite{Dittrich:2011ke}. This becomes especially clear in the 4D linearized theory where the 1--4 move generates four (lapse and shift) gauge degrees of freedom, the 2--3 move generates one propagating curvature (i.e.\ `graviton') degree of freedom, the 3--2 move produces the only non-trivial equations of motion for the curvature degrees of freedom and the 4--1 move removes four gauge degrees of freedom from the evolving hypersurface \cite{thesis,dh4}. Since the Regge action appears in the semiclassical limit of spin foam models \cite{Conrady:2008mk,Barrett:2009gg,Perez:2012wv}, all four Pachner evolution moves in 4D can likewise be expected to be relevant for a non-trivial spin foam dynamics. For instance, in \cite{Alesci:2010gb} a spin foam motivated regularization of the Hamiltonian constraint for 4D Loop Quantum Gravity is constructed which so far implements the 1--4 Pachner evolution move. However, while necessary for the full dynamics, a dynamics generated by solely applying 1--4 moves to a single 4--simplex only generates so-called stacked spheres \cite{Ambjorn:1996ny} which are flat geometries devoid of any internal degrees of freedom. In fact, since the 1--4 and 2--3 Pachner moves are of type I, an evolution generated {\it only} via these refining moves would (at least within the present formalism) never generate a non-trivial path integral, but only a factorization of physical states. States would be updated by multiplicative factors, as in theorem \ref{thm_I}, without integrating out any degrees of freedom. Therefore, all four Pachner evolution moves in 4D must be considered.\\~\\
(2) As regards Dirac observables on temporally varying discretizations, it follows from the results of section \ref{sec_dirac} that the (type I) refining 1--4 and 2--3 Pachner moves cannot induce any new post--observables. As proposed in \cite{Dittrich:2013xwa}, the newly added degrees of freedom can, instead, be interpreted as unexcited vacuum degrees of freedom. By contrast, the (type II) coarse graining 3--2 and 4--1 Pachner moves can be expected to generally project out finer degrees of freedom. Only those quantum post--observables survive a coarse graining move which also commute with the new quantum pre--constraints of this move. 
\\~\\
(3) As in lemma \ref{lema}, undoing a 1--4 Pachner move with a 4--1 move, or a 2--3 move with a 3--2 move (without distinction of `forward' and `backward' orientation and regularizing any divergences) in 4D Quantum Regge Calculus should result in an identity operation. Conversely, one can expect, that undoing a 4--1 Pachner move with a 1--4 move, or a 3--2 move with a 2--3 move as in lemma \ref{lemb}, will be equivalent to a projection onto the pre--constraints of the 4--1 and 3--2 moves, respectively.\\~\\
(4) The results of section \ref{sec_loc} also suggest a novel approach to (bubble) gauge divergences arising in the state sums of Quantum Regge Calculus \cite{williamsreview,Rocek:1982tj,ponzreg} and of spin foam quantum gravity models \cite{Bonzom:2013ofa,Bonzom:2010ar,Riello:2013bzw,Freidel:2002dw}. Gauge symmetry related divergences in the resulting state sums of the present formalism arise from constraints which are simultaneously pre-- and post--constraints and whose corresponding improper projectors are therefore doubly implemented. Presumably, this will also hold in Quantum Regge Calculus and thereby indirectly in spin foam models. For instance, as previously mentioned, the pre--constraints of the 3--1 Pachner evolution move in 3D Regge Calculus correspond to Hamiltonian and diffeomorphism constraints and are also post--constraints \cite{Dittrich:2011ke}. Similarly, for 4D perturbative Regge Calculus to linear order, the pre--constraints of the 4--1 Pachner move are automatically satisfied \cite{thesis,dh4}. In both cases, these constraints generate a vertex displacement symmetry \cite{Dittrich:2009fb,Dittrich:2011ke,Dittrich:2008pw,Bahr:2009ku,Bahr:2011xs} which corresponds to a diffeomorphism symmetry in triangulations. Since the corresponding orbits are non-compact, these moves can thus be expected to generate divergences in a state sum under the above circumstances. Similarly, one can also generate divergences in a state sum for 4D non-perturbative Quantum Regge Calculus by performing an invertible composition of Pachner moves discussed under (3): for example, undoing a 1--4 Pachner evolution move immediately with a 4--1 move should result in a divergence because the post--constraints of the 1--4 move will agree with the pre--constraints of the 4--1 move. All of these divergences can be regularized by dropping one instance of the doubly occurring improper projector or by inserting suitable (Faddeev-Popov) gauge fixing conditions. We emphasize, however, that further divergences unrelated to gauge symmetries may arise in the state sums.

 \appendix
 
 \section{Proofs of theorems \ref{thm_I}--\ref{thm_III}}\label{app_1}
 
 We begin with the proof of theorem \ref{thm_I}:
 
\begin{proof}
We have to impose the two sets of constraints (\ref{anh1c}) in the quantum theory. Clearly, $\hat{C}^k_n=\hat{p}^n_k$ is automatically implemented, while
\ba
{}^+\hat{C}^{k+1}_n\,{}^+\psi^{\rm phys}_{k+1}=\left(-i\hbar\f{\p}{\p x^n_{k+1}}-\f{\p S_{k+1}}{\p x^n_{k+1}}\right){}^+\psi^{\rm phys}_{k+1}=\left(-i\hbar\f{\p M_{k\rightarrow k+1}}{\p x^n_{k+1}}\right){}^+\psi^{\rm phys}_{k+1}\overset{!}{=}0\nn
\ea
implies that $M_{k\rightarrow k+1}(x^e_{k+1})$ cannot depend on the new degrees of freedom $x^n_{k+1}$.

Next, we have to ensure that $\fu^I_{k\rightarrow k+1}$ respects the post--constraints which already existed at step $k$. In (the appendix of) \cite{Dittrich:2013jaa} it is shown that, at the classical level, every post--constraint at $k$ is preserved and transformed as follows under momentum updating $\fh_k$ given above:
\ba
{}^+C^k(x^e_k,x^b_k,p^k_e,p^k_b)\big|_{p^k={}^+p^k}=0\overset{\fh_k}{\longmapsto}{}^+C^{k+1}\left(x^e_{k+1},x^b_{k+1},p^{k+1}_e-\f{\p S_{k+1}}{\p x^e_{k+1}},p^{k+1}_b\right)\Big|_{p^{k+1}={}^+p^{k+1}}=0,\nn
\ea
where $p={}^+p$ means that this holds for post--momenta. We can make use of the fact that 
\ba
\left(\hat{p}^{k+1}_e-\f{\p S_{k+1}}{\p x^n_{k+1}}\right)\psi(x_{k+1})=e^{iS_{k+1}/\hbar}\,\hat{p}^{k+1}_e\,e^{-iS_{k+1}/\hbar}\,\psi(x_{k+1})\nn
\ea
and that (in the position representation) $\hat{p}^{k+1}_b,\hat{p}^{k+1}_e$ are the derivative operators with respect to $x^b_{k+1},x^e_{k+1}$ acting on states at $k+1$, while $\hat{p}^{k}_b,\hat{p}^k_e$ are the derivative operators with respect to $x^b_k,x^e_k$ acting on states at $k$. For quantum post--constraints with a power series expansion (in the canonical variables) this yields 
\ba
&&{}^+\hat{C}^{k+1}\left(x^e_{k+1},x^b_{k+1},e^{iS_{k+1}/\hbar}\,\hat{p}^{k+1}_e\,e^{-iS_{k+1}/\hbar},\hat{p}^{k+1}_b\right){}^+\psi^{\rm phys}_{k+1}\q\q\q\q\q\q\q\q\q\q\q\q\q\q\q\nn\\
&&\q\q\q=e^{iS_{k+1}/\hbar}\,{}^+\hat{C}^k(x^e_k,x^b_k,\hat{p}^k_e,\hat{p}^k_e)\,e^{-iS_{k+1}/\hbar}\,M_{k\rightarrow k+1}\,e^{iS_{k+1}/\hbar}\,{}^+\psi^{\rm phys}_k\nn\\
&&\q\q\q=e^{iS_{k+1}/\hbar}\,{}^+\hat{C}^k\,M_{k\rightarrow k+1}\,{}^+\psi^{\rm phys}_k\overset{!}{=}0\label{conI}
\ea
because $e^{iS_{k+1}/\hbar}$ commutes with the $\hat{p}^{k+1}_b$. Thus, $\fu^I_{k\rightarrow k+1}$ preserves the quantum post--constraints of step $k$ provided $M_{k\rightarrow k+1}(x^e_{k+1})$ commutes with ${}^+\hat{C}^k$. This is possible if $M_{k\rightarrow k+1}$ is constant.

The map $\fu^I_{k\rightarrow k+1}$ is unitary because
\ba
\langle{}^+\psi^{\rm phys}_{k+1}\big|{}^+\phi^{\rm phys}_{k+1}\rangle_{\rm phys+}&=&\int \prod_{e,b,n}dx^e_{k+1}dx^b_{k+1}dx^n_{k+1}\left(\psi^{\rm kin}_{k+1}(x^n_{k+1},x^e_{k+1},x^b_{k+1})\right)^*\nn\\
&&\q\q\q\q\q\q\q\q\q\times{}^+\phi^{\rm phys}_{k+1}(x^n_{k+1},x^e_{k+1},x^b_{k+1})\nn\\
&\underset{(\ref{supI})}{=}&\int \prod_{e,b}dx^e_k\, dx^b_k \left(\psi^{\rm kin}_k(x^e_k,x^b_k)\right)^*{}^+\phi^{\rm phys}_{k}(x^e_{k},x^b_{k})\nn\\
&=&\langle{}^+\psi^{\rm phys}_{k}\big|{}^+\phi^{\rm phys}_{k}\rangle_{\rm phys+},\nn
\ea
where we have made use of the identification (and the two left equations in (\ref{anh1}, \ref{anh1b}))
\ba
\psi^{\rm kin}_k(x^e_k,x^b_k):=\int\prod_ndx^n_{k+1}(M^I_{k\rightarrow k+1})^*\,e^{-iS_{k+1}/\hbar}\,\psi^{\rm kin}_{k+1}(x^e_{k+1},x^b_{k+1},x^n_{k+1}).\nn
\ea 
Indeed, (\ref{conI}) entails that if $\psi^{\rm kin}_{k+1}$ does not satisfy any constraints, then neither does $\psi^{\rm kin}_k$.
\end{proof}

The last equation corresponds to the time reversed map $\fu_{k+1\rightarrow k}^{II}$ of the type II move (see section \ref{sec_comp}).

We continue with the proof of theorem \ref{thm_II}:

\begin{proof}
The constraints (\ref{case2c}) must be imposed in the quantum theory. The quantum constraints $\hat{C}^{k+1}_o$ are automatically satisfied by both (\ref{kupII}, \ref{supII}) because the physical post--states at $k+1$ do not depend on the gauge modes $x^o_{k+1}$. On the other hand, the propagator $K_{k\rightarrow f}$ (\ref{revpup}) must satisfy the new quantum pre--constraints ${}^-\hat{C}^k_o$ in the form (\ref{cond}). (For notational simplicity, we recombine the two sets of $\hat{C}^k_\alpha,{}^-\hat{C}^k_\beta$ into the single set ${}^-\hat{C}^k_o$.) To this end, we can make use of the fact that type II moves are the time reverse of type I such that in analogy to (\ref{kupI})
\ba
K_{k\rightarrow f}=K_{k+1\rightarrow f}(x_{k+1},x_f)\,M^{II}_{k\rightarrow k+1}\,e^{iS_{k+1}(x_{k}^e,x_{k}^o)/\hbar}.\nn
\ea
Using similar arguments to the type I move above, one finds ${}^-\hat{C}^k_o\,(K_{k\rightarrow f}(x_k,x_f))^*=0$ to be true, provided that $M_{k\rightarrow k+1}(x^e_k)$ does {\it not} depend on $x^o_k$.

We must check whether the move $k\rightarrow k+1$ preserves the other quantum post--constraints which already existed at step $k$. In the classical theory any post--constraint at $k$ is preserved and transformed into a post--constraint at step $k+1$ as follows (see the appendix of \cite{Dittrich:2013jaa})
\ba
{}^+C^k\left(x^b_k,x^e_k,p^k_b,p_e^k +\frac{\p S_{k+1}}{\p x^e_k}\right)\Big|_{p^k={}^+p^k}=\fh^*_k\,{}^+C^{k+1}(x^b_{k+1},x^e_{k+1},p^{k+1}_b,p_e^{k+1})\Big|_{p^{k+1}={}^+p^{k+1}}=0.\nn
\ea
In the quantum theory this means for constraints which admit a power series expansion
\ba
0={}^+\hat{C}^k \left(x^b_k,x^e_k,\hat{p}^k_b,\hat{p}_e^k +\frac{\p S_{k+1}}{\p x^e_k}\right)\,{}^+\psi^{\rm phys}_k=e^{-iS_{k+1}/\hbar}\,{}^+\hat{C}^k (x^b_k,x^e_k,\hat{p}^k_b,\hat{p}_e^k)\,e^{iS_{k+1}/\hbar}\,{}^+\psi^{\rm phys}_k\nn
\ea
Hence, by analogous arguments to the type I move, the type II move preserves all quantum post--constraints such that
\ba
{}^+\hat{C}^{k+1}(x^b_{k+1},x^e_{k+1},\hat{p}^{k+1}_b,\hat{p}^{k+1}_e){}^+\psi^{\rm phys}_{k+1}={}^+\hat{C}^{k+1}(x^b_{k+1},x^e_{k+1},\hat{p}^{k+1}_b,\hat{p}^{k+1}_e)\,\fu^{II}_{k\rightarrow k+1}({}^+\psi^{\rm phys}_{k})=0\nn
\ea
if $M_{k\rightarrow k+1}$ commutes with ${}^+\hat{C}^k$. (The ${}^+\hat{C}^k$ commute with $\prod_\alpha\delta(x'^\alpha_k-x^\alpha_k)$.) This is the case if $M_{k\rightarrow k+1}$ is constant. 

As regards (non--)unitarity of type II moves, one finds
\ba
\langle{}^+\psi^{\rm phys}_{k+1}|{}^+\phi^{\rm phys}_{k+1}\rangle_{\rm phys+}&\underset{(\ref{supII})}{=}&(2\pi\hbar)^{K_\alpha}\int\prod_{e,b}dx^e_{k+1}\,dx^b_{k+1}\,(\psi^{\rm kin}_{k+1}(x^e_{k+1},x^b_{k+1}))^*\nn\\
&&\times\int \prod_odx^o_k\,M^{II}_{k\rightarrow k+1}\,e^{iS_{k+1}/\hbar}\,\prod_\alpha\delta(x'^\alpha_k-x^\alpha_k)\,{}^+\phi^{\rm phys}_{k}(x_k)\nn\\
&=&(2\pi\hbar)^{K_\alpha}\int\prod_{o,e,b}dx^o_k\,dx^e_k\,dx^b_k\,\left(\mathbb{P}^A_k\,{}^-\mathbb{P}^B_k\,\psi^{\rm kin}_{k}(x^o_k,x^e_{k},x^b_{k})\right)^*\nn\\
&&\q\q\q\q\times\prod_\alpha\delta(x'^\alpha_k-x^\alpha_k)\,{}^+\phi^{\rm phys}_k(x_k)\nn\\
&=&(2\pi\hbar)^{K_\alpha}\int\prod_{o,e,b}dx^o_k\,dx^e_k\,dx^b_k\left(\psi^{\rm kin}_{k}(x^o_k,x^e_{k},x^b_{k})\right)^*\nn\\
&&\q\q\q\q\times{}^-\mathbb{P}^B_k\,\mathbb{P}^A_k\,\prod_\alpha\delta(x'^\alpha_k-x^\alpha_k)\,{}^+\phi^{\rm phys}_k(x_k)\nn\\
&\underset{(\ref{lem1})}{=}&\int\prod_{o,e,b}dx^o_k\,dx^e_k\,dx^b_k\left(\psi^{\rm kin}_{k}(x^o_k,x^e_{k},x^b_{k})\right)^*\,{}^-\mathbb{P}^B_k\,{}^+\phi^{\rm phys}_k(x_k)\nn\\
&=&\langle{}^+\psi^{\rm phys}_k|{}^-\mathbb{P}^B_k|\,{}^+\phi^{\rm phys}_k\rangle_{\rm phys+}\nn\\
&\neq&\langle{}^+\psi^{\rm phys}_k|{}^+\phi^{\rm phys}_k\rangle_{\rm phys+}.\label{nuII}
\ea
In line two we have made use of the identification (and the two left equations in (\ref{case2}, \ref{case2b}))
\ba
\mathbb{P}^A_k\,{}^-\mathbb{P}^B_k\psi^{\rm kin}_k(x^e_k,x^b_k,x^o_k):=(M^{II}_{k\rightarrow k+1})^*e^{-iS_{k+1}/\hbar}\,\psi^{\rm kin}_{k+1}(x^e_{k},x^b_{k})\nn
\ea
for a suitable kinematical state $\psi^{\rm kin}_k$. Indeed, the right hand side of the last expression is annihilated by both the $\hat{C}^k_\alpha$ and ${}^-\hat{C}^k_\beta$. In line three we have made use of the self-adjointness of the constraints (and the corresponding projectors) with respect to the kinematical inner product. In line four we have made use of the fact that ${}^+\phi^{\rm phys}_k$ contains a projector $\mathbb{P}^A_k$ and of (\ref{lem1}).

In conclusion, the map $\fu^{II}_{k\rightarrow k+1}$ is non-unitary in the presence of at least one ${}^-\hat{C}^k_\beta$ and unitary otherwise.
\end{proof}

Finally, the proof of theorem \ref{thm_III}:

\begin{proof}
Clearly, the quantum constraints $\hat{C}^k_n=\hat{p}^k_n$ and $\hat{C}^{k+1}_o=\hat{p}^{k+1}_o$ are automatically implemented because nothing depends on the auxiliary gauge variables $x^k_n,x^o_{k+1}$. Next, let us consider the $\kappa$ new post--constraints ${}^+C^{k+1}_\nu(x^e_{k+1},x^n_{k+1},p^{k+1}_n)$ and the $\kappa$ new pre--constraints ${}^-C^k_\nu(x^e_k,x^o_k,p^k_o)$ generated by the move. 
%
We begin with implementing the $\kappa$ quantum post--constraints ${}^+\hat{C}^{k+1}_\nu$. Assuming the latter admit a power series expansion in the canonical variables, we can utilize
\ba
e^{-iS_{k+1}/\hbar}\,{}^+\hat{C}^{k+1}_\nu (x^e_{k+1},x^n_{k+1},\hat{p}^{k+1}_n)\,e^{iS_{k+1}/\hbar}={}^+\hat{C}^{k+1}_\nu\left(x^e_{k+1},x^n_{k+1},\hat{p}^{k+1}_n+\f{\p S_{k+1}}{\p x^n_{k+1}}\right).\nn
\ea
This gives
\ba
{}^+\hat{C}^{k+1}_\nu\,{}^+\psi^{\rm phys}_{k+1}&=&(2\pi)^{\kappa_\alpha}\int\prod_odx^o_k {}^+\hat{C}^{k+1}_\nu (x^e_{k+1},x^n_{k+1},\hat{p}^{k+1}_n)\,M^{III}_{k\rightarrow k+1}\,e^{iS_{k+1}/\hbar}\nn\\
&&\q\q\q\q\q\q\q\times|\det[G^k_\alpha,\hat{C}^k_{\alpha'}]|\,\prod_\alpha\delta(G^k_\alpha)\,{}^+\psi^{\rm phys}_k(x_k)\nn
\ea
\ba
&=&(2\pi)^{\kappa_\alpha}\int\prod_odx^o_k\,|\det[G^k_\alpha,\hat{C}^k_{\alpha'}]|\,\prod_\alpha\delta(G^k_\alpha)\,e^{iS_{k+1}/\hbar}\nn\\
&&\q\q\q\q\q\times{}^+\hat{C}^{k+1}_\nu\left(x^e_{k+1},x^n_{k+1},\hat{p}^{k+1}_n+\f{\p S_{k+1}}{\p x^n_{k+1}}\right)\,M^{III}_{k\rightarrow k+1}\,{}^+\psi^{\rm phys}_k(x_k).\nn
\ea 
(The gauge fixing conditions commute with the ${}^+\hat{C}^{k}_\nu$.) Since ${}^+\psi^{\rm phys}_k$ does not depend on $x^n_{k+1}$, we need
\ba
{}^+\hat{C}^{k+1}_\nu\left(x^e_{k+1},x^n_{k+1},\hat{p}^{k+1}_n+\f{\p S_{k+1}}{\p x^n_{k+1}}\right)\,M^{III}_{k\rightarrow k+1}=0.\nn
\ea

Next, we need to ensure that the $\kappa$ quantum pre--constraints ${}^-\hat{C}^k_\nu$ annihilate the future propagator in the form (\ref{cond}):
\ba
{}^-\hat{C}^k_\nu\,(K_{k\rightarrow f})^*=\int\prod_ndx^n_{k+1}{}^-\hat{C}^k_\nu(x^e_k,x^o_k,\hat{p}^k_o)\,\left(M^{III}_{k\rightarrow k+1}\,e^{iS_{k+1}/\hbar}\,K_{k+1\rightarrow f}\right)^*\nn
\ea
Assuming the pre--constraints to admit a power series expansion, this is true if 
\ba
{}^-\hat{C}^k_\nu\left(x^e_k,x^o_k,\hat{p}^k_o-\f{\p S_{k+1}}{\p x^o_k}\right)\,(M^{III}_{k\rightarrow k+1})^*=0.\label{preIII}
\ea

Let us now check that the quantum type III move preserves all already existing post--constraints ${}^+\hat{C}^k$ at step $k$. Classically, these can be written in the form 
\ba
{}^+C^k\left(x^b_k,x^e_k,x^n_{k+1},p^k_b,p^k_e+\f{\p S_{k+1}}{\p x^e_k},\f{\p S_{k+1}}{\p x^n_{k+1}}\right)\Big|_{p^k={}^+p^k}=0\nn
\ea
and are preserved by the move (see the appendix in \cite{Dittrich:2013jaa}). Thus, suppose,
\ba
0&=&{}^+\hat{C}^k\left(x^b_k,x^e_k,x^n_{k+1},\hat{p}^k_b,\hat{p}^k_e+\f{\p S_{k+1}}{\p x^e_k},\f{\p S_{k+1}}{\p x^n_{k+1}}\right)\,{}^+\psi^{\rm phys}_k\label{IIIcon2}\\
&=&e^{-iS_{k+1}/\hbar}\,{}^+\hat{C}^{k+1}\left(x^b_{k+1},x^e_{k+1},x^n_{k+1},\hat{p}^{k+1}_b,\hat{p}^{k+1}_e,\hat{p}^{k+1}_n\right)\,e^{iS_{k+1}/\hbar}\,{}^+\psi^{\rm phys}_k,\nn
\ea
where we assume ${}^+\hat{C}^k$ to admit a power series expansion. Then, if $M^{III}_{k\rightarrow k+1}$ commutes with all ${}^+\hat{C}^{k+1}$, all quantum post--constraints are preserved by the quantum type III move
\ba
{}^+\hat{C}^{k+1}{}^+\psi^{\rm phys}_{k+1}&=&(2\pi)^{\kappa_\alpha}\int\prod_odx^o_k\,|\det[G^k_\alpha,\hat{C}^k_{\alpha'}]|\,\prod_\alpha\delta(G^k_\alpha)\,M^{III}_{k\rightarrow k+1}\nn\\
&&\q\q\q\q\times{}^+\hat{C}^{k+1}\!\!\left(x^b_{k+1},x^e_{k+1},x^n_{k+1},\hat{p}^{k+1}_b,\hat{p}^{k+1}_e,\hat{p}^{k+1}_n\right)\!e^{iS_{k+1}/\hbar}\,{}^+\psi^{\rm phys}_k\!=0\nn
\ea
because the gauge fixing conditions commute with ${}^+\hat{C}^{k+1}$.

Finally, we examine unitarity of $\fu^{III}_{k\rightarrow k+1}$,
\ba
\langle{}^+\psi^{\rm phys}_{k+1}|{}^+\phi^{\rm phys}_{k+1}\rangle_{\rm phys+}&=&(2\pi)^{\kappa_\alpha}\int\prod_{e,b,n}dx^e_{k+1}\,dx^b_{k+1}\,dx^n_{k+1}\,(\psi^{\rm kin}_{k+1}(x^e_{k+1},x^b_{k+1},x^n_{k+1}))^*\nn\\
&&\times\int \prod_odx^o_k\,M_{k\rightarrow k+1}\,e^{iS_{k+1}/\hbar}\,|\det[G^k_\alpha,\hat{C}^k_{\alpha'}]|\,\prod_\alpha\delta(G^k_\alpha)\,{}^+\phi^{\rm phys}_{k}(x_k)\nn\\
&=&(2\pi)^{\kappa_\alpha}\int\prod_{o,e,b}dx^o_k\,dx^e_k\,dx^b_k\left(\mathbb{P}^A_k\,{}^-\mathbb{P}^B_k\,\psi^{\rm kin}_{k}(x^o_k,x^e_{k},x^b_{k})\right)^*\nn\\
&&\q\q\q\q\times|\det[G^k_\alpha,\hat{C}^k_{\alpha'}]|\,\prod_\alpha\delta(G^k_\alpha)\,{}^+\phi^{\rm phys}_k(x_k)\nn\\
&\underset{(\ref{lem1III})}{=}&\int\prod_{o,e,b}dx^o_k\,dx^e_k\,dx^b_k\left(\psi^{\rm kin}_{k}(x^o_k,x^e_{k},x^b_{k})\right)^*{}^-\mathbb{P}^B_k\,{}^+\phi^{\rm phys}_k(x_k)\nn\\
&=&\langle{}^+\psi^{\rm phys}_k|{}^-\mathbb{P}^B_k|\,{}^+\phi^{\rm phys}_k\rangle_{\rm phys+}\neq\langle{}^+\psi^{\rm phys}_k|{}^+\phi^{\rm phys}_k\rangle_{\rm phys+},\nn
\ea
where in line two we have made use of the identification (and the left equations in (\ref{anh21}, \ref{anh21b}))
\ba
\mathbb{P}^A_k\,{}^-\mathbb{P}^B_k\,\psi^{\rm kin}_{k}(x^o_k,x^e_{k},x^b_{k}):=\int\prod_ndx^n_{k+1}\,(M^{III}_{k\rightarrow k+1})^*\,e^{-iS_{k+1}/\hbar}\,\psi^{\rm kin}_{k+1}(x^e_{k+1},x^b_{k+1},x^n_{k+1})\nn
\ea
for a suitable kinematical state $\psi^{\rm kin}_k$. Indeed, thanks to (\ref{preIII}), the right hand side of the last equation is annihilated by all pre--constraints $\hat{C}^k_\alpha,{}^-\hat{C}^k_\beta$. Hence, $\fu^{III}_{k\rightarrow k+1}$ is non-unitary if at least one ${}^-\hat{C}^k_\beta$ occurs and unitary otherwise.
\end{proof}

\section{Proofs of lemmas \ref{lema}--\ref{lemc}}\label{app_2}

We begin with the proof of lemma \ref{lema}:

\begin{proof}
All $K$ pre--constraints ${}^-\hat{C}^{k+1}_n$ of the type II move coincide with the $K$ post--constraints ${}^+\hat{C}^{k+1}_n$ of the type I move. From theorems \ref{thm_I} and \ref{thm_II} it follows that 
\ba
\fu^{II}_{k+1\rightarrow k}\cdot\fu^I_{k\rightarrow k+1}&=&(2\pi\hbar)^K\int \prod_ndx^n_{k+1}\delta(x'^n_{k+1}-x^n_{k+1})(M^{II}_{k\rightarrow k+1})^*e^{-iS_{k+1}/\hbar}\,M^I_{k\rightarrow k+1}\,e^{iS_{k+1}/\hbar}\nn\\
&=&(2\pi\hbar)^K(M^{II}_{k\rightarrow k+1})^*\,M^I_{k\rightarrow k+1}\nn
\ea
Since the type II move just reverses the type I move, its measure factor can be chosen identical to the measure factor of the type I move. This yields the stated result (up to phase).
\end{proof}

We continue with the proof of lemma \ref{lemb}:

\begin{proof}
Using the expressions in theorems \ref{thm_I} and \ref{thm_II}, one finds
\ba
\fu^{I}_{k+1 \rightarrow k}\cdot\fu^{II}_{k\rightarrow k+1}{}^+\psi^{\rm phys}_k&=&\left(\f{1}{2\pi\hbar}\right)^Ke^{-iS_{k+1}(x'^o_{k},x^e_k)/\hbar}\,(2\pi\hbar)^{K_\alpha}\int\prod_odx^o_k\,e^{iS_{k+1}(x^o_k,x^e_k)/\hbar}\nn\\
&&\q\q\q\q\q\times\prod_\alpha\delta(x'^\alpha_k-x^\alpha_k)\,{}^+\psi^{\rm phys}_k(x^o_k,x^e_k).\nn
\ea
Comparing this with 
\ba
&&(2\pi\hbar)^{K_\alpha}\,\mathbb{P}^A_k\,{}^-\mathbb{P}^B_k\,\prod_\alpha\delta(x'^\alpha_k-x^\alpha_k)\,{}^+\psi^{\rm phys}_k\nn\\
&=&\!\!\!\left(\f{1}{2\pi\hbar}\right)^K\!\!\!(2\pi\hbar)^{K_\alpha}\,e^{-iS_{k+1}(x^o_{k},x^e_k)/\hbar}\int\prod_ods^o\,e^{is^o\hat{p}^k_o} e^{iS_{k+1}(x^o_k,x^e_k)/\hbar}\,\prod_\alpha\delta(x'^\alpha_k-x^\alpha_k)\,{}^+\psi^{\rm phys}_k(x^o_k,x^e_k)\nn\\
&=&\left(\f{1}{2\pi\hbar}\right)^K(2\pi\hbar)^{K_\alpha}e^{-iS_{k+1}(x^o_{k},x^e_k)/\hbar}\int\prod_ods^o\,e^{iS_{k+1}(x^o_k+s^o,x^e_k)/\hbar}\,\prod_\alpha\delta(x'^\alpha_k-x^\alpha_k-s^\alpha)\nn\\
&&\q\q\q\q\q\q\q\q\q\q\q\times{}^+\psi^{\rm phys}_k(x^o_k+s^o,x^e_k)\nn\\
&=&\left(\f{1}{2\pi\hbar}\right)^K(2\pi\hbar)^{K_\alpha}e^{-iS_{k+1}(x^o_{k},x^e_k)/\hbar}\int\prod_ody^o\,e^{iS_{k+1}(y^o_k,x^e_k)/\hbar}\,\prod_\alpha\delta(x'^\alpha_k-y^\alpha_k)\,{}^+\psi^{\rm phys}_k(y^o_k,x^e_k)\nn
\ea
and using that ${}^+\psi^{\rm phys}_k$ contains a projector $\mathbb{P}^A_k$ and (\ref{lem1}) gives the claimed result.
\end{proof}

Lastly, the proof of lemma \ref{lemc}:

\begin{proof}
Using the expression in theorem \ref{thm_III} and noting that all $\kappa$ post--constraints of the type III move $k\rightarrow k+1$ are also pre--constraints for the reverse type III move $k+1\rightarrow k$ yields
\ba
&&\!\!\!\!\!\!\!\!\!\!\fu^{III}_{k+1 \rightarrow k}\cdot\fu^{III}_{k\rightarrow k+1}\,{}^+\psi^{\rm phys}_k\label{IIIcon3}\\
&\!\!\!\!=&\!\!\!\!(2\pi)^\kappa\!\!\!\int\prod_ndx^n_{k+1}|\det[G^{k+1}_\nu,{}^+\hat{C}^{k+1}_{\nu'}]|\,\prod_\nu\delta(G^{k+1}_\nu)\,(M^{III}_{k\rightarrow k+1}(x'^o_k,x^e_k,x^n_{k+1}))^*\,e^{-iS_{k+1}(x'^o_k,x^e_k,x^n_{k+1})/\hbar}\nn\\
&&\times\int\prod_odx^o_k\,M_{k\rightarrow k+1}^{III}(x^o_k,x^e_k,x^n_{k+1})\,e^{iS_{k+1}(x^o_k,x^e_k,x^n_{k+1})/\hbar}\,(2\pi)^{\kappa_\alpha}|\det[G^k_\alpha,\hat{C}^k_{\alpha'}]|\prod_\alpha\delta(G^k_\alpha)\,{}^+\psi^{\rm phys}_k.\nn
\ea
where $G^{k+1}_\nu(x^n_{k+1})$ are $\kappa$ gauge fixing conditions at $k+1$.

It follows from (\ref{preIII}) that the right hand side is, indeed, annihilated by ${}^-\hat{C}^k_\nu$, $\nu=1,\ldots,\kappa$. 
The right hand side and, in particular, the expression on the right hand side of (\ref{IIIcon}) contained in it thus includes both projectors $\mathbb{P}^A_k$ and ${}^-\mathbb{P}^B_k$ which commute in the case that the pre--constraints are linear in the momenta. 

However, we have also to ensure that (\ref{IIIcon3}) is still annihilated by all other post--constraints (\ref{IIIcon2}) of step $k$. In fact, using the expressions in both lines of (\ref{IIIcon2}), one can convince oneself that one can pull any other post--constraint ${}^+\hat{C}^k$ applied from the left through the integrals up to ${}^+\psi^{\rm phys}_k$ because the gauge conditions $G_\alpha(x^o_k)$ and the measure factor $M^{III}_{k\rightarrow k+1}$ commute with the post--constraints (\ref{IIIcon2}). We conclude that (\ref{IIIcon3}) is annihilated by all post--constraints at $k$ and all pre--constraints ${}^-\hat{C}^k_\nu$.

It follows that $\fu^{III}_{k+1 \rightarrow k}\cdot\fu^{III}_{k\rightarrow k+1}\,{}^+\psi^{\rm phys}_k={}^-\mathbb{P}^B_k\,{}^+f_k({}^+\psi^{\rm phys}_k)$, where ${}^+f_k:{}^+\ch^{\rm phys}_k\rightarrow{}^+\ch^{\rm phys}_k$ is some map from the post--physical Hilbert space at $k$ into itself. We are then free to require (\ref{IIIcon}) which is a condition on $M^{III}_{k\rightarrow k+1}$ and equivalent to ${}^+f_k$ being the identity. In this case, (\ref{lem1III}) gives the desired statement.
\end{proof}

\section*{Acknowledgements}
The author thanks Bianca Dittrich for discussion. Research at Perimeter Institute is supported by the Government of Canada through Industry Canada and by the Province of Ontario through the Ministry of Research and Innovation.

\bibliography{bibliography}{}
\bibliographystyle{utphys}

\end{document}